\documentclass[draftclsnofoot,onecolumn,11pt]{IEEEtran}
\usepackage{ifthen}
\newboolean{short_version}
\setboolean{short_version}{false} % false for full version/onecolumn version
\IEEEoverridecommandlockouts

\def\Figs{./figs/} % call figures (eps files) if needed
\usepackage[cmex10]{amsmath} % Use the [cmex10] option 
\usepackage[url,hyperrefblack,notheorems,IEEEtran]{research17} % customized LaTeX commands
\usepackage{lipsum} % print dummy text
\usepackage{balance} % balance the text in the final page under twocolumn formatting
\usepackage{amsbsy}
\usepackage{booktabs} % required for showing good \hline in matrix form --> \midrule
\usepackage{enumitem}
\usepackage{mathtools}
\usepackage{amsmath,amssymb,amsfonts,amsthm} % amsthm: typesetting theorems (AMS style)
\usepackage[lined,boxed,commentsnumbered,linesnumbered,ruled]{algorithm2e}
\usepackage{comment}
\usepackage{nicefrac} % displaying a nice fraction in in-line math
\usepackage{longtable}
\newtheorem{theorem}{\mytheoremname}
\newtheorem{lemma}[theorem]{\mylemmaname}
\newtheorem{corollary}[theorem]{\mycorollaryname}
\newtheorem{conjecture}[theorem]{\myconjecturename}
\newtheorem{proposition}[theorem]{\mypropositionname}

\newtheorem{definition}[theorem]{\mydefinitionname}
\newtheorem{remark}[theorem]{\myremarkname}
\newtheorem{example}[theorem]{\myexamplename}
\usepackage{array}
\usepackage{multirow}
\newcolumntype{C}[1]{>{\centering\arraybackslash}p{#1}}
\usepackage{tikz}
\usetikzlibrary{calc,shapes, patterns,decorations.text, decorations.pathreplacing}
\usepackage{afterpage} % execute command after the next page break
\usepackage{lscape} % for landscape table
\usepackage{pdflscape} %  attribute /Rotate for the landscape table
\usepackage{rotating} % rotate a long table
\usepackage{pbox} % for pbox{}
\usepackage[caption=false, font=footnotesize, subrefformat=parens, labelformat=parens]{subfig}
\allowdisplaybreaks

\newcommand{\Hdist}[2]{\dH\left(#1,#2\right)} % the (Hamming) distance function
\newcommand{\Hwt}[1]{\wH\left(#1\right)} % the (Hamming) weight function
\newcommand{\ConstrA}[1]{\Lambda_\textnormal{A}\left(#1\right)} % the lattice obtained from Construction A
\newcommand{\eConstrA}[1]{\Lambda_\textnormal{A}(#1)} 
\newcommand{\bigConstrA}[1]{\Lambda_\textnormal{A}\bigl(#1\bigr)} 
\newcommand{\GammaA}[1]{\Gamma_\textnormal{A}\left(#1\right)} % the nonlattice packing obtained from Construction A
\newcommand{\eGammaA}[1]{\Gamma_\textnormal{A}(#1)} 
 
\newcommand{\we}[1]{W_{#1}} % the weight enumerator of a code
\renewcommand{\Rationals}{\mathbb{Q}} % re-define the notation for reals
\renewcommand{\Reals}{\mathbb{R}} % re-define the notation for reals

\newcommand*{\Scale}[2][4]{\scalebox{#1}{\ensuremath{#2}}} % Scales an environment simply by a fraction
\makeatletter
\renewcommand*\env@matrix[1][*\c@MaxMatrixCols c]{%
  \hskip -\arraycolsep
  \let\@ifnextchar\new@ifnextchar
  \array{#1}}
\makeatother

\usepackage{todonotes}
\newcommand{\m}{\color{magenta}} % color magenta for reviewer 1 of 2nd round or our modifications
\begin{document}

\title{Formally Unimodular Packings\\ for the Gaussian Wiretap Channel}
% \title{The Secrecy Gain of Formally Unimodular Lattices\\ on the Gaussian Wiretap Channel} % IZS title

\author{Maiara F.~Bollauf, Hsuan-Yin Lin, and {\O}yvind Ytrehus
% \IEEEauthorblockA{Simula UiB, N--5006 Bergen, Norway\\             
%             Emails: \{maiara, lin, oyvindy\}@simula.no}}
%\author{IEEE Publication Technology,~\IEEEmembership{Staff,~IEEE,}
        % <-this % stops a space
\thanks{All authors are affiliated to Simula UiB, N--5006 Bergen, Norway. Their respective e-mail addresses are maiara@simula.no, lin@simula.no, and oyvindy@simula.no.}% <-this % stops a space
\thanks{This paper was published partially in the proceedings of the International Zurich Seminar on Information and Communication (IZS), Zurich, Switzerland, March, 2022~\cite{BollaufLinYtrehus22_1}.}}%

\maketitle

\begin{abstract}
  This paper introduces the family of \emph{lattice-like packings}, which generalizes \emph{lattices}, consisting of packings possessing \emph{periodicity} and \emph{geometric uniformity}. The subfamily of \emph{formally unimodular (lattice-like) packings} is further investigated. It can be seen as a generalization of the \emph{unimodular} and \emph{isodual lattices}, and the Construction A formally unimodular packings obtained from \emph{formally self-dual codes} are presented. Recently, lattice coding for the Gaussian wiretap channel has been considered. A measure called \emph{secrecy function} was proposed to characterize the eavesdropper's probability of correctly decoding. The aim is to determine the global maximum value of the secrecy function, called \emph{(strong) secrecy gain}.

  We further apply lattice-like packings to coset coding for the Gaussian wiretap channel and show that the family of formally unimodular packings shares the same secrecy function behavior as unimodular and isodual lattices. We propose a universal approach to determine the secrecy gain of a Construction A formally unimodular packing obtained from a formally self-dual code. From the weight distribution of a code, we provide a necessary condition for a formally self-dual code such that its Construction A formally unimodular packing is secrecy-optimal. Finally, we demonstrate that formally unimodular packings/lattices can achieve higher secrecy gain than the best-known unimodular lattices.
\end{abstract}

\begin{IEEEkeywords}
  Lattices, nonlattice packings, Construction~A, secrecy gain, Gaussian wiretap channel.
\end{IEEEkeywords}

\section{Introduction}
\label{sec:introduction}

\IEEEPARstart{I}{n} recent years, \emph{physical layer security}, that only utilizes the resources at the physical layers of the transmission parties and provides \emph{information-theoretically unbreakable security}, has been recognized as an appealing technique for safeguarding confidential data in 5G and beyond 5G wireless communication systems (see~\cite{WuKhisti-etal18_1,Bloch-etal21_1,CostaOggierCampelloBelfioreViterbo17_1} and references therein). The root of this line of research goes back to the epochal work \emph{wiretap channel (WTC) model} introduced by Aaron Wyner~\cite{Wyner75_1}, which showed that reliable and secure communication can be achieved simultaneously without the need of an additional cryptographic layer on top of the communication protocol.

Since then, substantial research efforts have been devoted to developing practical codes for reliable and secure data transmission over WTCs. Potential candidates for practical wiretap code constructions include low density parity check (LDPC) codes~\cite{ThangarajDihidarCalderbankMcLaughlinMerolla07_1}, polar codes~\cite{AnderssonRathiThobabenKliewerSkoglund10_1,MahdavifarVardy11_1}, and \emph{lattices}~\cite{BelfioreOggier10_1,OggierSoleBelfiore16_1,LingLuzziBelfioreStehle14_1}. In \cite{BelfioreOggier10_1,OggierSoleBelfiore16_1} it was shown that a lattice-based coset coding approach can provide secure and reliable communication over the Gaussian WTC. In particular, it was shown that for Gaussian WTC, the so-called~\emph{secrecy function} expressed in terms of the \emph{theta series} of a lattice (see the precise definition in Section~\ref{sec:secrecy-function_GU-packings}) can be considered as a quality criterion of good wiretap lattice codes: to minimize the eavesdropper's probability of correct decoding, one needs to maximize the secrecy function, and the corresponding maximum value is referred to as \emph{(strong) secrecy gain}. Moreover, Ling \emph{et al.} have also proposed another design criterion for wiretap lattice codes, called \emph{flatness factor}, to quantify how much confidential information can leak to Eve in terms of mutual information~\cite{LingLuzziBelfioreStehle14_1}. To guarantee secrecy-goodness, both the criteria of secrecy gain and the flatness factor require small theta series of the designed Eve's lattice $\Lambda_\textnormal{e}$ at a particular point.

In~\cite{BelfioreSole10_1}, Belfiore and Sol{\'{e}} discovered that the secrecy functions of \emph{unimodular} lattices have a symmetry point. The value of the secrecy function at this point is called the \emph{weak secrecy gain}. Based on this, the authors conjectured that for unimodular lattices, the secrecy gain is achieved at the symmetry point of its secrecy function. I.e., the strong secrecy gain of a unimodular lattice is equivalent to its weak secrecy gain. Finding good unimodular lattices that attain large secrecy gain is of practical importance. The Belfiore and Sol{\'{e}} conjecture has also been extended to \emph{isodual lattices} in~\cite{OggierSoleBelfiore16_1}, which  includes unimodular lattices as a special case. In~\cite{Ernvall-Hytonen12_1}, a novel technique was proposed to verify or disprove the Belfiore and Sol{\'{e}} conjecture for a given unimodular lattice. Using this method, the conjecture is validated for all known \emph{extremal unimodular} lattices in even dimensions less than $80$. In another work~\cite{LinOggier13_1}, the authors use a similar method as~\cite{Ernvall-Hytonen12_1} to classify the best unimodular lattices in dimensions from $8$ to $23$. For unimodular lattices obtained by Construction~A from binary \emph{doubly even self-dual} codes (the so-called \emph{type II} codes, where their weights of all the codewords are multiple of four) up to dimensions $40$, their secrecy gains are also shown to be achieved at their symmetry points~\cite{Pinchak13_1}. Recently, the analysis of secrecy gain has been extended to the family of \emph{$\ell$-modular lattices}~\cite{HouLinOggier14_1,LinOggierSole15_1,OggierBelfiore18_1}, which is equivalent to the isodual lattices for the case of $\ell=1$. It is expected that one can achieve a better secrecy gain by using an $\ell$-modular lattice with a higher parameter $\ell$.

In coding theory, the notion of \emph{formally self-dual codes} apply for either linear or nonlinear codes,\footnote{In our context, a \emph{formally-self dual} code has the weight enumerator satisfying \eqref{eq:MacWilliams-identity-fsd} in Section~\ref{sec:definitions-preliminaries}.} and it is known that this broader class of codes possesses several better features than isodual and self-dual linear codes. For instance, the Nordstrom-Robinson code is formally self-dual~\cite[Ch.~19]{MacWilliamsSloane77_1}, and it has a larger minimum Hamming distance than any isodual codes of length $16$. In our earlier work~\cite{BollaufLinYtrehus22_1}, a new and wider family of lattices, referred to as \emph{formally unimodular lattices}, that consists of lattices having the same theta series as their dual, has been introduced. It was shown that formally unimodular lattices have the same symmetry point as unimodular and isodual lattices. In this work, we further define a larger family of packings (either lattice or nonlattice packings), called \emph{formally unimodular packings}, that contains formally unimodular lattices and also enjoys the same secrecy function properties as unimodular and isodual lattices. Our main contributions are summarized as follows:
\begin{itemize}
\item Due to the assumptions of \emph{periodicity} and \emph{geometric uniformity} for general packings (not necessarily lattices), we introduce the concept of \emph{lattice-like packings}, where their volumes and theta series are well defined, and they have congruent Voronoi regions.
  
\item We show how to construct the formally unimodular packings via Construction A from geometrically uniform formally self-dual codes that contain the all-zero vector (see Theorem~\ref{thm:Funimodular_FSDcode}). Moreover, we briefly discuss how to obtain Construction A lattice-like packings from codes through Gray map.
  
\item The coset coding scheme, originally for two nested lattices $\Lambda_{\textnormal{e}}\subset\Lambda_{\textnormal{b}}$, is generalized to the setup where $\Lambda_\textnormal{b}$ is a lattice and its subset $\Gamma_\textnormal{e}$ is a lattice-like packing. The challenge to construct such coset decomposition for this setup is briefly discussed. Furthermore, we derive analytical results on the eavesdropper's probability of correct decision.  Under certain assumption of our proposed lattice-like packings, we show that the performance of secrecy gain can be determined by the theta series of $\Gamma_\textnormal{e}$ (see Theorem~\ref{thm:2nd-bound_Eve-success-probability}). Construction~A lattice-like packings do satisfy this particular assumption (see Remark~\ref{rem:ConstrA-packings_2nd-bound_Eve-Pc}). Hence, the secrecy gain criterion keeps unchanged even we consider Construction~A lattice-like packings for Eve.

\item We show that the formally unimodular packings possess the same secrecy function behavior as the isodual lattices (see Theorem~\ref{thm:weak_secrecy}). I.e., both formally unimodular packings and isodual lattices have the symmetry point at $1$.

\item A universal approach to determine the strong secrecy gain of formally unimodular packings is provided (see Theorem~\ref{thm:inv_secrecy-function_WeightEnumerator}). For Construction A formally unimodular packings obtained from \emph{even} formally self-dual codes, we also provide a sufficient condition to verify the Belfiore and Sol{\'{e}} conjecture on the secrecy gain (see Theorem~\ref{thm:strong-secrecy-gain_unimodular-lattices}).\footnote{A code is called \emph{even} if all of its codewords have even weight, otherwise the code is \emph{odd}.}
  
\item By using the weight distribution of a code, we provide a new necessary condition to verify the secrecy-optimality of a Construction A formally unimodular packing obtained from a formally self-dual code (see Theorem~\ref{thm:necessary-condition_optimal-FSDcode}). Naturally, one would anticipate that a formally self-dual code with a large minimum Hamming distance and a low number of low-weight codewords should lead to high secrecy gain of the corresponding Construction A formally unimodular packing. However, we present two counterexamples to disprove this observation.

\item To have a more thorough secrecy performance comparison between Construction A formally unimodular packings obtained from formally self-dual codes, we provide a systematic approach through tail-biting the rate $\nicefrac{1}{2}$ binary convolutional codes to construct isodual codes (see Section~\ref{sec:isodual_coco}).
\end{itemize}

Finally, numerical results are presented to validate our theoretical findings for strong secrecy gain of Construction A packings/lattices obtained from formally self-dual codes. For dimensions up to $70$, we note that formally unimodular packings/lattices have better secrecy gain than the best-known unimodular lattices described in the literature, e.g.,~\cite{LinOggier13_1}. In addition to applying good formally self-dual codes from the literature, we also demonstrate high secrecy gains of the newly presented isodual codes by tail-biting the rate $\nicefrac{1}{2}$ convolutional codes.

%%%%%%%%%%%%%%%%%%%%%%%%%%%%%%%%%%%%%%%%%%%%%%%%%%%%%%%%%%%%%%%%%%%%%%%%%%%%%%%%%%%%%%%%%%%%%%%%%%%%%%%%%%%%%%%%%%%%%%%%% 
\section{Definitions and Preliminaries}
\label{sec:definitions-preliminaries}

\subsection{Notation}
\label{sec:notation}

We denote by $\Integers$, $\Rationals$, and $\Reals$ the set of integers, rationals, and reals, respectively. Moreover, $\Integers_{\geq 0}$ denote the nonnegative integers, and $[a:b]\eqdef\{a,a+1,\ldots,b\}$ for $a,b\in \Integers$, $a \leq b$. A binary field is denoted by $\Field_2\eqdef\{0,1\}$. Vectors are boldfaced, e.g., $\vect{x}$. $\inner{\vect{x}}{\vect{y}}$ denotes the inner product between vectors $\vect{x}$ and $\vect{y}$ over $\Field_2$ or $\Reals$. $\oplus$ represents the element-wise addition over $\Field_2$. Matrices and sets are represented by capital sans serif letters and calligraphic uppercase letters, respectively, e.g., $\mat{X}$ and $\set{X}$. The Hamming, the Lee, and the Euclidean are the distance metrics attributed to codes over $\Field_2$, codes over $\Integers_4\eqdef\{0,1,2,3\}$, and packings, respectively. Any binary $(n,M)$ or $(n,M,d)$ code $\code{C}$ (linear or nonlinear) is a subset of $\Field_2^n$ with $M$ codewords and minimum Hamming distance $d$. In case $\code{C}$ is a linear subspace of dimension $k$ of $\Field_2^n$, then $\code{C}$ is said to be linear, with parameters $[n,k]$ or $[n,k,d]$. The Hamming weight of a codeword $\vect{x}\in\Field^n$ is denoted by $\Hwt{\vect{x}}$, and $\enorm{\vect{x}}$ represents the Euclidean norm of a vector $\vect{x}\in\Reals^n$. $\phi\colon\Field^n_2 \rightarrow \Integers^n$ is defined as the natural embedding, i.e., the elements of $\Field_2$ are mapped to the respective integer by $\phi$ element-wisely.

%$\phi(x)$ is the remainder of the division of $x$ by $2$.

\subsection{Codes and Packings}
\label{sec:codes-packings}

This paper focuses on the relation between binary codes and packings. Such binary codes can be linear or not, and the packings can be lattices or not. We start by presenting the definitions relevant to our study.
    
The \emph{weight enumerator} of an $(n,M)$ code $\code{C} \subseteq \Field_2^n$ is
\begin{IEEEeqnarray}{c}
  W_\code{C}(x,y)=\sum_{\vect{c}\in\code{C}} x^{n-\Hwt{\vect{c}}}y^{\Hwt{\vect{c}}}=\sum_{w=0}^n A_w(\code{C})\, x^{n-w}y^w,
  \label{eq:weight-enumerator}
\end{IEEEeqnarray}
where $A_w(\code{C})\eqdef\card{\{\vect{c}\in\code{C}\colon\Hwt{\vect{c}}=w\}}$, $w\in[0:n]$. 

For an $[n,k]$ code $\code{C}$, %we can also define  
the \emph{dual code} of $\code{C}$ is the $[n,n-k]$ code $\dual{\code{C}}\eqdef\{\vect{u} \in \Field_2^n \colon \inner{\vect{u}}{\vect{v}}=0,\forall\,\vect{v}\in\code{C}\}$. 
%Some particular cases of formally self-dual codes can be written in terms of $\dual{\code{C}}$.
\begin{definition}[Self-dual and isodual] Let $\code{C}$ be an $[n,k]$ code. Then
  \begin{itemize}
  \item $\code{C}$ is said to be self-dual if $\code{C}=\dual{\code{C}}$.
  \item If there is a permutation $\pi$ of coordinates such that $\code{C}=\pi(\dual{\code{C}})$, $\code{C}$ is called isodual.
  %\item A code $\code{C}$ is \emph{formally self-dual} if it is equal to its MacWilliams' identity, i.e.,  $W_{\code{C}}(x,y)=\frac{1}{2^{n-k}}W_{{\code{C}}}(x+y,x-y)$, which for linear codes means that $W_{\code{C}}(x,y) = W_{\dual{\code{C}}}(x,y)$.
  %\item A code $\code{C}$ is \emph{formally self-dual} if $\code{C}$ and $\dual{\code{C}}$ have the same weight enumerator, i.e., $W_\code{C}(x,y)=W_{\dual{\code{C}}}(x,y)$.
  \end{itemize}
\end{definition}
For an $[n,k]$ code $\code{C}$, the relation between $W_\code{C}(x,y)$ and $W_{\dual{\code{C}}}(x,y)$ is characterized by the well-known \emph{MacWilliams identity} (see, e.g.,~\cite[Th.~1, Ch.~5]{MacWilliamsSloane77_1}):
\begin{IEEEeqnarray}{c}
  W_{\code{C}}(x,y)=\frac{1}{|\dual{\code{C}}|}W_{\dual{\code{C}}}(x+y,x-y)=\frac{1}{2^{n-k}}W_{\dual{\code{C}}}(x+y,x-y).
  \label{eq:MacWilliams-identity-linear}
\end{IEEEeqnarray}
% Since our results are concerning codes that satisfy~\eqref{eq:MacWilliams-identity-fsd}, the concept of dual of a nonlinear code (or transform of a code) does not play an active role here. 
\begin{remark}
  The MacWilliams identity as expressed in~\eqref{eq:MacWilliams-identity-linear} relies on the notion of duality, which was defined above only for linear codes. This concept can be generalized for nonlinear codes in terms of the transform of a code. To simplify the presentation we omit further details on transforms, since our focus is on formally self-dual codes only, in which case the transform is implicitly defined. For the generalization, we refer the reader to~\cite[pp.~132--141]{MacWilliamsSloane77_1}.
\end{remark}

\begin{definition}[Formally self-dual]
  \label{def:FSD-codes}
  Given an $(n,M)$ code $\code{C}$, we say it is formally self-dual if and only if % it is equal to its MacWilliams identity, i.e., 
  \begin{IEEEeqnarray}{c}
    W_{\code{C}}(x,y)=\frac{1}{M}W_{{\code{C}}}(x+y,x-y).
    \label{eq:MacWilliams-identity-fsd}
  \end{IEEEeqnarray}
\end{definition}

Next, we prove that the code size $M$ of a formally self-dual code should be $2^{\nicefrac{n}{2}}$.
\begin{proposition}
  \label{prop:code-size_n_over_2}
  If an $(n,M)$ code is formally self-dual, then $M=2^{\nicefrac{n}{2}}$.
\end{proposition}
\begin{IEEEproof}
  Because $\code{C}$ is formally self-dual, then
  \begin{IEEEeqnarray*}{rCl}
    W_{\code{C}}(x,y)& \stackrel{\eqref{eq:MacWilliams-identity-fsd}}{=} &\frac{1}{M}W_{{\code{C}}}(x+y,x-y)    
    \stackrel{\eqref{eq:MacWilliams-identity-fsd}}{=}\frac{1}{M}\cdot\frac{1}{M}W_{{\code{C}}}(x+y+x-y,x+y-x+y)
    \\
    & = &\frac{1}{M^2}W_{{\code{C}}}(2x,2y)=\frac{2^n}{M^2}W_{{\code{C}}}(x,y),
  \end{IEEEeqnarray*}
  which implies that $\frac{2^n}{M^2} = 1$ and $M = 2^{\nicefrac{n}{2}}$.
\end{IEEEproof}

Thus, similar to self-dual and isodual codes, the weight enumerator $W_\code{C}(x,y)$ of a formally self-dual code satisfies~\cite[eq.~(7), p.~599]{MacWilliamsSloane77_1}
\begin{IEEEeqnarray}{c}
  W_\code{C}(x,y)=W_{\code{C}}\left(\frac{x+y}{\sqrt{2}},\frac{x-y}{\sqrt{2}}\right).  %MacWilliams&Sloane, p.599, eq. (7)  
  \IEEEeqnarraynumspace\label{eq:FSD_MacWilliams-identity}
\end{IEEEeqnarray}
In general, formally self-dual codes constitute a broader family of codes than isodual and self-dual codes.

A \emph{packing} $\Gamma \subset \mathbb{R}^n$ is an infinite discrete set. %An isometry $T: \Reals^n \to \Reals^n$, is a transformation that preserves Euclidean distance, i.e., $\|T({\bm x}) - T({\bm y}) \|^2 = \|{\bm x} - {\bm y} \|^2$, $\forall\,\vect{x},\vect{y}\in\Reals^n$. 
An isometry $T$ is a transformation that preserves distance, with respect to a certain metric. We call a packing $\Gamma$ \emph{geometrically uniform} if for any two elements $\vect{x},\vect{x}'\in\Gamma$, there exists an isometry $T_{\vect{x},\vect{x}'}$ such that $\vect{x}'=T_{\vect{x},\vect{x}'}(\vect{x})$ and $T_{\vect{x},\vect{x}'}(\Gamma)\eqdef\{T_{\vect{x},\vect{x}'}(\vect{x})\colon\vect{x}\in\Gamma\}=\Gamma$. A geometrically uniform packing $\Gamma$ is also \emph{distance-invariant}, which means that the number of elements of $\Gamma$ at a distance $d$ from an element $\vect{x}\in\Gamma$ is independent of ${\bm x}$.

The \emph{Voronoi region} of a point $\vect{x}$ in a packing $\Gamma \subset \Reals^n$ is defined as
\begin{IEEEeqnarray*}{c}
  \set{V}_{\Gamma}({\vect{x}})\eqdef\bigl\{\vect{y} \in \Reals^{n}\colon \|{\vect{y}-\vect{x}}\|^2 \leq \|{\vect{y}-\vect{x}'\|}^2,\,\forall\,\vect{x}' \in \Gamma\bigr\}.
\end{IEEEeqnarray*}
All Voronoi regions of geometrically uniform packings have the same shape~\cite{Forney91_1}, that we will refer to as $\set{V}({\Gamma})$.

A packing that has a group structure is called a \emph{lattice}, i.e., a (full rank) lattice $\Lambda$ is a discrete additive subgroup of $\mathbb{R}^{n}$, which is generated as
\begin{IEEEeqnarray*}{c}
  \Lambda=\{\vect{\lambda}=\vect{u}\mat{L}_{n\times n}\colon{\bm u}=(u_1,\ldots,u_n)\in\Integers^n\},
\end{IEEEeqnarray*}
where the $n$ rows of $\mat{L}$ form a lattice basis. The \emph{volume} of $\Lambda$ is $\vol{\Lambda} = \ecard{\det(\mat{L})}$. A lattice is an example of a geometrically uniform packing.

% Throughout this paper, we will particularly work with full rank lattices, i.e., $m=n$.

For lattices, the analogue of the weight enumerator is the \emph{theta series}.
\begin{definition}[Theta series]
  \label{def:theta-series}
  Let $\Lambda \subset \Reals^n$ be a lattice, its theta series is given by
  \begin{IEEEeqnarray}{c}
    \Theta_\Lambda(z) = \sum_{{\bm \lambda} \in \Lambda} q^{\norm{\vect{\lambda}}^2},
    \label{eq:theta_series}
  \end{IEEEeqnarray}
  where $q\eqdef e^{i\pi z}$ and $\Im{z} > 0$. % The theta series of an integer lattice is absolutely and uniformly convergent \cite[p.~71]{Gunning:62}
\end{definition}

The theta series converges uniformly absolutely for all $z$ such that $\Im{z} > 0$~\cite[Lemma~2.2., pp.~39--40]{Ebeling13_1}. Note that sometimes the theta series of a lattice can be expressed in terms of the \emph{Jacobi theta functions} defined as follows.
\begin{IEEEeqnarray*}{rCl}
  \vartheta_2(z)& \eqdef &\sum_{m\in\Integers} q^{\bigl(m+\frac{1}{2}\bigr)^2}=\Theta_{\Integers + \frac{1}{2}}(z),
  \nonumber\\
  % & = & 2q^{1/4}(1+q^2+q^6+q^{12}+ q^{20}+\dots)
  % \nonumber\\
  \vartheta_3(z)& \eqdef &\sum_{m \in \Integers} q^{m^2}=\Theta_{\Integers}(z), ~ ~ \vartheta_4(z) \eqdef \sum_{m\in \Integers} (-q)^{m^2}.
  % \label{eq:Jacobi-theta-functions}
  % & = & 1 + 2q + 2q^4 +  2q^9 + 2q^{16} + 2q^{25} + \dots
  % \nonumber\\
  % \nonumber\\
  % & =  &1 - 2q + 2q^4 -  2q^9 + 2q^{16} - 2q^{25} + \dots
\end{IEEEeqnarray*}

\begin{example}
  Consider $\Lambda=\Integers^n$. Then,
  \begin{IEEEeqnarray*}{rCl}
    \Theta_{\Integers^n}(z) & = &\sum_{\vect{\lambda} \in \Integers^n} q^{\norm{\vect{\lambda}}^2} = %\sum_{\lambda \in \Integers^n} q^{\| \lambda \|^2} =
    \sum_{{\bm \lambda} \in \Integers^n} q^{\lambda_1^2 + \dots + \lambda_n^2} \\
    & = &\biggl(\sum_{\lambda_1\in \Integers} q^{\lambda_1^2}\biggr)\cdots\biggl(\sum_{\lambda_n\in\Integers}q^{\lambda_n^2}\biggr) % = \Biggl(\sum_{m \in \Integers} q^{m^2}\Biggr)^n
    = \vartheta_3^n(z).\hfill\exampleend
  \end{IEEEeqnarray*}
\end{example}

Some theta series identities~\cite[p.~104]{ConwaySloane99_1} will be useful in the course of this paper, such as
\begin{IEEEeqnarray}{rCl}
  \vartheta_3\biggl(\frac{-1}{z}\biggr)& = &\biggl(\frac{z}{i}\biggr)^{\frac{1}{2}}\vartheta_3(z),\quad\vartheta_4\biggl(\frac{-1}{z}\biggr)=\biggl(\frac{z}{i}\biggr)^{\frac{1}{2}}\vartheta_2(z),\label{eq:mw_t3t4}
  \\[1mm]
  \vartheta_3(z)+\vartheta_2(z)& = &\vartheta_3\biggl(\frac{z}{4}\biggr),\quad\vartheta_3(z)-\vartheta_2(z)  =  \vartheta_4\biggl(\frac{z}{4}\biggr),\label{eq:add_sub}
  \\[1mm]
  \vartheta_3^2(z)+\vartheta_4^2(z)& = &2\vartheta^2_3(2z),\quad\vartheta_3^2(z)-\vartheta_4^2(z)=2\vartheta^2_2(2z). \label{eq:identity_squares}
\end{IEEEeqnarray}

For a lattice $\Lambda$, we have similar concepts to self-dual and isodual codes. If a lattice $\Lambda$ has generator matrix $\mat{L}$, then the lattice $\Lambda^\star\subset\mathbb{R}^n$ generated by $\trans{\bigl(\inv{\mat{L}}\bigr)}$ is called the \emph{dual lattice} of $\Lambda$.

\begin{remark}
  \label{rm:volume}
  $\vol{\Lambda^\star}=\inv{\vol{\Lambda}}$.
\end{remark}

% Here, we also introduce \emph{formally unimodular} lattices.
\begin{definition}[Unimodular and isodual lattices]
  \label{def:uni-iso-fum}
  A lattice $\Lambda \subset\Reals^n$ is said to be integral if the inner product of any two lattice vectors is an integer. 
  \begin{itemize}
  \item An integral lattice such that $\Lambda = \Lambda^\star$ is a unimodular lattice.
  \item A lattice $\Lambda$ is called isodual if it can be obtained from its dual $\Lambda^\star$ by (possibly) a rotation or reflection.
    % \item A lattice $\Lambda$ is \emph{formally unimodular} if it has the same theta series as its dual, i.e., $\Theta_{\Lambda}(z)=\Theta_{\Lambda^\star}(z)$.
  \end{itemize}  
\end{definition}

Analogously, the spirit of the MacWilliams identity can be captured by the \emph{Jacobi's formula}~\cite[eq.~(19), Ch.~4]{ConwaySloane99_1} for a lattice $\Lambda$:
\begin{IEEEeqnarray}{c}
  \Theta_{\Lambda}(z)=\vol{\Lambda^\star}\Bigl(\frac{i}{z}\Bigr)^{\frac{n}{2}}\Theta_{\Lambda^\star}\Bigl(-\frac{1}{z}\Bigr).
  \label{eq:Jacobi-formula-lattice}
\end{IEEEeqnarray}

A packing $\Gamma$ is said to be \emph{periodic} if it is a union of translates of a lattice, i.e., $\Gamma=\bigcup_{j=1}^{K}(\vect{u}_j+\Lambda)$, where $\Lambda$ is a lattice and $\vect{u}_1, \ldots, \vect{u}_K \in \Reals^n$. Given a periodic packing, we define its \textit{volume} as $\vol{\Gamma}=\nicefrac{\vol{\Lambda}}{K}$.

In general, there is no systematic way to derive the theta series of a nonlattice packing. However, it is known that the theta series of periodic packings $\Gamma$ (not necessarily lattices) can be expressed as follows. 
\begin{proposition}[{\cite{OdlyzkoSloane80_1}}]
  \label{prop:thm_thetanonl} 
  Let $\Gamma$ be a periodic packing. Then
  \begin{IEEEeqnarray}{c}
    \Theta_\Gamma(z)=\frac{1}{K} \sum_{j=1}^{K}  \sum_{\ell=1}^{K}  \sum_{{\bm \lambda} \in \Lambda} q^{\enorm{\vect{\lambda}+\vect{u}_j-\vect{u}_\ell}^2} = \Theta_\Lambda(z) + \frac{2}{K} \sum_{j< \ell}\sum_{{\bm \lambda} \in \Lambda} q^{\|{\bm \lambda}+{\bm u}_j-{\bm u}_\ell \|^2}.
    \label{eq:Theta_NL}
  \end{IEEEeqnarray}
  If $\Gamma$ is distance-invariant, then the theta series in~\eqref{eq:Theta_NL} reduces to
  \begin{IEEEeqnarray}{c}
    \Theta_\Gamma(z) = \sum_{j=1}^{K}\sum_{{\bm \lambda} \in \Lambda} q^{\|{\bm \lambda}+{\bm u}_j-{\bm u}_1 \|^2}.
    \label{eq:tf_edp}
  \end{IEEEeqnarray}
\end{proposition}

Throughout this paper we work with lattices and a slightly larger class, that we denote by lattice-like packings.
\begin{definition}
  \label{def:lattice-like}(Lattice-like packing)
  A packing $\Gamma$ is said to be lattice-like if the following two properties hold: 
  \begin{description}[leftmargin=0cm]
  \item[\textit{Periodicity}:] $\Gamma$ is periodic, so it has its volume and theta series well defined.
  \item[\textit{Geometric uniformity}:] $\Gamma$ is assumed to be geometrically uniform, so that it has congruent Voronoi regions~\cite[Th.~1]{Forney91_1}. 
  \end{description}
\end{definition}

\begin{figure}[t!]
  \centering
  \Scale[0.8]{
    \begin{tikzpicture}[thick,
    set/.style = { circle, minimum size = 3cm}]
    
\begin{scope}
    \clip (0.8,0) circle(3.2cm);
    \clip (3.2,0) circle(3.2cm);
    \fill[orange!15](1.2,0) circle(3.2cm);
\end{scope}

\begin{scope}
    \clip (3.4,0) circle(2.2cm);
    \clip (2,-0.8) circle(1.3cm);
    \fill[orange!40, very thick, pattern=north west lines, pattern color=orange](2,-0.8) circle(1.3cm);
\end{scope}
 
% Circles outline
\draw[orange!70, very thick] (0.8,0) circle(3.2cm);
\draw[orange!70, very thick] (3.2,0) circle(3.2cm);
\draw[violet!70, very thick] (3.4,0) circle(2.2cm);
\draw[red!100, dashed] (0.8,0) circle(3.7cm);
\draw[teal!100, very thick] (-3.8,-4) rectangle(8,4);
\draw[black!70, very thick] (2,0.8) circle(1.3cm);
\draw[black!70, very thick] (2,-0.8) circle(1.3cm);

%\node[set,label={85:\scriptsize{$\Gamma$~geometrically uniform}}] at (-1.9,0.5) {};
\node[set,label={85:\scriptsize{\color{orange!100}{\textbf{Geometrically}}}}] at (-1.58,0.45) {};
\node[set,label={85:\scriptsize{\color{orange!100}{\textbf{uniform}}}}] at (-1.0,0.25) {};
\node[set,label={25:\scriptsize{\color{violet!100}\textbf{Construction A}}}] at (2.15,-0.65) {};
\node[set,label={25:\scriptsize{\color{black!100}{\textbf{Lattice}}}}] at (0.7,0.2) {};
\node[set,label={25:\scriptsize{\color{black!100}{\textbf{Formally}}}}] at (0.3,-1.65) {};
\node[set,label={25:\scriptsize{\color{black!100}{\textbf{unimodular}}}}] at (0.3,-1.85) {};
\node[set,label={75:\scriptsize{\color{red!100}{\textbf{Distance-invariant}}}}] at (-0.85,1.68) {};
\node[set,label={25:\scriptsize{\color{orange!100}\textbf{Periodic}}}] at (2.8,1.4) {};
\node[set,label={30:{\color{teal!100}{\textbf{Packing}}}}] at (5,2.6) {};

\end{tikzpicture}}
  % \vspace{-1ex}
  \caption{Classification of packings. The shaded area represents lattice-like packings, while the hashed area represents the formally unimodular lattice-like packings from Construction A we consider in this paper.}
  \label{fig:venn_packings_lattices}
  % \vspace{-2ex}
\end{figure}
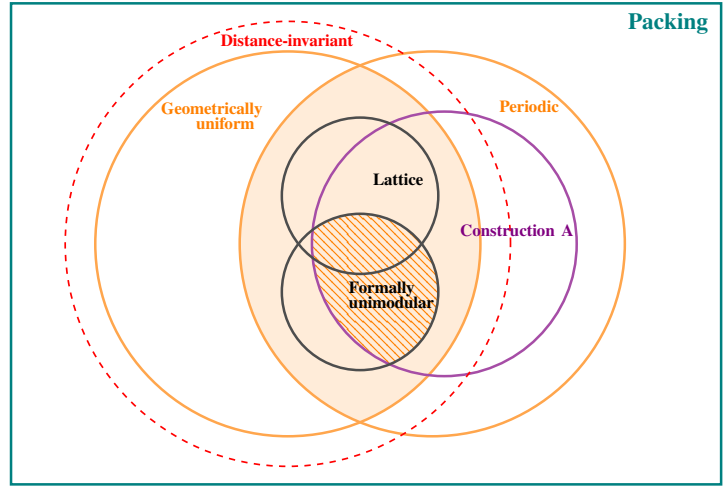

Lattice-like packings are illustrated by the shaded area in Figure~\ref{fig:venn_packings_lattices}. Observe that a lattice is an example of a lattice-like packing. Consequently, the family of lattice-like packings can be seen as a generalization of lattices. In Example~\ref{ex:ex_ConstrC-LatticeLike_n2}, we present an example of a lattice-like packing.

\begin{example} 
  \label{ex:ex_ConstrC-LatticeLike_n2}
   Consider a $\Gamma=\bigcup_{\vect{c}\in\code{C}}\bigl(\vect{c}+4\Integers^2\bigr)$, where $\code{C}=\{(0,0),(1,1)\}$. $\Gamma$ is clearly periodic and it is also geometrically uniform according to~\cite[Th.~2]{BollaufZamirCosta19_1}. Figure~\ref{fig:ex_ConstrC-LatticeLike_n2} illustrates such a packing and its congruent Voronoi regions.
  \begin{figure}[t!]
    \centering
    % \Scale[0.8]{
    \definecolor{ccqqww}{rgb}{0.8,0.0,0.4}
\definecolor{ttqqqq}{rgb}{0.2,0.0,0.0}

\begin{tikzpicture}[line cap=round,line join=round,x=1.0cm,y=1.0cm,scale=0.8]
\pgfplotsset{every tick label/.append style={font=\scriptsize}}
\begin{axis}[
x=1.0cm,y=1.0cm,
axis lines=middle,
xmin=-4.95,
xmax=5.2,
ymin=-2.0,
ymax=5.5,
xtick={-4.0,-3.0,...,5.0},
ytick={-1.0,...,5.0},]
\clip(-4.95,-2.) rectangle (5.2,5.5);
\fill[line width=1.25pt,fill=black,fill opacity=0.10000000149011612] (-2.,-1.) -- (-1.,-2.) -- (2.,-1.) -- (-1.,2.) -- cycle;
\fill[line width=1.25pt,fill=black,fill opacity=0.10000000149011612] (2.,3.) -- (3.,2.) -- (2.,-1.) -- (-1.,2.) -- cycle;
\fill[line width=1.25pt,fill=black,fill opacity=0.10000000149011612] (3.,2.) -- (6.,-1.) -- (3.,-2.) -- (2.,-1.) -- cycle;
\fill[line width=1.25pt,fill=black,fill opacity=0.10000000149011612] (2.,-1.) -- (3.,-2.) -- (2.,-5.) -- (-1.,-2.) -- cycle;
\fill[line width=1.25pt,fill=black,fill opacity=0.10000000149011612] (-2.,3.) -- (-1.,2.) -- (-2.,-1.) -- (-5.,2.) -- cycle;
\fill[line width=1.25pt,fill=black,fill opacity=0.10000000149011612] (-5.,2.) -- (-2.,-1.) -- (-5.,-2.) -- (-6.,-1.) -- cycle;
\fill[line width=1.25pt,fill=black,fill opacity=0.10000000149011612] (-2.,-1.) -- (-1.,-2.) -- (-2.,-5.) -- (-5.,-2.) -- cycle;
\fill[line width=1.25pt,fill=black,fill opacity=0.10000000149011612] (6.,-1.) -- (7.,-2.) -- (6.,-5.) -- (3.,-2.) -- cycle;
\fill[line width=1.25pt,fill=black,fill opacity=0.10000000149011612] (6.,3.) -- (7.,2.) -- (6.,-1.) -- (3.,2.) -- cycle;
\fill[line width=1.25pt,fill=black,fill opacity=0.10000000149011612] (3.,-2.) -- (6.,-5.) -- (3.,-6.) -- (2.,-5.) -- cycle;
\fill[line width=1.25pt,fill=black,fill opacity=0.10000000149011612] (-1.,-2.) -- (2.,-5.) -- (-1.,-6.) -- (-2.,-5.) -- cycle;
\fill[line width=1.25pt,fill=black,fill opacity=0.10000000149011612] (2.,3.) -- (3.,2.) -- (6.,3.) -- (3.,6.) -- cycle;
\fill[line width=1.25pt,fill=black,fill opacity=0.10000000149011612] (-2.,3.) -- (-1.,2.) -- (2.,3.) -- (-1.,6.) -- cycle;
\fill[line width=1.25pt,fill=black,fill opacity=0.10000000149011612] (6.,-1.) -- (7.,2.) -- (10.,-1.) -- (7.,-2.) -- cycle;
\fill[line width=1.25pt,fill=black,fill opacity=0.10000000149011612] (-6.,3.) -- (-5.,2.) -- (-6.,-1.) -- (-9.,2.) -- cycle;
\fill[line width=1.25pt,fill=black,fill opacity=0.10000000149011612] (-6.,3.) -- (-5.,6.) -- (-2.,3.) -- (-5.,2.) -- cycle;
\fill[line width=1.25pt,fill=black,fill opacity=0.10000000149011612] (2.,3.) -- (3.,6.) -- (2.,7.) -- (-1.,6.) -- cycle;
\fill[line width=1.25pt,fill=black,fill opacity=0.10000000149011612] (-2.,3.) -- (-1.,6.) -- (-2.,7.) -- (-5.,6.) -- cycle;
\fill[line width=1.25pt,fill=black,fill opacity=0.10000000149011612] (-10.003727100560011,2.990140806096375) -- (-9.,2.) -- (-5.9963670880482125,2.996367088048212) -- (-8.993750391249689,5.993750391249689) -- cycle;
\fill[line width=1.25pt,fill=black,fill opacity=0.10000000149011612] (3.,6.) -- (6.,7.) -- (7.,6.) -- (6.,3.) -- cycle;
\fill[line width=1.25pt,fill=black,fill opacity=0.10000000149011612] (-5.,-2.) -- (-2.,-5.) -- (-5.,-6.) -- (-6.,-5.) -- cycle;
\draw[line width=4.pt,fill=black,fill opacity=0.10000000149011612] (-11.413734177728347,5.3957631365267185) -- (-9.016553723493242,5.3957631365267185);
\draw[line width=4.pt,fill=black,fill opacity=0.10000000149011612] (-11.401748275457173,4.940298850222049) -- (-9.004567821222068,4.940298850222049);
\draw [line width=1.25pt] (-2.,-1.)-- (-1.,-2.);
\draw [line width=1.25pt] (-1.,-2.)-- (2.,-1.);
\draw [line width=1.25pt] (-1.,2.)-- (-2.,-1.);
\draw [line width=1.25pt] (2.,3.)-- (3.,2.);
\draw [line width=1.25pt] (3.,2.)-- (2.,-1.);
\draw [line width=1.25pt] (2.,-1.)-- (-1.,2.);
\draw [line width=1.25pt] (-1.,2.)-- (2.,3.);
\draw [line width=1.25pt] (3.,2.)-- (6.,-1.);
\draw [line width=1.25pt] (6.,-1.)-- (3.,-2.);
\draw [line width=1.25pt] (3.,-2.)-- (2.,-1.);
\draw [line width=1.25pt] (2.,-1.)-- (3.,2.);
\draw [line width=1.25pt] (2.,-1.)-- (3.,-2.);
\draw [line width=1.25pt] (3.,-2.)-- (2.,-5.);
\draw [line width=1.25pt] (2.,-5.)-- (-1.,-2.);
\draw [line width=1.25pt] (-1.,-2.)-- (2.,-1.);
\draw [line width=1.25pt] (-2.,3.)-- (-1.,2.);
\draw [line width=1.25pt] (-1.,2.)-- (-2.,-1.);
\draw [line width=1.25pt] (-2.,-1.)-- (-5.,2.);
\draw [line width=1.25pt] (-5.,2.)-- (-2.,3.);
\draw [line width=1.25pt] (-5.,2.)-- (-2.,-1.);
\draw [line width=1.25pt] (-2.,-1.)-- (-5.,-2.);
\draw [line width=1.25pt] (-5.,-2.)-- (-6.,-1.);
\draw [line width=1.25pt] (-6.,-1.)-- (-5.,2.);
\draw [line width=1.25pt] (-2.,-1.)-- (-1.,-2.);
\draw [line width=1.25pt] (-1.,-2.)-- (-2.,-5.);
\draw [line width=1.25pt] (-2.,-5.)-- (-5.,-2.);
\draw [line width=1.25pt] (-5.,-2.)-- (-2.,-1.);
\draw [line width=1.25pt] (6.,-1.)-- (7.,-2.);
\draw [line width=1.25pt] (7.,-2.)-- (6.,-5.);
\draw [line width=1.25pt] (6.,-5.)-- (3.,-2.);
\draw [line width=1.25pt] (3.,-2.)-- (6.,-1.);
\draw [line width=1.25pt] (6.,3.)-- (7.,2.);
\draw [line width=1.25pt] (7.,2.)-- (6.,-1.);
\draw [line width=1.25pt] (6.,-1.)-- (3.,2.);
\draw [line width=1.25pt] (3.,2.)-- (6.,3.);
\draw [line width=1.25pt] (3.,-2.)-- (6.,-5.);
\draw [line width=1.25pt] (6.,-5.)-- (3.,-6.);
\draw [line width=1.25pt] (3.,-6.)-- (2.,-5.);
\draw [line width=1.25pt] (2.,-5.)-- (3.,-2.);
\draw [line width=1.25pt] (-1.,-2.)-- (2.,-5.);
\draw [line width=1.25pt] (2.,-5.)-- (-1.,-6.);
\draw [line width=1.25pt] (-1.,-6.)-- (-2.,-5.);
\draw [line width=1.25pt] (-2.,-5.)-- (-1.,-2.);
\draw [line width=1.25pt] (2.,3.)-- (3.,2.);
\draw [line width=1.25pt] (3.,2.)-- (6.,3.);
\draw [line width=1.25pt] (6.,3.)-- (3.,6.);
\draw [line width=1.25pt] (3.,6.)-- (2.,3.);
\draw [line width=1.25pt] (-2.,3.)-- (-1.,2.);
\draw [line width=1.25pt] (-1.,2.)-- (2.,3.);
\draw [line width=1.25pt] (2.,3.)-- (-1.,6.);
\draw [line width=1.25pt] (-1.,6.)-- (-2.,3.);
\draw [line width=1.25pt] (6.,-1.)-- (7.,2.);
\draw [line width=1.25pt] (7.,2.)-- (10.,-1.);
\draw [line width=1.25pt] (10.,-1.)-- (7.,-2.);
\draw [line width=1.25pt] (7.,-2.)-- (6.,-1.);
\draw [line width=1.25pt] (-6.,3.)-- (-5.,2.);
\draw [line width=1.25pt] (-5.,2.)-- (-6.,-1.);
\draw [line width=1.25pt] (-6.,-1.)-- (-9.,2.);
\draw [line width=1.25pt] (-9.,2.)-- (-6.,3.);
\draw [line width=1.25pt] (-6.,3.)-- (-5.,6.);
\draw [line width=1.25pt] (-5.,6.)-- (-2.,3.);
\draw [line width=1.25pt] (-2.,3.)-- (-5.,2.);
\draw [line width=1.25pt] (-5.,2.)-- (-6.,3.);
\draw [line width=1.25pt] (2.,3.)-- (3.,6.);
\draw [line width=1.25pt] (3.,6.)-- (2.,7.);
\draw [line width=1.25pt] (2.,7.)-- (-1.,6.);
\draw [line width=1.25pt] (-1.,6.)-- (2.,3.);
\draw [line width=1.25pt] (-2.,3.)-- (-1.,6.);
\draw [line width=1.25pt] (-1.,6.)-- (-2.,7.);
\draw [line width=1.25pt] (-2.,7.)-- (-5.,6.);
\draw [line width=1.25pt] (-5.,6.)-- (-2.,3.);
\draw [line width=1.25pt] (-10.003727100560011,2.990140806096375)-- (-9.,2.);
\draw [line width=1.25pt] (-9.,2.)-- (-5.9963670880482125,2.996367088048212);
\draw [line width=1.25pt] (-5.9963670880482125,2.996367088048212)-- (-8.993750391249689,5.993750391249689);
\draw [line width=1.25pt] (-8.993750391249689,5.993750391249689)-- (-10.003727100560011,2.990140806096375);
\draw [line width=1.25pt] (3.,6.)-- (6.,7.);
\draw [line width=1.25pt] (6.,7.)-- (7.,6.);
\draw [line width=1.25pt] (7.,6.)-- (6.,3.);
\draw [line width=1.25pt] (6.,3.)-- (3.,6.);
\draw [line width=1.25pt] (-5.,-2.)-- (-2.,-5.);
\draw [line width=1.25pt] (-2.,-5.)-- (-5.,-6.);
\draw [line width=1.25pt] (-5.,-6.)-- (-6.,-5.);
\draw [line width=1.25pt] (-6.,-5.)-- (-5.,-2.);
\begin{scriptsize}
\draw [fill=black] (-9.016553723493242,5.3957631365267185) circle (2.5pt);
\draw[color=black] (-8.96861011440854,5.677431839899343) node {$n = 3$};
\draw [fill=black] (-9.80362797263377,4.940298850222049) circle (2.5pt);
\draw[color=black] (-9.759679664306125,5.2219675535946735) node {$i = 1$};
\draw [fill=ttqqqq] (13.,5.) circle (3.5pt);
\draw [fill=ttqqqq] (12.,4.) circle (3.5pt);
\draw [fill=ttqqqq] (0.,0.) circle (3.5pt);
\draw [fill=ttqqqq] (1.,1.) circle (3.5pt);
\draw [fill=ttqqqq] (-3.,1.) circle (3.5pt);
\draw [fill=ttqqqq] (4.,0.) circle (3.5pt);
\draw [fill=ttqqqq] (-4.,0.) circle (3.5pt);
\draw [fill=ttqqqq] (-3.,-3.) circle (3.5pt);
\draw [fill=ttqqqq] (1.,-3.) circle (3.5pt);
\draw [fill=ttqqqq] (5.,-3.) circle (3.5pt);
\draw [fill=ttqqqq] (5.,1.) circle (3.5pt);
\draw [fill=ttqqqq] (-4.,4.) circle (3.5pt);
\draw [fill=ttqqqq] (-3.,5.) circle (3.5pt);
\draw [fill=ttqqqq] (0.,4.) circle (3.5pt);
\draw [fill=ttqqqq] (1.,5.) circle (3.5pt);
\draw [fill=ttqqqq] (4.,4.) circle (3.5pt);
\draw [fill=ttqqqq] (5.,5.) circle (3.5pt);
\draw [fill=ttqqqq] (8.,4.) circle (3.5pt);
\draw [fill=ccqqww] (-8.,4.) circle (3.5pt);
\draw [fill=ccqqww] (12.,4.) circle (3.5pt);
\draw [fill=ccqqww] (13.054794766825841,4.9823096629932415) circle (3.5pt);
\draw [fill=ccqqww] (13.008465258525913,0.9979719491995062) circle (3.5pt);
\draw [fill=ccqqww] (12.,0.) circle (3.5pt);
\draw [fill=ccqqww] (-8.,0.) circle (3.5pt);
\draw [fill=ccqqww] (-10.990220040836448,5.005474417143205) circle (3.5pt);
\draw [fill=ccqqww] (-12.,4.) circle (3.5pt);
\draw [fill=ccqqww] (-10.990220040836448,0.9748071950495426) circle (3.5pt);
\draw [fill=ccqqww] (-12.,0.) circle (3.5pt);
\draw [fill=ccqqww] (-8.,-4.) circle (3.5pt);
\draw [fill=ccqqww] (12.,-4.) circle (3.5pt);
\end{scriptsize}
\end{axis}
\end{tikzpicture} 
    % \vspace{-1ex}
    \caption{A lattice-like packing $\Gamma=\bigcup_{\vect{c}\in\code{C}}\bigl(\vect{c}+4\Integers^2\bigr)\subset\Reals^2$, where $\code{C}=\{(0,0),(1,1)\}$.}
    \label{fig:ex_ConstrC-LatticeLike_n2}
    % \vspace{-2ex}
  \end{figure}
\end{example}

 In Section~\ref{sec:further-connections_codes-packings}, we will show how to construct lattice-like packings.
 
The theta series of a lattice-like packing is obtained according to~\eqref{eq:tf_edp}, due to the geometric uniformity. 
Mimicking the definition of formally self-dual codes, we define formally unimodular packings. 

\begin{definition}[Formally unimodular lattice-like packings]
  We say that a lattice-like packing $\Gamma$ is formally unimodular if and only if
  \begin{IEEEeqnarray}{c}
    \Theta_{\Gamma}(z)=\vol{\Gamma}\biggl(\frac{i}{z}\biggr)^{\frac{n}{2}}\Theta_{\Gamma}\biggl(-\frac{1}{z}\biggr).
    \label{eq:Jacobi-formula-FU-packings}
  \end{IEEEeqnarray}
\end{definition}
The class is represented by a inner circle in Figure~\ref{fig:venn_packings_lattices}, while the hashed area represents packings obtained via Construction A, to be discussed in a sequel. The latter are the ones we will work closely in this paper. Note that in our prior work~\cite{BollaufLinYtrehus22_1}, we have introduced a smaller class of \emph{formally unimodular lattices}, where we only considered the lattices that have the same theta series as their dual and, thus, satisfy~\eqref{eq:Jacobi-formula-FU-packings}. Throughout out this paper, we will usually call formally unimodular lattice-like packings by \emph{formally unimodular packings} for simplicity.

Analogous to formally self-dual codes, we show that $\vol{\Gamma}=1$ if $\Gamma$ is formally unimodular.
\begin{proposition}
  \label{prop:volume_FU-lattices}
  If a lattice-like packing $\Gamma$ is formally unimodular, then $\vol{\Gamma}=1$.
\end{proposition}
\begin{IEEEproof}
  By applying~\eqref{eq:Jacobi-formula-FU-packings} twice, we have that
  \begin{IEEEeqnarray*}{rCl}
    \Theta_{\Gamma}(z)& = &\vol{\Gamma}\biggl(\frac{i}{z}\biggr)^{\frac{n}{2}}\Theta_{\Gamma}\Bigl(-\frac{1}{z}\Bigr)
    = \vol{\Gamma}\biggl(\frac{i}{z}\biggr)^{\frac{n}{2}}\cdot\vol{\Gamma}(-i z)^{\frac{n}{2}}\Theta_{\Gamma}(z) \nonumber \\
    & = & \vol{\Gamma}^2 \Theta_{\Gamma}(z). 
  \end{IEEEeqnarray*}
  Hence, $\vol{\Gamma}^2=1$, which implies that  $\vol{\Gamma}=1$.
\end{IEEEproof}
Consequently, analogous to unimodular and isodual lattices, the theta series of a formally unimodular packing satisfies
\begin{IEEEeqnarray*}{c}
  \Theta_{\Gamma}(z)=\biggl(\frac{i}{z}\biggr)^{\frac{n}{2}}\Theta_{\Gamma}\Bigl(-\frac{1}{z}\Bigr).
  \label{eq:FSD_Jacobi-formula}
\end{IEEEeqnarray*}

Table~\ref{tab:table_dictionary} is a summary of the notations and connections between codes and lattice-like packings that we will refer to throughout this paper.

\begin{table*}[t]
  \centering
  \caption{Dictionary of Codes and Lattice-like Packings} %,~$\gamma_n$ refers to the Hermite's constant~\cite[p.~20, (47)]{ConwaySloane99_1}}
  \label{tab:table_dictionary}
  \vskip -2.0ex
  \Scale[1]{\begin{IEEEeqnarraybox}[
    \IEEEeqnarraystrutmode
    \IEEEeqnarraystrutsizeadd{3.5pt}{4.0pt}]{V/c/V/c/V}
    \IEEEeqnarrayrulerow\\
    & \textnormal{Code}~\code{C} \subseteq \Field_2^n
    && \textnormal{Lattice-like Packing}~\Gamma \subset \mathbb{R}^n
    & \\
    \hline\hline
    & \textnormal{Linear code} && \textnormal{Lattice} &
    \\*\IEEEeqnarrayrulerow\\
    & \textnormal{Hamming distance} && \textnormal{Euclidean distance} &
    \\*\IEEEeqnarrayrulerow\\
    & \textnormal{Weight enumerator:} && \textnormal{Theta series:} &
    \\
    & W_\code{C}(x,y)=\sum_{\vect{c}\in\code{C}} x^{n-\Hwt{\vect{c}}}y^{\Hwt{\vect{c}}} && \Theta_\Gamma(z) = \sum_{{\bm \lambda} \in \Gamma} q^{\norm{\vect{\lambda}}^2} &    
    \\*\IEEEeqnarrayrulerow\\
    % & \textnormal{Coding gain:} && \textnormal{Lattice coding gain:} &
    % \\
    % & \frac{k}{n}d_{\textnormal{min}}(\code{C})~\textnormal{\cite[pp.~15--18]{LinCostello04_1}} && \frac{d_{\textnormal{min}}^2(\Gamma)}{\vol{\Gamma}^{\nicefrac{n}{2}}}~\textnormal{\cite[p.~20]{ConwaySloane90_1}\footnotemark} &
    % \\*\IEEEeqnarrayrulerow\\
    & \textnormal{Formally self-dual code } \code{C}: && \textnormal{Formally lattice-like unimodular packing } \Gamma: &
    \\
    & W_{\code{C}}(x,y)=\frac{1}{2^{\nicefrac{n}{2}}}W_{{\code{C}}}(x+y,x-y) && \Theta_{\Gamma}(z)=\Bigl(\frac{i}{z}\Bigr)^{\frac{n}{2}}\Theta_{\Gamma}\Bigl(-\frac{1}{z}\Bigr) &
    \\*\IEEEeqnarrayrulerow\\
    & \textnormal{Self-dual code} && \textnormal{Unimodular lattice} &
    \\*\IEEEeqnarrayrulerow\\
    & \textnormal{Isodual code} && \textnormal{Isodual lattice} &
    \\*\IEEEeqnarrayrulerow
  \end{IEEEeqnarraybox}}
\end{table*}
%\footnotetext[1]{This quantity is also called the \emph{Hermite's constant}.}

% \begin{remark}
%   For packings, the relations among unimodular, isodual, and formally unimodular lattices are given according to the diagram in~\ref{fig:venn_packings_lattices}
% \end{remark}

\subsection{Further Connections Between Codes and Packings}
\label{sec:further-connections_codes-packings}

To get lattice-like packings, we consider~\emph{Construction A} packings from codes, not necessarily linear.
\begin{definition}[{Construction A~\cite[p.~137]{ConwaySloane99_1}}]
  \label{def:ConstructionA}
  Let $\code{C}$ be an $(n,M)$ code, then a Construction A packing is defined as
  \begin{IEEEeqnarray*}{c}
    \GammaA{\code{C}}\eqdef\frac{1}{\sqrt{2}}\left(\phi(\code{C}) + 2\Integers^n\right)=\frac{1}{\sqrt{2}}\bigcup_{\vect{c}\in\code{C}}\bigl(\phi(\vect{c})+2\Integers^n\bigr).
  \end{IEEEeqnarray*}
\end{definition}

\begin{remark}\leavevmode
  \label{rem:GU_GammaA-codes}
  \begin{itemize}
  \item $\GammaA{\code{C}}=\ConstrA{\code{C}}$ is a lattice if and only if the code $\code{C}$ is an $[n,k]$ code. Also, $\vol{\ConstrA{\code{C}}}=2^{\nicefrac{n}{2}-k}$.
  \item $\GammaA{\code{C}}$ is, by definition, a periodic packing for any $(n,M)$ code $\code{C}$, and $\vol{\GammaA{\code{C}}}=\nicefrac{2^{\nicefrac{n}{2}}}{M}$.    
  \item % Since the minimum squared-Euclidean distance between the coset representatives $\vect{c} + 2\Integers^n$ and $\vect{c}' + 2\Integers^n$, for any $\vect{c}, \vect{c}' \in \code{C}$, is equal to the Hamming distance between $\vect{c}$ and $\vect{c}'$~\cite[p.~1255]{Forney91_1}, the following holds: 
    If and only if $\code{C}$ is geometrically uniform (with respect to the Hamming metric), then  $\GammaA{\code{C}}$ is geometrically uniform (with respect to the Euclidean metric), and consequently, a lattice-like packing.\footnote{For completeness, we briefly provide the proof here. If $\GammaA{\code{C}}$ is geometrically uniform, then for any $\vect{x}=\vect{c}+2\vect{z},\vect{x}'=\vect{c}'+2\vect{z}'\in\GammaA{\code{C}}$, $\vect{c},\vect{c}'\in\code{C}$ (ignoring the factor $\nicefrac{1}{\sqrt{2}}$), there exists an isometry $T^{\textnormal{E}}_{\vect{x},\vect{x}'}$ such that $\vect{x}'=T^{\textnormal{E}}_{\vect{x},\vect{x}'}(\vect{x})$. Let $\vect{z}=\vect{z}'=\vect{0}$. Since $\enorm{\vect{x}-\vect{x}'}^2=\Hdist{\vect{c}}{\vect{c}'}$, the isometry $T^{\textnormal{E}}_{\vect{c}+2\vect{z},\vect{c}'+2\vect{z}'}\equiv T^{\textnormal{E}}_{\vect{c},\vect{c}'}$ also acts an isometry of $\code{C}$, which proves that $\code{C}$ is geometrically uniform. Conversely, if for any $\vect{c},\vect{c}'\in\code{C}$, there exists an isometry $T^{\textnormal{H}}_{\vect{c},\vect{c}'}$ such that $\vect{c}'=T^{\textnormal{H}}_{\vect{c},\vect{c}'}(\vect{c})$. Now, consider any $\vect{x}=\vect{c}+2\vect{z},\vect{x}'=\vect{c}'+2\vect{z}'\in\GammaA{\code{C}}$. We define $T^{\textnormal{E}}_{\vect{x},\vect{x}'}=T^{\textnormal{H}}_{\vect{c},\vect{c}'}$ if $\vect{z}=\vect{z}'$, and otherwise let $T^{\textnormal{E}}_{\vect{x},\vect{x}'}({\vect{x}})=T^{\textnormal{H}}_{\vect{c},\vect{c}'}(\vect{c})+2\vect{z}'=\vect{c}'+2\vect{z}'=\vect{x}'$. Since $\enorm{\vect{c}+2\vect{z}-(\vect{c}'+2\vect{z})}^2=\Hdist{\vect{c}}{\vect{c}'}$ and $T^{\textnormal{E}}_{\vect{x},\vect{x}'}$ is a composition of isometries, $\GammaA{\code{C}}$ is geometrically uniform.}
  \end{itemize}
\end{remark}

Example~\ref{ex:ex_ConstrA-LatticeLike-packing_n2} presents a lattice-like Construction A packing.
\begin{example} 
  \label{ex:ex_ConstrA-LatticeLike-packing_n2}
  Consider a $\GammaA{\code{C}}= \bigcup_{\vect{c}\in\code{C}}\bigl(\vect{c}+2\Integers^2\bigr)$, where $\code{C} = \{(1,0),(0,1)\}$ as in Figure~\ref{fig:ex_ConstrA-LatticeLike-packing_n2}. Here, we dropped the scalar $\nicefrac{1}{\sqrt{2}}$ for illustration purposes. Observe that $\GammaA{\code{C}}$ is not a lattice, since $\code{C}$ is a nonlinear code.
  \begin{figure}[t!]
    \centering
    % \Scale[0.8]{
    \input{\Figs/voronoi-lattice-like-2.tex}
  % \vspace{-1ex}
    \caption{A lattice-like packing $\GammaA{\code{C}}\subset\Reals^2$, where $\code{C}=\{(0,1),(1,0)\}$.}
    \label{fig:ex_ConstrA-LatticeLike-packing_n2}
    % \vspace{-2ex}
  \end{figure}
\end{example}

\section{Theta Series of Construction A Lattice-like Packings}
\label{sec:theta-series_ConstrA-packings}

Lemma~\ref{lem:ThetaSeries_WeightDistribution_ConstructionA} gives a connection between the weight enumerator $W_{\code{C}}(x,y)$ of a linear code $\code{C}$ and a lattice $\ConstrA{\code{C}}$. % can be established.
\begin{lemma}[{\cite[Th.~3, Ch.~7]{ConwaySloane99_1}}]
  \label{lem:ThetaSeries_WeightDistribution_ConstructionA}
  Consider an $[n,k]$ code $\code{C}$ with $W_{\code{C}}(x,y)$, then the theta series of $\ConstrA{\code{C}}$ is given by
  \begin{IEEEeqnarray*}{c}
    \Theta_{\Lambda_\textnormal{A}(\code{C})}(z) = W_{\code{C}}(\vartheta_3(2z), \vartheta_2(2z)).
  \end{IEEEeqnarray*}
\end{lemma}

\begin{remark}
  \label{rmk:FSD-codes_ConstructionA}
  It follows immediately from Lemma~\ref{lem:ThetaSeries_WeightDistribution_ConstructionA} that if an $[n,\nicefrac{n}{2}]$ code $\code{C}$ is formally self-dual then $\ConstrA{\code{C}}$ is a formally unimodular lattice.
\end{remark}

%Lemma~\ref{lem:ThetaSeries_WeightDistribution_ConstructionA} gives the theta series of a Construction A lattice $\ConstrA{\code{C}}$ in terms of the weight enumerator of the linear code $\code{C}$. 
We will now show that the property in Lemma~\ref{lem:ThetaSeries_WeightDistribution_ConstructionA} holds for lattice-like Construction A packings $\GammaA{\code{C}}$, conditioned on some assumptions on the underlying $(n,M)$ code $\code{C}$.

\begin{lemma}
  \label{lem:theta-series_non-lattice}
  Let $\code{C}$ be an $(n,M)$ geometrically uniform code where $\vect{0}\in\code{C}$. Also, let $\Gamma_{\textnormal{A}}(\code{C}) = \frac{1}{\sqrt{2}} (\phi(\code{C}) + 2\Integers^{n})$  be a lattice-like packing generated via Construction A. Then its theta series is
  \begin{IEEEeqnarray*}{c}
    \Theta_{\Gamma_{\textnormal{A}}(\code{C})}(z) =  W_{\code{C}}(\vartheta_3(2z), \vartheta_2(2z)).
  \end{IEEEeqnarray*}
\end{lemma}
\begin{IEEEproof}
  We aim to apply the result from Proposition~\ref{prop:thm_thetanonl} to $\Gamma_{\textnormal{A}}(\code{C})$. Since $\code{C}$ is geometrically uniform and $\vect{0}\in\code{C}$, we can fix ${\vect{u}}_1 = \vect{0}$ as a representative. Since $\code{C}$ is geometrically uniform, $\GammaA{\code{C}}$ is also a geometrically uniform packing from Remark~\ref{rem:GU_GammaA-codes}. Hence, we can apply \eqref{eq:tf_edp}. We will also disregard the scalar $\nicefrac{1}{\sqrt{2}}$ in $\Gamma_{\textnormal{A}}(\code{C})$ for now. Therefore, we can rewrite
  \begin{IEEEeqnarray*}{c}
    \Theta_{\GammaA{\code{C}}}(z) = \sum_{k=1}^{\ecard{\code{C}}}\sum_{\vect{\lambda}\in 2\Integers^n} q^{\enorm{\vect{\lambda}+{\vect{u}_k}}^2} =\sum_{k=1}^{M}\sum_{\vect{z}\in\Integers^n} q^{\enorm{2\vect{z}+\vect{u}_k}^2}.
  \end{IEEEeqnarray*}
  
  The $i$-th coordinate of the vector $2{\bm z}+{\bm u}_k$ can only assume the following values
  \begin{IEEEeqnarray*}{c}
    (2\vect{z}+\vect{u}_{k})_i  =
    \begin{cases}
      2 z_i                 & \textnormal{if } u_{k_i}=0,
      \\
      2 (z_i + \frac{1}{2}) & \textnormal{if } u_{k_i}=1.
    \end{cases}
  \end{IEEEeqnarray*}
  
  The theta series associated to each of these cases are $ \Theta_{2\Integers}(z) = \vartheta_{3}(4z)$ and $\Theta_{2\bigl(\Integers+\frac{1}{2}\bigr)}(z) = \vartheta_{2}(4z)$.  Hence, if we fix a $\vect{u}_k = (u_{k_1}, \dots, u_{k_n}) \in \code{C}$, we obtain
  \begin{IEEEeqnarray*}{c}
   \displaystyle\sum_{{\bm z} \in \Integers^n} q^{\|2{\bm z}+{\bm u}_k \|^2}  =  \underbrace{\biggl(\sum_{z_1\in \Integers} q^{(2z_1 + u_{k_1})^2}\biggr)}_{\vartheta_3(4z_1) \textnormal{ or }\vartheta_2(4z_1)}\cdots\underbrace{\biggl(\sum_{z_n \in\Integers}q^{(2z_n+u_{k_n})^2}\biggr)}_{\vartheta_3(4z_n)\textnormal{ or }\vartheta_2(4z_n)}  = \vartheta_3(4z)^{n-w}\vartheta_2(4z)^{w},
  \end{IEEEeqnarray*}
  for $z\in \Integers$ and $w=\Hwt{\vect{u}_k}$. Therefore, we can simply write,
% Running through all coordinates of the vector ${\bm u}_k$ we have that
 \begin{IEEEeqnarray}{c}
   \Theta_{\Gamma_{\textnormal{A}}(\code{C})}(z)=\sum_{k=1}^{M}\sum_{\vect{z}\in\Integers^n} q^{\enorm{2\vect{z}+\vect{u}_k}^2} = \sum_{w=0}^n A_w(\code{C})\vartheta_3^{n-w}(4z)\vartheta_2^{w}(4z),\label{eq:theta_nr}
 \end{IEEEeqnarray}
 where $A_w(\code{C})$ is defined as in~\eqref{eq:weight-enumerator}. Note that a property of theta series is that $\Theta_{\alpha\Gamma}(z)=\Theta_{\Gamma}(\alpha^2 z)$ for some $\alpha>0$. Hence, by considering the factor $\nicefrac{1}{\sqrt{2}}$, \eqref{eq:theta_nr} becomes $\Theta_{\Gamma_{\textnormal{A}}(\code{C})}(z) = W_{\code{C}}(\vartheta_3(2z), \vartheta_2(2z))$, as we wanted to demonstrate. 
\end{IEEEproof}   

\begin{theorem}
  \label{thm:Funimodular_FSDcode}
  Consider an $(n,2^{\nicefrac{n}{2}})$ formally self-dual code $\code{C} \subseteq \Field_2^n$ that is geometrically uniform and $\vect{0}\in\code{C}$, then 
  \begin{IEEEeqnarray}{c}
    \Theta_{\GammaA{\code{C}}}(z)=W_{\code{C}}(\vartheta_3(2z),\vartheta_2(2z)) = \left(\frac{i}{z} \right)^{\nicefrac{n}{2}} \Theta_{\GammaA{\code{C}}}\left(-\frac{1}{z}\right),
    \label{eq:MacW-nonl-lattice}
  \end{IEEEeqnarray}
  and $\GammaA{\code{C}}$ is a formally unimodular packing.
\end{theorem}

\begin{IEEEproof}
  Expanding the left-hand side of (\ref{eq:MacW-nonl-lattice}), considering that $\code{C}$ is formally self-dual and \eqref{eq:MacWilliams-identity-fsd} holds, we get
  \begin{IEEEeqnarray}{rCl}
    W_{\code{C}}(\vartheta_3(2z),\vartheta_2(2z))& = &\frac{1}{2^{\nicefrac{n}{2}}} W_{\code{C}}\left(\vartheta_3(2z) + \vartheta_2(2z), \vartheta_3(2z)-\vartheta_2(2z)\right) \nonumber \\
    & \stackrel{\eqref{eq:add_sub}}{=} & \frac{1}{2^{\nicefrac{n}{2}}} W_{\code{C}} (\vartheta_3(\nicefrac{z}{2}), \vartheta_4(\nicefrac{z}{2}))
    \nonumber \\
    & \stackrel{\eqref{eq:mw_t3t4}}{=} &  \frac{1}{2^{\nicefrac{n}{2}}} W_{\code{C}} \left( \left(\frac{z}{2i}\right)^{-1/2} \vartheta_3\left(\frac{-2}{z} \right), \left(\frac{z}{2i}\right)^{-1/2} \vartheta_2\left(\frac{-2}{z} \right) \right)
    \nonumber \\
    & = & \frac{1}{2^{\nicefrac{n}{2}}} \left(\frac{2i}{z}\right)^{n/2} W_{\code{C}} \left( \vartheta_3\left(\frac{-2}{z} \right), \vartheta_2\left(\frac{-2}{z} \right) \right) \label{eq:justify}
    \\
    & = & \frac{2^{n/2}}{2^{\nicefrac{n}{2}}} \left(\frac{i}{z}\right)^{n/2} W_{\code{C}} \left( \vartheta_3\left(2 \cdot \frac{-1}{z} \right), \vartheta_2\left(2\cdot \frac{-1}{z} \right) \right)
    \nonumber \\
    & \stackrel{\textnormal{Lemma~\ref{lem:theta-series_non-lattice}}}{=} & \vol{\GammaA{\code{C}}} \left(\frac{i}{z} \right)^{\nicefrac{n}{2}} \Theta_{\GammaA{\code{C}}}\left(-\frac{1}{z}\right) =  \left(\frac{i}{z} \right)^{\nicefrac{n}{2}} \Theta_{\GammaA{\code{C}}}\left(-\frac{1}{z}\right),\nonumber
  \end{IEEEeqnarray}
  where in~\eqref{eq:justify}, the factor $\left(\frac{z}{2i}\right)^{-1/2}$ appears exactly $n$ times, because the sum of the degrees of each term of $W_{\code{C}}(x,y)$ is $n$. Since $\Theta_{\GammaA{\code{C}}}(z)=W_{\code{C}}(\vartheta_3(2z),\vartheta_2(2z))$, the proof is complete.
\end{IEEEproof}

We present now how to construct geometrically uniform codes where $\vect{0}\in\code{C}$. 

Consider $\Integers_4 = \{0,1,2,3\}$, the ring of integers modulo 4, and let $\code{C}_4\subseteq\Integers_4$ be a \emph{linear} code over $\Integers_4$, i.e., $\code{C}_4$ is an additive subgroup of $\Integers_4^n$. Then, we construct the binary code $\code{C}_{\textnormal{g}}$ as the binary image of $\code{C}_4$ under the Gray map $\psi\colon\Integers_4\rightarrow\Field_2\times\Field_2$, which maps 
\begin{IEEEeqnarray*}{c}
  0 \mapsto (0,0),\quad 1 \mapsto (0,1),\quad 2 \mapsto (1,1),\quad 3 \mapsto (1,0).
\end{IEEEeqnarray*}
This mapping can be naturally extended such that $\psi:\Integers_4^n \rightarrow \Field_2^{2n}$ and we define $\code{C}_{\textnormal{g}}=\psi(\code{C}_4)$.

We review several useful properties of binary codes obtained from Gray map.
\begin{remark}\leavevmode
  \label{rem:properties_Gray-map_linear-code_over-Z4}
  \begin{itemize}
  \item The code $\code{C}_{\textnormal{g}}$ is not necessarily linear, but it is geometrically uniform with respect to the Hamming metric~\cite{ForneySloaneTrott92_1}, which implies that the set of Hamming distances from a fixed codeword in $\code{C}_{\textnormal{g}}$ coincides with the set of Lee distances from a fixed codeword in $\code{C}_4$.
  \item If $\code{C}_4$ is also formally self-dual over $\Integers_4$ with respect to the Lee metric, then $\code{C}_{\textnormal{g}}$ is formally self-dual with respect to the Hamming metric~\cite[Th.~1]{Dougherty12_1}.
  \end{itemize}
\end{remark}
Hence, we can then draw the following conclusions regarding the packing $\GammaA{\code{C}_\textnormal{g}}=\frac{1}{\sqrt{2}}\left(\phi(\code{C}_{\textnormal{g}}) + 2\Integers^n\right):$
\textit{i)} when $\code{C}_{\textnormal{g}}$ is nonlinear, $\GammaA{\code{C}_{\textnormal{g}}}$ is a nonlattice packing (see Definition~\ref{def:ConstructionA}) and \textit{ii)} since $\code{C}_{\textnormal{g}}$ is geometrically uniform and ${\vect{0}} \in \code{C}_{\textnormal{g}}$, then $\GammaA{\code{C}_{\textnormal{g}}}$ is also a geometrically uniform packing. Therefore, the results of both Lemma~\ref{lem:theta-series_non-lattice} and Theorem~\ref{thm:Funimodular_FSDcode} apply.

The Nordstrom-Robinson code~\cite[Ch.~2.~{\S}8]{MacWilliamsSloane77_1} is a well known example of a nonlinear code and it can be constructed from the $\Integers_4$-linear \emph{octacode} through Gray map.
    
\begin{example} 
  \label{ex:nr_properties}
  The octacode $\code{O}_8\subseteq\Integers_4^8$ has the following generator matrix
  \begin{equation*}
    \mat{G} = \begin{pmatrix}
      1 & 0 & 0 & 0 & 2 & 1 & 1 & 1 \\
      0 & 1 & 0 & 0 & 1 & 2 & 1 & 3 \\
      0 & 0 & 1 & 0 & 1 & 3 & 2 & 1 \\
      0 & 0 & 0 & 1 & 1 & 1 & 3 & 2
    \end{pmatrix}.
  \end{equation*}
  The code $\code{N}_{16}=\psi(\code{O}_8)$ of length $16$ is the Nordstrom-Robinson code, which is the unique (up to translation or permutation) $(16, 256, 6)$ binary code and it is optimal in the sense that no $(16,M,6)$ binary code can have $M>256$ codewords~\cite{Snover73_1}. Moreover, since the octacode $\code{O}_8$ is linear and self-dual over $\Integers_4$, $\code{N}_{16}=\psi(\code{O}_8)$ is geometrically uniform, formally self-dual, and contains $\vect{0}$. Its weight enumerator is~\cite[p.~74]{MacWilliamsSloane77_1}
  \begin{IEEEeqnarray*}{c}
    \label{eq:wef_NR-code}
    W_{\code{N}_{16}}(x,y)=x^{16}+112x^{10}y^6+30x^8y^8+112x^6y^{10}+y^{16}.
  \end{IEEEeqnarray*}
  
  The packing $\Gamma_{\textnormal{A}}(\code{N}_{16})$ is lattice-like and has the following theta series, according to Lemma~\ref{lem:theta-series_non-lattice},
  \begin{IEEEeqnarray*}{rCl}
    \Theta_{\Gamma_{\textnormal{A}}(\code{N}_{16})}(z) & = &  W_{\code{N}_{16}}(\vartheta_3(2z), \vartheta_2(2z)) \\
    & = & 1 + 32 q^2+ 7168 q^3 +8160 q^4+ 258048 q^5+ 127360 q^6 + 2709504 q^7+\cdots.  %1016288 q^8+ 15769600 q^9+  4564416 q^{10}+ \dots 
    % & = &  \left(\frac{i}{z} \right)^{\nicefrac{n}{2}} \Theta_{\GammaA{\code{N}_{16}}}\left(-\frac{1}{z}\right).
  \end{IEEEeqnarray*}
\end{example}

\section{Secrecy Function of Lattice-like Packings}
\label{sec:secrecy-function_GU-packings}

\subsection{Coset Coding for Lattice-Like Packings }
\label{sec:coset-encoding-lattice-like-packings}

The coset coding idea proposed in Wyner's seminal paper~\cite{Wyner75_1} for linear codes has been applied to lattices on the Gaussian WTC, see \emph{e.g.}~\cite{BelfioreSole10_1,OggierSoleBelfiore16_1}. In this section we generalize the approach of ~\cite{BelfioreSole10_1,OggierSoleBelfiore16_1} to lattice-like packings. 

Geometric uniformity and periodicity are the properties that allow us to generalize the notion of secrecy function to a lattice-like packing. Indeed, if $\Gamma$ is lattice-like, then its volume and theta series are well-defined, and all Voronoi regions have the same shape. This indicates that \emph{these} packings, for most purposes, behave ``like lattices'' when used in the coset coding scheme for the Gaussian wiretap channel, such as in the discussion on error probability for coset decoding, for example. 

The classical coset coding strategy is done via two nested lattices $\Lambda_\textnormal{e}\subset\Lambda_\textnormal{b}$. It requires that $\Lambda_\textnormal{b}$ is partitioned into a union of disjoint cosets of the form as
\begin{IEEEeqnarray}{c}  
  \Lambda_\textnormal{b}=\bigcup_{j=1}^{2^k}\bigl(\vect{v}_j+\Lambda_\textnormal{e}\bigr),
  \label{eq:expression_coset-decomposition}
\end{IEEEeqnarray}
where $\vect{v}_j$, $j\in[1:2^k]$, are some vectors that do not belong to $\Lambda_{\textnormal{e}}$. For lattices, the guarantee of a coset decomposition as in~\eqref{eq:expression_coset-decomposition} for two nested lattices has been discussed to some extent~\cite[Ch.~4]{Robinson15_1},~\cite[Ch.~8]{Zamir14_1}. However, for lattice-like packings, conditions that guarantee a coset decomposition like~\eqref{eq:expression_coset-decomposition} are not yet investigated in the literature. An example explaining the challenges of coset decomposition is presented below.
\begin{example}
  \label{ex:gu_lattice}
  Consider $\code{C}=\{\vect{c}_1,\vect{c}_2,\vect{c}_3,\vect{c}_4\}=\{(0,0,0,0),(1,1,0,0),(1,0,1,0),(1,0,0,1)\}\subseteq\Field_2^4$. Observe that $\code{C}$ is nonlinear, but geometrically uniform. Indeed, given any $\vect{c}, \vect{c}'\in\code{C}$, there exists an isometry $T_{\vect{c},\vect{c}'} = P_{\vect{c},\vect{c}'}\circ Q_{\vect{c},\vect{c}'}$, where $P_{\vect{c},\vect{c}'}$ is a permutation and $Q_{\vect{c},\vect{c}'}$ is a translation, such that $T_{\vect{c},\vect{c}'}(\vect{c})=\vect{c}'$ and $T_{\vect{c},\vect{c}'}(\code{C})=\code{C}$. Hence, we can write $\Gamma_{\textnormal{e}} = \GammaA{\code{C}}=\bigcup_{\ell=1}^4 ({\bm c}_{\ell}+2\Integers^4)$ (disregarding the scalar $\nicefrac{1}{\sqrt{2}}$) as a lattice-like packing. We want to write $\Gamma_{\textnormal{b}}$ as a union of cosets of $\Gamma_{\textnormal{e}}$ like~\eqref{eq:expression_coset-decomposition}, i.e., 
  \begin{IEEEeqnarray*}{c}
    \Gamma_{\textnormal{b}} = \bigcup_{j=1}^{2^k} (\vect{b}_j+\Gamma_{\textnormal{e}}) = \bigcup_{j=1}^{2^k}\bigcup_{\ell=1}^4 \bigl(\vect{b}_j\oplus\vect{c}_{\ell}+2\Integers^4\bigr) = \bigcup_{i=1}^{4\cdot 2^k} \bigl(\vect{a}_{i}+2\Integers^4\bigr),
  \end{IEEEeqnarray*}
  where $\vect{a}_i=\vect{b}_j\oplus\vect{c}_\ell$, $j\in[1:2^k],\ell\in[1:4]$. The choices of $\vect{b}_j$ would impact the characteristics of $\Gamma_{\textnormal{b}}$ and the possibilities are, for example:  
  \begin{description}
  \item[i) Lattice:]  Let $\code{B}=\{\vect{b}_1,\vect{b}_2,\vect{b}_3,\vect{b}_4\} = \{(0,0,0,0), (1,1,1,1), (0,0,0,1),(1,1,1,0)\}\subseteq\Field_2^4$. Then, $\Gamma_{\textnormal{b}}= \bigcup_{i=1}^{16}(\vect{a}_i+2\Integers^4)$, where $\code{A}=\{\vect{a}_i\}_{i=1}^{16}$ is the linear code $\Field_2^4$, and $\Lambda_{\textnormal{b}}=\GammaA{\Field_2^4}$ is a lattice. 
    
  \item[ii) Nonlattice:] Let $\code{B}=\{\vect{b}_1,\vect{b}_2\} = \{(0,0,0,0), (1,1,1,0)\}$. Then, $\Gamma_{\textnormal{b}} = \bigcup_{i=1}^{8} (\vect{a}_i+2\Integers^4)$, where $\code{A}=\{\vect{a}_i\}_{i=1}^{8}=\{(0,0,0,0), (1,1,0,0), (1,0,1,0), (1,0,0,1), (1,1,1,0), (0,0,1,0), (0,1,0,0), (0,1,1,1)\}$. Because $\code{A}$ is nonlinear, $\GammaA{\code{A}}$ is not a lattice. Moreover, since $\code{A}$ is not distance-invariant, and therefore, not geometrically uniform, $\GammaA{\code{A}}$ is also not a lattice-like packing.
    
  \item[iii) Indecomposable:] Let $\code{B}=\{\vect{b}_1,\vect{b}_2\} = \{(0,0,0,0), (0,1,1,0)\}$. Then, $\Gamma_{\textnormal{b}} = \bigcup_{j=1}^{2} (\vect{b}_j+\Gamma_{\textnormal{e}})$ can not be written as a union of disjoint cosets since, for example, a binary vector $(1,0,1,0)\in\code{A}=\{\vect{a}_i\}_{i=1}^{8}$, can be represented in two different ways as follows.
    \begin{IEEEeqnarray*}{rCl}
      (1,0,1,0) = \underbrace{(0,0,0,0)}_{\vect{b}_1}\oplus \underbrace{(1,0,1,0)}_{\in\code{C}} = \underbrace{(0,1,1,0)}_{\vect{b}_2} \oplus \underbrace{(1,1,0,0)}_{\in\code{C}}.
    \end{IEEEeqnarray*}
    Hence, there exists a vector $\vect{a}\in\code{A}$ that can not be uniquely written as $\vect{a}=\vect{b}+\vect{c}$ over $\Field_2$, $\vect{b}\in\code{B}$, $\vect{c}\in\code{C}$. Therefore, this leads to that $(\vect{b}_1+\Gamma_{\textnormal{e}})\cap(\vect{b}_2+\Gamma_{\textnormal{e}})\neq\emptyset$, which prevents this pair of packings from being used in our context.
  \end{description}    
  % The secrecy function of $\Gamma_{\textnormal{e}}$ is invariant on the choices of ${\bm u}_j$, but the resulting constellation should be a lattice in our application. The performance of the secrecy function is shown in Figure~\ref{fig:secrecy_function_d4} and the (strong) secrecy gain is $1$. For this particular example, the notion of weak secrecy gain is not useful, since the symmetry refers to the minima and not the maxima. Notice that $\code{C}$ is not formally self-dual.
  
  % \begin{figure}[t!]
  %   \centering
  %   \input{\Figs/P4D.tex}
  %   %   \vspace{-1ex}
  %   \caption{The secrecy function of $\GammaA{\code{C}}$, as a function of $\tau$ in dB.}
  %   \label{fig:secrecy_function_d4}
  %   %   \vspace{-2ex}
  % \end{figure}
  
\end{example}

 Our main objective is the following: given a lattice-like packing $\Gamma_{\textnormal{e}}$, we aim to find a lattice $\Lambda_{\textnormal{b}}$ such that $\Gamma_{\textnormal{e}} \subseteq \Lambda_{\textnormal{b}}$. Examples are described below with our previous examples of lattice-like packings.

\begin{example}
  \label{ex:Voronoi-regions_Gamme_e-Lambda_b}
   We consider here the lattice-like packings presented in Examples~\ref{ex:ex_ConstrC-LatticeLike_n2} and~\ref{ex:ex_ConstrA-LatticeLike-packing_n2}. For the first construction, $\Lambda_{\textnormal{b}} = \bigcup_{i=1}^{4}(\vect{a}_i+4\Integers^n)$, where $\vect{a}_i \in \{(0,0),(1,1),(2,2),(3,3)\}$. For the second one, $\Lambda_{\textnormal{b}} = \GammaA{\code{A}}$, where $\code{A} = \{(0,0),(1,0),(0,1),(1,1)\}$. Their corresponding Voronoi regions are represented in Figure~\ref{fig:voronoi_lattice_sub}. The points in black represent $\Gamma_{\textnormal{e}}$, while the lattice points of $\Lambda_{\textnormal{b}}$ are the union of the black and orange points.
  \begin{figure}[t!]
    \centering 
    \begin{minipage}[t]{7cm} 
      \centering 
      \definecolor{ffzztt}{rgb}{1.0,0.6,0.2}
\definecolor{ccqqww}{rgb}{0.8,0.0,0.4}
\definecolor{ttqqqq}{rgb}{0.2,0.0,0.0}

\begin{tikzpicture}[line cap=round,line join=round,x=1.0cm,y=1.0cm]
\pgfplotsset{every tick label/.append style={font=\scriptsize}}
\begin{axis}[
x=1.0cm,y=1.0cm,
axis lines=middle,
xmin=-2.5,
xmax=2.5,
ymin=-2.5,
ymax=2.5,
xtick={-2.0,-1.0,...,2.0},
xticklabels={$-2~~$,$-1~~$,$0$,$1$,$2$},
ytick={-2.0,-1.0,...,2.0},]
\clip(-2.5,-2.5) rectangle (2.5,2.5);
\fill[line width=1.25pt,fill=black,fill opacity=0.10000000149011612] (-2.,-1.) -- (-1.,-2.) -- (2.,-1.) -- (-1.,2.) -- cycle;
\fill[line width=1.25pt,color=ffzztt,fill=ffzztt,fill opacity=0.25] (-0.5,1.5) -- (-1.5,0.5) -- (0.5,-1.5) -- (1.5,-0.5) -- cycle;
\draw[line width=3.pt,fill=black,fill opacity=0.10000000149011612] (-14.338294331895177,2.087654109682274) -- (-11.941113877660072,2.087654109682274);
\draw[line width=3.pt,fill=black,fill opacity=0.10000000149011612] (-14.326308429624001,1.6321898233776042) -- (-11.929127975388894,1.6321898233776042);
\draw [line width=1.25pt] (-2.,-1.)-- (-1.,-2.);
\draw [line width=1.25pt] (-1.,-2.)-- (2.,-1.);
\draw [line width=1.25pt] (2.,-1.)-- (-1.,2.);
\draw [line width=1.25pt] (-1.,2.)-- (-2.,-1.);
\draw [line width=1.25pt,color=ffzztt] (-0.5,1.5)-- (-1.5,0.5);
\draw [line width=1.25pt,color=ffzztt] (-1.5,0.5)-- (0.5,-1.5);
\draw [line width=1.25pt,color=ffzztt] (0.5,-1.5)-- (1.5,-0.5);
\draw [line width=1.25pt,color=ffzztt] (1.5,-0.5)-- (-0.5,1.5);
\begin{scriptsize}
\draw [fill=black] (-11.941113877660072,2.087654109682274) circle (2.5pt);
\draw[color=black] (-11.89317026857537,2.3693228130548984) node {$n = 3$};
\draw [fill=black] (-12.728188126800598,1.6321898233776042) circle (2.5pt);
\draw[color=black] (-12.684239818472955,1.913858526750229) node {$i = 1$};
\draw [fill=ttqqqq] (13.,5.) circle (3.5pt);
\draw [fill=ttqqqq] (12.,4.) circle (3.5pt);
\draw [fill=ttqqqq] (0.,0.) circle (3.5pt);
\draw [fill=ttqqqq] (1.,1.) circle (3.5pt);
\draw [fill=ttqqqq] (-3.,1.) circle (3.5pt);
\draw [fill=ttqqqq] (4.,0.) circle (3.5pt);
\draw [fill=ttqqqq] (-4.,0.) circle (3.5pt);
\draw [fill=ttqqqq] (-3.,-3.) circle (3.5pt);
\draw [fill=ttqqqq] (1.,-3.) circle (3.5pt);
\draw [fill=ttqqqq] (5.,-3.) circle (3.5pt);
\draw [fill=ttqqqq] (5.,1.) circle (3.5pt);
\draw [fill=ttqqqq] (-4.,4.) circle (3.5pt);
\draw [fill=ttqqqq] (-3.,5.) circle (3.5pt);
\draw [fill=ttqqqq] (0.,4.) circle (3.5pt);
\draw [fill=ttqqqq] (1.,5.) circle (3.5pt);
\draw [fill=ttqqqq] (4.,4.) circle (3.5pt);
\draw [fill=ttqqqq] (5.,5.) circle (3.5pt);
\draw [fill=ttqqqq] (8.,4.) circle (3.5pt);
\draw [fill=ccqqww] (-7.,-3.) circle (3.5pt);
\draw [fill=ccqqww] (-4.,-4.) circle (3.5pt);
\draw [fill=ccqqww] (0.,-4.) circle (3.5pt);
\draw [fill=ccqqww] (4.,-4.) circle (3.5pt);
\draw [fill=ccqqww] (8.,-4.) circle (3.5pt);
\draw [fill=ccqqww] (9.,-3.) circle (3.5pt);
\draw [fill=ccqqww] (8.,0.) circle (3.5pt);
\draw [fill=ccqqww] (9.,1.) circle (3.5pt);
\draw [fill=ccqqww] (9.,5.) circle (3.5pt);
\draw [fill=ccqqww] (-7.,1.) circle (3.5pt);
\draw [fill=ccqqww] (-7.,5.) circle (3.5pt);
\draw [fill=ccqqww] (-8.,4.) circle (3.5pt);
\draw [fill=ccqqww] (12.,4.) circle (3.5pt);
\draw [fill=ccqqww] (13.054794766825841,4.9823096629932415) circle (3.5pt);
\draw [fill=ccqqww] (13.008465258525913,0.9979719491995062) circle (3.5pt);
\draw [fill=ccqqww] (12.,0.) circle (3.5pt);
\draw [fill=ccqqww] (-8.,0.) circle (3.5pt);
\draw [fill=ccqqww] (-10.990220040836448,5.005474417143205) circle (3.5pt);
\draw [fill=ccqqww] (-12.,4.) circle (3.5pt);
\draw [fill=ccqqww] (-10.990220040836448,0.9748071950495426) circle (3.5pt);
\draw [fill=ccqqww] (-12.,0.) circle (3.5pt);
\draw [fill=ccqqww] (-8.,-4.) circle (3.5pt);
\draw [fill=ccqqww] (12.,-4.) circle (3.5pt);
\draw [fill=ffzztt] (-1.,-1.) circle (2.5pt);
\draw [fill=ffzztt] (-2.,2.) circle (2.5pt);
\draw [fill=ffzztt] (-1.,3.) circle (2.5pt);
\draw [fill=ffzztt] (2.,2.) circle (2.5pt);
\draw [fill=ffzztt] (3.,-1.) circle (2.5pt);
\draw [fill=ffzztt] (2.,-2.) circle (2.5pt);
\draw [fill=ffzztt] (-2.,-2.) circle (2.5pt);
\draw [fill=ffzztt] (-5.,-1.) circle (2.5pt);
\draw [fill=ffzztt] (-6.,-2.) circle (2.5pt);
\draw [fill=ffzztt] (3.,3.) circle (2.5pt);
\end{scriptsize}
\end{axis}
\end{tikzpicture}  
      % \caption{} 
    \end{minipage} 
    \hspace{0.5cm} 
    \begin{minipage}[t]{7cm} 
      \centering 
      \definecolor{ffzztt}{rgb}{1.0,0.6,0.2}

\begin{tikzpicture}[line cap=round,line join=round,x=1.0cm,y=1.0cm]
\pgfplotsset{every tick label/.append style={font=\scriptsize}}
\begin{axis}[
x=1.0cm,y=1.0cm,
axis lines=middle,
xmin=-2.5,
xmax=2.8107991966790844,
ymin=-2.5,
ymax=2.4333544233183186,
xtick={-2.0,-1.0,...,2.0},
xticklabels={$-2~~$,$-1~~$,$0$,$1$,$2$},
ytick={-2.0,-1.0,...,2.0},]
\clip(-2.5,-2.5) rectangle (2.8107991966790844,2.4333544233183186);
\fill[line width=1.25pt,fill=black,fill opacity=0.10000000149011612] (0.,0.) -- (1.,-1.) -- (2.,0.) -- (1.,1.) -- cycle;
\fill[line width=1.25pt,fill=black,fill opacity=0.10000000149011612] (5.,5.) -- (6.,4.) -- (5.,3.) -- (4.,4.) -- cycle;
\fill[line width=1.25pt,color=ffzztt,fill=ffzztt,fill opacity=0.25] (0.5,0.5) -- (1.5,0.5) -- (1.5,-0.5) -- (0.5,-0.5) -- cycle;
\draw[line width=3.pt,fill=black,fill opacity=0.10000000149011612] (-9.839619515014483,1.994423132655485) -- (-7.698491267878704,1.994423132655485);
\draw[line width=3.pt,fill=black,fill opacity=0.10000000149011612] (-9.8824420799572,1.512669277049936) -- (-7.74131383282142,1.512669277049936);
\draw [line width=1.25pt] (0.,0.)-- (1.,-1.);
\draw [line width=1.25pt] (1.,-1.)-- (2.,0.);
\draw [line width=1.25pt] (2.,0.)-- (1.,1.);
\draw [line width=1.25pt] (1.,1.)-- (0.,0.);
\draw [line width=1.25pt] (5.,5.)-- (6.,4.);
\draw [line width=1.25pt] (6.,4.)-- (5.,3.);
\draw [line width=1.25pt] (5.,3.)-- (4.,4.);
\draw [line width=1.25pt] (4.,4.)-- (5.,5.);
\draw [line width=1.25pt,color=ffzztt] (0.5,0.5)-- (1.5,0.5);
\draw [line width=1.25pt,color=ffzztt] (1.5,0.5)-- (1.5,-0.5);
\draw [line width=1.25pt,color=ffzztt] (1.5,-0.5)-- (0.5,-0.5);
\draw [line width=1.25pt,color=ffzztt] (0.5,-0.5)-- (0.5,0.5);
\begin{scriptsize}
\draw [fill=black] (-9.48276480715852,1.994423132655485) circle (2.5pt);
\draw[color=black] (-9.411393865587327,2.2460057016939388) node {$n = -2$};
\draw [fill=black] (-9.8824420799572,1.512669277049936) circle (2.5pt);
\draw[color=black] (-9.807502591307447,1.7642518460883894) node {$i = -3$};
\draw [fill=black] (-3.,-6.) circle (3.5pt);
\draw [fill=black] (-4.,-5.) circle (3.5pt);
\draw [fill=black] (-1.,4.) circle (3.5pt);
\draw [fill=black] (-2.,3.) circle (3.5pt);
\draw [fill=black] (0.,3.) circle (3.5pt);
\draw [fill=black] (1.,4.) circle (3.5pt);
\draw [fill=black] (2.,3.) circle (3.5pt);
\draw [fill=black] (3.,4.) circle (3.5pt);
\draw [fill=black] (4.,3.) circle (3.5pt);
\draw [fill=black] (-2.,1.) circle (3.5pt);
\draw [fill=black] (-1.,2.) circle (3.5pt);
\draw [fill=black] (0.,1.) circle (3.5pt);
\draw [fill=black] (1.,2.) circle (3.5pt);
\draw [fill=black] (2.,1.) circle (3.5pt);
\draw [fill=black] (3.,2.) circle (3.5pt);
\draw [fill=black] (4.,1.) circle (3.5pt);
\draw [fill=black] (4.,-1.) circle (3.5pt);
\draw [fill=black] (3.,0.) circle (3.5pt);
\draw [fill=black] (1.,0.) circle (3.5pt);
\draw [fill=black] (-1.,0.) circle (3.5pt);
\draw [fill=black] (-2.,-1.) circle (3.5pt);
\draw [fill=black] (0.,-1.) circle (3.5pt);
\draw [fill=black] (2.,-1.) circle (3.5pt);
\draw [fill=black] (-1.,-2.) circle (3.5pt);
\draw [fill=black] (1.,-2.) circle (3.5pt);
\draw [fill=black] (3.,-2.) circle (3.5pt);
\draw [fill=black] (-2.,-3.) circle (3.5pt);
\draw [fill=black] (0.,-3.) circle (3.5pt);
\draw [fill=black] (2.,-3.) circle (3.5pt);
\draw [fill=black] (4.,-3.) circle (3.5pt);
\draw [fill=black] (-3.,2.) circle (3.5pt);
\draw [fill=black] (-3.,4.) circle (3.5pt);
\draw [fill=black] (0.,5.) circle (3.5pt);
\draw [fill=black] (2.,5.) circle (3.5pt);
\draw [fill=black] (4.,5.) circle (3.5pt);
\draw [fill=black] (5.,4.) circle (3.5pt);
\draw [fill=black] (5.,2.) circle (3.5pt);
\draw [fill=black] (5.,0.) circle (3.5pt);
\draw [fill=black] (6.,1.) circle (3.5pt);
\draw [fill=black] (6.,-1.) circle (3.5pt);
\draw [fill=black] (-3.,0.) circle (3.5pt);
\draw [fill=black] (-4.,1.) circle (3.5pt);
\draw [fill=black] (-4.,3.) circle (3.5pt);
\draw [fill=black] (-4.,5.) circle (3.5pt);
\draw [fill=black] (-2.,5.) circle (3.5pt);
\draw [fill=ffzztt] (0.,0.) circle (2.5pt);
\draw [fill=ffzztt] (-1.,1.) circle (2.5pt);
\draw [fill=ffzztt] (-1.,-1.) circle (2.5pt);
\draw [fill=ffzztt] (1.,-1.) circle (2.5pt);
\draw [fill=ffzztt] (1.,1.) circle (2.5pt);
\draw [fill=ffzztt] (2.,0.) circle (2.5pt);
\draw [fill=ffzztt] (2.,2.) circle (2.5pt);
\draw [fill=ffzztt] (2.,-2.) circle (2.5pt);
\draw [fill=ffzztt] (0.,-2.) circle (2.5pt);
\draw [fill=ffzztt] (-2.,-2.) circle (2.5pt);
\draw [fill=ffzztt] (-2.,2.) circle (2.5pt);
\draw [fill=ffzztt] (-3.,1.) circle (2.5pt);
\draw [fill=ffzztt] (-3.,-1.) circle (2.5pt);
\draw [fill=ffzztt] (0.,2.) circle (2.5pt);
\draw [fill=ffzztt] (1.,-3.) circle (2.5pt);
\draw [fill=ffzztt] (-1.,-3.) circle (2.5pt);
\draw [fill=ffzztt] (-3.,-2.) circle (2.5pt);
\end{scriptsize}
\end{axis}
\end{tikzpicture}  
      % \caption{} 
    \end{minipage} 
    \caption{Voronoi regions of two lattice-like packings, $\Gamma_{\textnormal{e}}\subset\Lambda_{\textnormal{b}}$.}
    \label{fig:voronoi_lattice_sub}
  \end{figure}
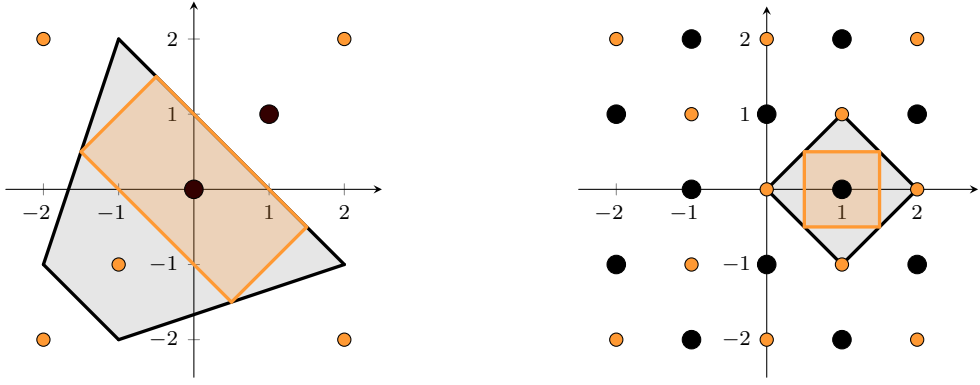
\end{example}

Our work is focused on Construction A packings. Hence, consider a geometrically uniform code $\code{C}\subseteq\Field_2^n$ where $\card{\code{C}}=M$ and $\Gamma_{\textnormal{e}}=\GammaA{\code{C}} = \bigcup_{\ell=1}^{M}(\vect{c}_{\ell}+2\Integers^n)$, $\vect{c}_{\ell}\in\code{C}$, we are interested in some conditions on $\vect{b}_j$, $j\in [1:2^k]$, such that
\begin{IEEEeqnarray}{c}
  \Gamma_{\textnormal{b}}=\bigcup_{j=1}^{2^k}\bigcup_{\ell=1}^{M}(\vect{b}_j\oplus {\bm c}_{\ell}+2\Integers^n)
  =\bigcup_{i=1}^{M\cdot 2^k}(\vect{a}_i+2\Integers^n)=\GammaA{\code{A}}  
  \label{eq:Lambda_b_binary-case}
\end{IEEEeqnarray} 
is a lattice, and those cosets $\{\vect{a}_i+2\Integers^n\}$ are disjoint, where $\code{A}\eqdef\{\vect{a}_i\colon\vect{a}_i=\vect{b}_j\oplus\vect{c}_\ell,\,j\in [1:2^k],\,\ell\in [1:M]\}\subseteq \Field_2^n$. Since Construction A packings are lattices if and only if the underlying code is linear, we will work directly with binary codes of $\code{A}$ and $\code{C}$, such that $\Gamma_\textnormal{e}=\GammaA{\code{C}}$ is a lattice-like packing and $\Lambda_\textnormal{b}=\GammaA{\code{A}}$ is a lattice. Here, to simplify the analysis further, $\code{B}$ is assumed to be linear.

Now, consider a geometrically uniform code $\code{C}=\{\vect{c}_1,\dots,\vect{c}_{M}\}\subseteq\Field_2^n$. Let us choose $\code{A}=\espn{\code{C}}$, the smallest linear code that contains $\code{C}$. We first show that it is always possible to find a code $\code{B}$ such that~\eqref{eq:Lambda_b_binary-case} holds. Observe that since $\code{A}=\{\vect{a}_i\colon\vect{a}_i=\vect{b}_j\oplus\vect{c}_\ell\}=\espn{\code{C}}$ would be linear, we have $\vect{b}\oplus\tilde{\vect{c}}=\vect{c}\oplus\hat{\vect{c}}$ for all distinct vectors $\vect{c},\hat{\vect{c}},\tilde{\vect{c}}\in\code{C}$. Thus, $\vect{b} = \vect{c}\oplus\hat{\vect{c}}\oplus\tilde{\vect{c}}$. In addition, we have to guarantee that the representation $\vect{a}_i = \vect{b_j}\oplus\vect{c_\ell}$ is unique, so the code $\code{B}$ and $\code{C}$ have to satisfy the following proposition adapted from~\cite[Prop.~1]{Dinitz06_1}. Here, we define $\code{C}-\code{C}\eqdef\{\vect{c}_1\oplus\vect{c}_2\colon\vect{c}_1,\vect{c}_2\in\code{C}\}$, as the operations are defined over $\Field_2^n$.
\begin{proposition}[{\cite[Prop.~1]{Dinitz06_1}}]
  \label{prop:unique-decomposition_cosets}
  Given $\code{A}\subseteq\Field_2^n$ a linear code and $\code{B},\code{C}\subseteq\Field_2^n$ two codes (linear or not) where $\vect{0}\in\code{B},\code{C}$. Then, every element $\vect{a}\in\code{A}$ can be uniquely written as $\vect{a} = \vect{b} \oplus \vect{c}$, with $\vect{b}\in\code{B}$ and $\vect{c}\in\code{C}$, if and only if $(\code{B}-\code{B}) \cap (\code{C} -\code{C}) = \{\vect{0}\}$ and $\card{\code{A}}=\card{\code{B}}\card{\code{C}}$.
\end{proposition}

\begin{example}
  \label{ex:decomposition-examples}
  We continue with Example~\ref{ex:gu_lattice}. Let us first choose
  \begin{IEEEeqnarray*}{rCl}
    \code{A}& = &\espn{\{(1,1,0,0),(1,0,1,0),(1,0,0,1)\}}
    \\
    & = &\{(0,0,0,0),(1,1,0,0),(0,0,1,1),(1,0,0,1),(0,1,0,1),(1,0,1,0),(0,1,1,0),(1,1,1,1)\}.
  \end{IEEEeqnarray*}
  Then, $\vect{b}$ can be $(1,1,0,0)\oplus(1,0,1,0)\oplus(1,0,0,1)=(1,1,1,1)$. Since we want $\code{B}$ to be linear, $\code{B}=\{(0,0,0,0),(1,1,1,1)\}$. In addition, it can be seen that
  \begin{IEEEeqnarray*}{rCl}
    (\code{B}-\code{B})\cap(\code{C}-\code{C})& = &\{(0,0,0,0),(1,1,1,1)\}\cap\{(0,0,0,0),(1,1,0,0),
  \nonumber\\
  &&\quad \>(1,0,1,0),(1,0,0,1),(0,1,1,0),(1,0,0,1),(0,0,1,1)\}=\vect{0},\IEEEeqnarraynumspace
  \end{IEEEeqnarray*}
  which satisfies the condition of Proposition~\ref{prop:unique-decomposition_cosets}. Let $\code{A}=\Field_2^n=\spn{\code{C}\cup\{(1,0,0,0)\}}$, and in this case, $\vect{b}$ can be $(1,1,1,1)$ or $(1,1,0,0)\oplus(1,0,1,0)\oplus(1,0,0,0)=(1,1,1,0)$. We can stop looking for more $\vect{b}$ vectors as we want $\code{B}$ to be linear. Hence, $\code{B} = \{(0,0,0,0),(1,1,1,1),(1,1,1,0),(0,0,0,1)\}$, and one can check that this $\code{B}$ also satisfies Proposition~\ref{prop:unique-decomposition_cosets}.
\end{example}

We remark here that the construction for the coset decomposition as in~\eqref{eq:Lambda_b_binary-case} is an interesting direction to study. However, in this work, we mainly focus on the analysis of secrecy gain of lattice-like packings.

\subsection{Probability Analysis}
\label{sec:prob-analysis}

Based on the discussion of the previous subsection, we consider two packings, $\Gamma_\textnormal{e}\subset\Gamma_\textnormal{b}$, where $\Gamma_\textnormal{e}$ is a lattice-like packing and $\Gamma_\textnormal{b}=\Lambda_\textnormal{b}$ is a lattice. This will be our assumption from now on. $\Lambda_\textnormal{b}$ is designed to ensure reliability for a legitimate receiver Bob and required to have a good coding gain.\footnote{Quantified in terms of the \textit{Hermite's constant}, defined as $\gamma_n = \frac{d_{\textnormal{min}}^2(\Gamma)}{\vol{\Gamma}^{\nicefrac{n}{2}}}$~\textnormal{\cite[p.~20]{ConwaySloane90_1}}.} The packing $\Gamma_\textnormal{e}$ is aimed to increase the eavesdropper Eve's confusion, so it should be chosen to  minimize $P_{\textnormal{c,e}}$, the eavesdropper's success probability of correctly guessing the transmitted message, or (almost) equivalently, to maximize Eve's equivocation conditioned on the channel output.

We start by writing $\Lambda_{\textnormal{b}} = \bigcup_{j=1}^{2^k} (\vect{u}_j+\Gamma_{\textnormal{e}})$. A lattice coset coding scheme works as follows: Alice wants to transmit a message $\vect{s}=(s_1,\ldots,s_{k}) \in \{0,1\}^k$ to Bob. The message ${\bm s}$ is mapped to a coset ${\vect{u}_j}={\bm u}_{j(s)}$. Alice selects a \emph{random} vector ${\bm r} \in \Gamma_{\textnormal{e}}$ and transmits a codeword $\vect{x}={\bm u}_j+{\bm r}$ over the Gaussian WTC. Bob, the legitimate receiver, is assumed to have a channel of sufficient quality to enable correct decoding, while the eavesdropper Eve, has an inferior SNR, i.e., $\sigma_{\textnormal{b}}^2 \ll \sigma_{\textnormal{e}}^2$.

When a message $\vect{x}\in\Reals^n$ is transmitted over an additive Gaussian channel with variance $\sigma^2$, the decoder makes the correct decision if and only if the received vector $\vect{y}=\vect{x}+\vect{h}$ at the destination lies in $\set{V}(\vect{x})$, where $\vect{h}$ is a vector that has $n$ independent and identically distributed Gaussian random variables, each has mean $0$ and variance $\sigma^2$. This gives the  probability of correct decoding:
\begin{IEEEeqnarray*}{c}
  \frac{1}{(2\pi\sigma^2)^{\nicefrac{n}{2}}} \int_{\set{V}(\vect{x})} e^{-\frac{\enorm{\vect{y}-\vect{x}}^2}{2\sigma^2}}\dd{\vect{y}}.
\end{IEEEeqnarray*}

Suppose a lattice vector $\vect{x} = \vect{u}_j+\vect{r} \in \Lambda_{\textnormal{b}}$, $j\in[1:2^k]$, is transmitted, where a random vector $\vect{r}$ is chosen from $\Gamma_\textnormal{e}$. Then, the  probability of correctly guessing the transmitted message, $P_{\textnormal{c}}$, can be shown to be bounded from above by~\cite[Appendix~B]{OggierSoleBelfiore16_1} % (ignoring boundary effects)
\begin{IEEEeqnarray}{c}
  P_{\textnormal{c}}\leq\frac{1}{(2\pi\sigma^2)^{\nicefrac{n}{2}}}\sum_{\vect{r}\in\Gamma_{\textnormal{e}}}\int_{\set{V}_{\Lambda_{\textnormal{b}}}(\vect{x} + \vect{r})} e^{-\frac{\enorm{\vect{y}-\vect{x}}^2}{2\sigma^2}}\dd{\vect{y}}.
  \label{eq:pc_general}
\end{IEEEeqnarray}

Since all the Voronoi regions of a lattice are independent on the choice of $\vect{x}\in\Lambda_{\textnormal{b}}$ and are equal to $\set{V}(\Lambda_\textnormal{b})$, \eqref{eq:pc_general} indicates that $P_{\textnormal{c,e}}$ in an additive Gaussian channel with variance $\sigma_\textnormal{e}^2$ is bounded from above by
\begin{IEEEeqnarray*}{rCl}
  P_{\textnormal{c,e}}& \leq &\frac{1}{(2\pi\sigma^2_\textnormal{e})^{\nicefrac{n}{2}}}\sum_{\vect{r}\in \Gamma_{\textnormal{e}}}\int_{\set{V}_{\Lambda_{\textnormal{b}}}(\vect{x} + \vect{r})} e^{-\frac{\enorm{\vect{y}-\vect{x}}^2}{2\sigma_{\textnormal{e}}^2}}\dd\vect{y}
  \nonumber \\
  % & \stackrel{\textnormal{\cite[eq.~(41)]{OggierSoleBelfiore16_1}}}{\leq} &
  & = &
  \frac{1}{(2\pi\sigma^2_\textnormal{e})^{\nicefrac{n}{2}}}\sum_{\vect{r}\in\Gamma_{\textnormal{e}}}\int_{\set{V}(\Lambda_{\textnormal{b}})} e^{-\frac{\enorm{\vect{w}+\vect{r}}^2}{2\sigma_{\textnormal{e}}^2}}\dd{\vect{w}},\IEEEeqnarraynumspace\IEEEyesnumber\label{eq:1st-bound_Eve-success-probability}
  % \nonumber \\
  % & \stackrel{\textnormal{\cite[eq.~(45)]{OggierSoleBelfiore16_1}}}{\leq} & \frac{1}{(\sqrt{2 \pi}\sigma_\textnormal{e})^n} \vol{\Lambda_{\textnormal{b}}}  \sum_{{\bm x} \in \Gamma_{\textnormal{e}}} e^{-\frac{\norm{\vect{x}}^2}{2\sigma_\textnormal{e}^2}}.
\end{IEEEeqnarray*}
where~\eqref{eq:1st-bound_Eve-success-probability} holds by using the change of variable $\vect{w}=\vect{y}-\vect{x}-\vect{r}$.

In~\cite[eq.~(45)]{OggierSoleBelfiore16_1}, the authors use the Poisson summation formula for lattices to further bound $P_{\textnormal{c,e}}$ for the setting of a pair of nested lattices $\Lambda_{\textnormal{e}}\subset\Lambda_{\textnormal{b}}$. However, we need to generalize this bound in order to adapt it to our setup, where $\Lambda_\textnormal{b}$ is a lattice and its subset $\Gamma_\textnormal{e}$ is a lattice-like packing. To this end, we first introduce the following two useful lemmas.
\begin{lemma}[{Poisson Summation Formula~\cite[Th.~2.3]{Ebeling13_1},~\cite[Appendix~C]{OggierSoleBelfiore16_1}}] %~\cite[Lemma]{OdlyzkoSloane80_1}
  \label{lem:Poisson-summation}
  Let $\Lambda\subset\Reals^n$ be an arbitrary lattice and $f\colon \Reals^n\rightarrow\Complex$ be a % rapidly decreasing smooth
  function satisfying
  \begin{enumerate}
  \item $\int_{\Reals^n}\abs{f(\vect{t})}\dd\vect{t}<\infty$,
  \item The infinite series $\sum_{\vect{\lambda}\in\Lambda}\abs{f(\vect{\lambda}+\vect{u})}$ converges uniformly for all $\vect{u}$ belonging to a compact subset of $\Reals^n$,
  \item The infinite series $\sum_{\vect{\lambda}^\star\in\Lambda^\star}\bigabs{\hat{f}(\vect{\lambda}^\star)}$ converges, where
    \begin{IEEEeqnarray*}{c}
      \hat{f}(\vect{\lambda}^\star)\eqdef\int_{\Reals^n}f(\vect{t})e^{-2\pi i\inner{\vect{t}}{\vect{\lambda}^\star}}\dd\vect{t}.
    \end{IEEEeqnarray*}
  \end{enumerate}
  Then,
  \begin{IEEEeqnarray}{c}
    \sum_{\vect{\lambda} \in \Lambda} f(\vect{\lambda}) = \frac{1}{\vol{\Lambda}} \sum_{\vect{\lambda}^\star\in\Lambda^\star}\hat{f}(\vect{\lambda}^\star).\label{eq:Poisson-summation-formula}
  \end{IEEEeqnarray}
  % where $\hat{f}(\vect{\lambda}^\star)\eqdef\int_{\Reals^n} f(\vect{t}) e^{-2\pi i\inner{\vect{t}}{\vect{\lambda}^\star}}\dd\vect{t}$
\end{lemma}

\begin{lemma}[{\cite[p.~462]{OdlyzkoSloane80_1}}]
  \label{lem:one-useful-identity}
  For any $\alpha\in\Complex$ such that its real part $\Re{\alpha} > 0$, we have
  \begin{IEEEeqnarray}{c}
    \frac{1}{(2\pi\alpha)^{\nicefrac{n}{2}}} \int_{\Reals^n} e^{-\frac{\norm{\vect{t}}^2}{2\alpha}-2\pi i\inner{\vect{t}}{\vect{\lambda}^\star}} \dd\vect{t} = e^{-2\pi^2\alpha\norm{\vect{\lambda}^\star}^2}.\label{eq:one-useful-identity}
  \end{IEEEeqnarray}
\end{lemma}

\begin{theorem}
  \label{thm:2nd-bound_Eve-success-probability}
   Let $\Gamma_{\textnormal{e}}=\bigcup_{j=1}^K(\vect{u}_j+\Lambda_\textnormal{e})$ be a lattice-like packing such that for all $j\in [1:K]$, $\inner{\vect{u}_j}{\lambda^\ast}\in\frac{1}{2}\Integers,\,\forall\,\vect{\lambda}^\ast\in\Lambda_{\textnormal{e}}^\star$. Then,
  \begin{IEEEeqnarray*}{c}
  P_{\textnormal{c,e}}\leq\frac{\vol{\Lambda_\textnormal{b}}}{(2\pi\sigma^2_\textnormal{e})^{\nicefrac{n}{2}}}\sum_{\vect{r}\in\Gamma_{\textnormal{e}}}e^{-\frac{\enorm{\vect{r}}^2}{2\sigma^2_\textnormal{e}}}.
\end{IEEEeqnarray*}
  % \begin{IEEEeqnarray}{c}
  %   \frac{1}{(2\pi\sigma_{\textnormal{e}}^2)^{\nicefrac{n}{2}}}\sum_{\vect{r} \in \Gamma_{\textnormal{e}}}\int_{\set{V}(\Lambda_{\textnormal{b}})} e^{-\frac{\enorm{\vect{w}+\vect{r}}^2}{2\sigma_{\textnormal{e}}^2}}\dd{\vect{w}}
  %   \leq\frac{\vol{\Lambda_{\textnormal{b}}}}{(2 \pi\sigma_\textnormal{e}^2)^{\nicefrac{n}{2}}}\sum_{\vect{r}\in\Gamma_{\textnormal{e}}} e^{-\frac{\norm{\vect{r}}^2}{2\sigma_\textnormal{e}^2}}.\label{eq:2nd-bound_Eve-success-probability}
  % \end{IEEEeqnarray}
\end{theorem}
\begin{IEEEproof}
  Since by definition, $\Gamma_{\textnormal{e}}$ is periodic, it can be expressed as $\Gamma_{\textnormal{e}}=\bigcup_{j=1}^{K}(\vect{u}_j+\Lambda_\textnormal{e})$ for a lattice $\Lambda_\textnormal{e}\subset\Reals^n$, and some translating vectors $\vect{u}_j\in\Reals^n$, $j\in [1:K]$. From~\eqref{eq:1st-bound_Eve-success-probability}, one needs to calculate
  
  \begin{IEEEeqnarray}{rCl}
    \frac{1}{(2\pi\sigma^2_\textnormal{e})^{\nicefrac{n}{2}}}\sum_{\vect{r}\in\Gamma_{\textnormal{e}}} \int_{\set{V}(\Lambda_\textnormal{b})} e^{-\frac{\enorm{\vect{w}+\vect{r}}^2}{2\sigma_{\textnormal{e}}^2}}\dd\vect{w}
    & = &\int_{\set{V}(\Lambda_\textnormal{b})}\frac{1}{(2\pi\sigma^2_\textnormal{e})^{\nicefrac{n}{2}}}\sum_{\vect{r}\in\Gamma_{\textnormal{e}}} e^{-\frac{\norm{\vect{w}+\vect{r}}^2}{2\sigma_{\textnormal{e}}^2}} \dd\vect{w}
    \nonumber\\
    & = & \int_{\set{V}(\Lambda_\textnormal{b})}\frac{1}{(2\pi\sigma^2_\textnormal{e})^{\nicefrac{n}{2}}}\sum_{j=1}^K\sum_{\vect{\lambda}\in\Lambda_\textnormal{e}} e^{-\frac{\norm{\vect{w}+\vect{u}_j+\vect{\lambda}}^2}{2\sigma_{\textnormal{e}}^2}}\dd\vect{w}.
    \IEEEeqnarraynumspace\label{eq:equality_1st-bound_Eve-success-probability}
  \end{IEEEeqnarray}

  % To get~\eqref{eq:2nd-bound_Eve-success-probability}, 
  Define $z\eqdef\frac{i}{2\pi\sigma_\textnormal{e}^2}$ with $\bigIm{\nicefrac{i}{2\pi\sigma_\textnormal{e}^2}}=\Im{z}>0$. We will apply Lemma~\ref{lem:Poisson-summation} (Poisson summation formula) twice to~\eqref{eq:equality_1st-bound_Eve-success-probability}. We first let $f(\vect{\lambda})\eqdef e^{i\pi z\enorm{\vect{w}+\vect{u}_j+\vect{\lambda}}^2}$ for fixed vectors $\vect{w}$ and $\vect{u}_j$. Then, according to Lemma~\ref{lem:Poisson-summation}, we have
  \begin{IEEEeqnarray*}{rCl}
    \sum_{\vect{\lambda}\in\Lambda_\textnormal{e}} e^{i\pi z\enorm{\vect{w}+\vect{u}_j+\vect{\lambda}}^2}& = &\sum_{\vect{\lambda}\in\Lambda_\textnormal{e}} f(\vect{\lambda})
    \\
    & \stackrel{\eqref{eq:Poisson-summation-formula}}{=} & \frac{1}{\vol{\Lambda_\textnormal{e}}}\sum_{\vect{\lambda}^\star\in\Lambda_{\textnormal{e}}^\star} \hat{f}(\vect{\lambda}^\star)
    \nonumber \\
    & = & \frac{1}{\vol{\Lambda_{\textnormal{e}}}}\sum_{\vect{\lambda}^\star\in\Lambda_{\textnormal{e}}^\star}\int_{\Reals^n} e^{i\pi z\enorm{\vect{w}+\vect{u}_j+\vect{t}}^2}e^{-2\pi i\inner{\vect{t}}{\vect{\lambda}^\star}} \dd\vect{t}
    \\
    & \stackrel{\substack{(a)}}{=} &\frac{1}{\vol{\Lambda_{\textnormal{e}}}}\sum_{\vect{\lambda}^\star\in\Lambda_{\textnormal{e}}^\star} \int_{\Reals^n}\ope^{i\pi z\enorm{\vect{\rho}}^2-2\pi i\inner{\vect{\rho}-\vect{w}-\vect{u}_j}{\vect{\lambda}^\star}}\dd\vect{\rho}
    \\
    & = & \frac{1}{\vol{\Lambda_\textnormal{e}}}\sum_{\vect{\lambda}^\star\in\Lambda_{\textnormal{e}}^\star} e^{2\pi i\inner{\vect{w}+\vect{u}_j}{\vect{\lambda}^\star}}\int_{\Reals^n}e^{i\pi z\enorm{\vect{\rho}}^2-2\pi i\inner{\vect{\rho}}{\vect{\lambda}^\star}} \dd\vect{\rho}
    \\
    & \stackrel{\substack{(b)}}{=} & \frac{(2\pi\sigma_{\textnormal{e}}^2)^{\nicefrac{n}{2}}}{\vol{\Lambda_\textnormal{e}}}\sum_{\vect{\lambda}^\star\in\Lambda_\textnormal{e}^\star} e^{2\pi i\inner{\vect{w}+\vect{u}_j}{\vect{\lambda}^\star}}e^{\frac{\pi}{i z}\enorm{\vect{\lambda}^\star}^2}
    \\
    & = &\frac{(2 \pi \sigma_{\textnormal{e}}^2)^{n/2}}{\vol{\Lambda_\textnormal{e}}}\sum_{\vect{\lambda}^\star\in \Lambda_\textnormal{e}^\ast}\ope^{2\pi i\inner{\vect{w}}{\vect{\lambda}^\star}}\ope^{2\pi i\inner{\vect{u}_j}{\vect{\lambda}^\star}}\ope^{\frac{\pi}{i z}\enorm{\vect{\lambda}^\star}^2}
    \\
    & = &\frac{(2 \pi\sigma_{\textnormal{e}}^2)^{n/2}}{\vol{\Lambda_\textnormal{e}}}\sum_{\vect{\lambda}^\star\in\Lambda_\textnormal{e}^\star}\bigl[\cos{(2\pi\inner{\vect{w}}{\vect{\lambda}^\star})}+i\sin{(2\pi\inner{\vect{w}}{\vect{\lambda}^\star})}\bigr]\ope^{2\pi i\inner{\vect{u}_j}{\vect{\lambda}^\star}}\ope^{\frac{\pi}{i z}\enorm{\vect{\lambda}^\star}^2}
    \\
    & \stackrel{(c)}{=} &\frac{(2 \pi \sigma_{\textnormal{e}}^2)^{n/2}}{\vol{\Lambda_\textnormal{e}}}\left[\sum_{\vect{\lambda}^\star\in\Lambda_\textnormal{e}^\star}\cos{(2\pi\inner{\vect{w}}{\vect{\lambda}^\star})}\ope^{2\pi i\inner{\vect{u}_j}{\vect{\lambda}^\star}}e^{\frac{\pi}{i z}\enorm{\vect{\lambda}^\star}^2}\right.
    \nonumber\\
    &&\hspace*{2.5cm}-\>\left.2\sum_{\vect{\lambda}^\star }\sin{(2\pi\inner{\vect{w}}{\vect{\lambda}^\star})}\sin{(2\pi\inner{\vect{u}_j}{\vect{\lambda}^\star})}e^{\frac{\pi}{i z}\enorm{\vect{\lambda}^\star}^2}\right]
    \\
    & \stackrel{(d)}{=} &\frac{(2\pi\sigma_{\textnormal{e}}^2)^{n/2}}{\vol{\Lambda_\textnormal{e}}}\sum_{\vect{\lambda}^\star\in\Lambda_\textnormal{e}^\star}\cos{(2\pi\inner{\vect{w}}{\vect{\lambda}^\star})}\ope^{2\pi i\inner{\vect{u}_j}{\vect{\lambda}^\star}}e^{\frac{\pi}{i z}\enorm{\vect{\lambda}^\star}^2}
    \\
    & \stackrel{(e)}{\leq} &\frac{(2 \pi \sigma_{\textnormal{e}}^2)^{n/2}}{\vol{\Lambda_\textnormal{e}}}\sum_{\vect{\lambda}^\star\in\Lambda_\textnormal{e}^\ast} e^{2\pi i\inner{\vect{u}_j}{\vect{\lambda}^\star}}e^{\frac{\pi}{i z}\enorm{\vect{\lambda}^\star}^2},
    \IEEEeqnarraynumspace\IEEEyesnumber\label{eq:f-ft_Poisson-summation-formula_Lambda_dual}
  \end{IEEEeqnarray*}
  where $(a)$ follows by making the change of variable $\vect{\rho} = \vect{w} + \vect{u}_j+\vect{t}$; $(b)$ holds by choosing $\alpha=\frac{i}{2\pi z}=\sigma^2_\textnormal{e}>0$ in~\eqref{eq:one-useful-identity}; $(c)$ is due to the fact that for any lattice point $\vect{\lambda}^\star\in\Lambda_{\textnormal{e}}^\star$, there always exists a lattice point $-\vect{\lambda}^\star\in\Lambda_{\textnormal{e}}^\star$ such that
  \begin{IEEEeqnarray*}{rCl}
    \IEEEeqnarraymulticol{3}{l}{%
      \sum_{\vect{\lambda}^\star\in\Lambda_\textnormal{e}^\star}i\sin{(2\pi\inner{\vect{w}}{\vect{\lambda}^\star})}\ope^{2\pi i\inner{\vect{u}_j}{\vect{\lambda}^\star}}}\nonumber\\*\quad%
    & = &\sum_{\vect{\lambda}^\star}\bigl[i\sin{(2\pi\inner{\vect{w}}{\vect{\lambda}^\star})}\ope^{2\pi i\inner{\vect{u}_j}{\vect{\lambda}^\star}}+i\sin{(2\pi\inner{\vect{w}}{-\vect{\lambda}^\star})}\ope^{2\pi i\inner{\vect{u}_j}{-\vect{\lambda}^\star}}\bigr]
    \\
    & = &\sum_{\vect{\lambda}^\star}i\sin{(2\pi\inner{\vect{w}}{\vect{\lambda}^\star})}\bigl[\ope^{2\pi i\inner{\vect{u}_j}{\vect{\lambda}^\star}}-\ope^{-2\pi i\inner{\vect{u}_j}{\vect{\lambda}^\star}}\bigr]
    \\
    & = &\sum_{\vect{\lambda}^\star}i\sin{(2\pi\inner{\vect{w}}{\vect{\lambda}^\star})}2i\sin{(2\pi\inner{\vect{u}_j}{\lambda})}
    =-2\sum_{\vect{\lambda}^\star}\sin{(2\pi\inner{\vect{w}}{\vect{\lambda}^\star})}\underbrace{\sin{(2\pi\inner{\vect{u}_j}{\vect{\lambda}^\star})}}_{= 0 \textnormal{ as }\inner{\vect{u}_j}{\vect{\lambda}^\star}\in\frac{1}{2}\Integers}=0;
  \end{IEEEeqnarray*}
  and $(d)$ is trivial as $\cos{(2\pi\inner{\vect{w}}{\vect{\lambda}^\star})}\leq 1$.
  
  Secondly, let $g(\vect{\lambda}^\star) = e^{2\pi i \inner{\vect{u}_j}{\vect{\lambda}^\star}+\frac{\pi}{i z}\enorm{\vect{\lambda}^\star}^2}$. Then, applying Lemma~\ref{lem:Poisson-summation} again, we obtain\footnote{As stated in the conditions of Lemma~\ref{lem:Poisson-summation}, we only need to verify the absolute convergence for the function $g(\cdot)$. Since $\eabs{g(\vect{\lambda}^\star)}=\bigabs{e^{2\pi i\inner{\vect{u}_j}{\vect{\lambda}^\star}+\frac{\pi}{i z}}\enorm{\vect{\lambda}^\star}^2}=\bigabs{e^{\frac{\pi}{i z}\enorm{\vect{\lambda}^\star}^2}}$, it indicates that the absolute convergence behavior of $g(\cdot)$ is exactly the same as the function $f(\vect{t}^\star)$ in~\cite[p.~5706, 1st column]{OggierSoleBelfiore16_1}, and hence the Poisson summation formula applies.}
  \begin{IEEEeqnarray*}{rCl}
    \sum_{\vect{\lambda}^\star\in\Lambda_{\textnormal{e}}^\star} e^{2\pi i\inner{\vect{u}_j}{\vect{\lambda}^\star}} e^\frac{\pi}{i z}\enorm{\vect{\lambda}^\star}^2& = &\sum_{\vect{\lambda}^\star\in\Lambda_{\textnormal{e}}^\star} g(\vect{\lambda}^\star)
    \\
    & \stackrel{\eqref{eq:Poisson-summation-formula}}{=} & \frac{1}{\vol{\Lambda_{\textnormal{e}}^\star}}\sum_{\vect{\lambda}\in\Lambda_{\textnormal{e}}}\hat{g}(\vect{\lambda})
    \\
    & = &\frac{1}{\vol{\Lambda_{\textnormal{e}}^\star}}\sum_{\vect{\lambda}\in\Lambda_{\textnormal{e}}}\int_{\Reals^n}e^{2\pi i \inner{\vect{u}_j}{\vect{t}} +\frac{\pi}{i z}}\enorm{\vect{t}}^2e^{-2\pi i\inner{\vect{t}}{\vect{\lambda}}}\dd\vect{t}
    \\
    & = &\frac{1}{\vol{\Lambda_{\textnormal{e}}^\star}}\sum_{\vect{\lambda}\in\Lambda_{\textnormal{e}}}\int_{\Reals^n}e^{-2\pi i \inner{\vect{t}}{\vect{\lambda}-\vect{u}_j} +\frac{\pi}{i z}}\enorm{\vect{t}}^2\dd\vect{t}
    \\
    & \stackrel{(i)}{=} &\frac{\bigl(2\pi\frac{1}{4\pi^2\sigma^2_\textnormal{e}}\bigr)^{\nicefrac{n}{2}}}{\vol{\Lambda_{\textnormal{e}}^\star}}\sum_{\vect{\lambda}\in\Lambda_{\textnormal{e}}}e^{-2\pi^2\frac{1}{4\pi^2\sigma^2_\textnormal{e}}\enorm{\vect{\lambda}-\vect{u}_j}^2}
    \\
    & = &\frac{\vol{\Lambda_{\textnormal{e}}}}{(2\pi\sigma^2_\textnormal{e})^{\nicefrac{n}{2}}}\sum_{\vect{\lambda}\in\Lambda_{\textnormal{e}}}e^{-\frac{\enorm{\vect{\lambda}-\vect{u}_j}^2}{2\sigma^2_\textnormal{e}}},
    \IEEEeqnarraynumspace\IEEEyesnumber\label{eq:g-ft_Poisson-summation-formula_Lambda}
  \end{IEEEeqnarray*}
  where $(i)$ follows by choosing $\alpha=\frac{z}{2\pi i}=\frac{1}{4\pi^2\sigma_\textnormal{e}^2}>0$  in~\eqref{eq:one-useful-identity}.

  Finally, using~\eqref{eq:f-ft_Poisson-summation-formula_Lambda_dual} and~\eqref{eq:g-ft_Poisson-summation-formula_Lambda} together, \eqref{eq:equality_1st-bound_Eve-success-probability} becomes
  \begin{IEEEeqnarray*}{rCl}
    \int_{\set{V}(\Lambda_\textnormal{b})} \frac{1}{(2\pi\sigma^2_\textnormal{e})^{\nicefrac{n}{2}}} \sum_{j=1}^K \sum_{\vect{\lambda}\in\Lambda_\textnormal{e}} e^{-\frac{\norm{{\vect{w}}+\vect{u}_j+\vect{\lambda}}^2}{2\sigma_{\textnormal{e}}^2}}\dd{\vect{w}}
    & \stackrel{\eqref{eq:f-ft_Poisson-summation-formula_Lambda_dual}}{\leq} &\int_{\set{V}(\Lambda_\textnormal{b})}\frac{1}{\vol{\Lambda_\textnormal{e}}}\sum_{j = 1}^{K}\sum_{\vect{\lambda}^\star\in\Lambda^\star} e^{2\pi i \inner{\vect{u}_j}{\vect{\lambda}^\star}} e^{-2\pi^2\sigma_{\textnormal{e}}^2\enorm{\vect{\lambda}^\star}^2} {\dd{\vect{w}}}
    \\
    & = &\frac{\vol{\Lambda_\textnormal{b}}}{\vol{\Lambda_\textnormal{e}}}\sum_{j = 1}^{K}\sum_{\vect{\lambda}^\star\in\Lambda^\star} e^{2\pi i\inner{\vect{u}_j}{\vect{\lambda}^\star}} e^{-2\pi^2\sigma_{\textnormal{e}}^2\enorm{\vect{\lambda}^\star}^2}
    \\
    & \stackrel{\eqref{eq:g-ft_Poisson-summation-formula_Lambda}}{=} &\frac{\vol{\Lambda_\textnormal{b}}}{(2\pi\sigma^2_\textnormal{e})^{\nicefrac{n}{2}}}\sum_{j = 1}^{K}\sum_{\vect{\lambda}\in\Lambda_{\textnormal{e}}}e^{-\frac{\enorm{\vect{\lambda}-\vect{u}_j}^2}{2\sigma^2_\textnormal{e}}}
    \\
    & = &\frac{\vol{\Lambda_\textnormal{b}}}{(2\pi\sigma^2_\textnormal{e})^{\nicefrac{n}{2}}}\sum_{\vect{r}\in\Gamma_{\textnormal{e}}}e^{-\frac{\enorm{\vect{r}}^2}{2\sigma^2_\textnormal{e}}}.
  \end{IEEEeqnarray*}
  % This completes the proof of~\eqref{eq:2nd-bound_Eve-success-probability}.
  Now, from~\eqref{eq:1st-bound_Eve-success-probability}, % and Theorem~\ref{thm:2nd-bound_Eve-success-probability},
  we can obtain a similar result as~\cite[eq.~(45)]{OggierSoleBelfiore16_1} for a lattice-like packing $\Gamma_\textnormal{e}$ which is a subset of a lattice $\Lambda_\textnormal{b}$. I.e.,
  \begin{IEEEeqnarray*}{c}
    P_{\textnormal{c,e}}\leq\frac{\vol{\Lambda_\textnormal{b}}}{(2\pi\sigma^2_\textnormal{e})^{\nicefrac{n}{2}}}\sum_{\vect{r}\in\Gamma_{\textnormal{e}}}e^{-\frac{\enorm{\vect{r}}^2}{2\sigma^2_\textnormal{e}}}.
  \end{IEEEeqnarray*}
\end{IEEEproof}

Therefore, minimizing  the upper bound on $P_{\textnormal{c,e}}$ is equivalent to minimizing
\begin{IEEEeqnarray*}{c}
  \sum_{\vect{r}\in\Gamma_\textnormal{e}} e^{-\frac{\enorm{\vect{r}}^2}{2 \sigma_{\textnormal{e}}^2}} = \Theta_{\Gamma_\textnormal{e}} (z),
\end{IEEEeqnarray*}
subject to $\nicefrac{\ecard{\Lambda_\textnormal{b}}}{\ecard{\Gamma_\textnormal{e}}}=2^k$. Note that since $\Gamma_\textnormal{e}$ is a lattice-like packing, its theta series $\Theta_{\Gamma_\textnormal{e}}(z)$ is well-defined. Moreover, as $\Im{z}>0$, we consider only the positive values of $\tau\eqdef-i z=\nicefrac{1}{2\pi\sigma_{\textnormal{e}}^2}>0$ for $\Theta_{\Gamma_\textnormal{e}}(z)$. In summary, the lattice coset coding scheme is aimed at finding a lattice-like packing $\Gamma_{\textnormal{e}}$ such that $\Theta_{\Gamma_\textnormal{e}}(z)$ is minimized, which motivates the definition of~\emph{secrecy function} below. % Definition~\ref{def:secrecy_function}.
It is worth mentioning that in~\cite{LingLuzziBelfioreStehle14_1}, the authors also pointed out that minimizing the theta series of $\Gamma_{\textnormal{e}}$ leads to a small \emph{flatness factor}, a criterion that directly relates to the mutual information leakage to the eavesdropper, instead of the success probability. Therefore, the optimization of $\Theta_{\Gamma_\textnormal{e}}(z)$ is of interest in both scenarios.

\begin{remark}
  \label{rem:ConstrA-packings_2nd-bound_Eve-Pc}
  Construction A packings $\Gamma_{\textnormal{e}}=\GammaA{\code{C}} = \frac{1}{\sqrt{2}}\bigcup_{\vect{c}\in\code{C}}\bigl(\phi(\vect{c})+2\Integers^n\bigr)$ satisfy the condition imposed by Theorem~\ref{thm:2nd-bound_Eve-success-probability}. Notice that $\vect{u}_j = \phi(\vect{c}) \in \code{C} \subseteq \Field_2^n$, $\Lambda_{\textnormal{e}}=2\Integers^n$ and $\Lambda_{\textnormal{e}}^\star = \frac{1}{2} \Integers^n$. Therefore, it can be verified that $2\inner{\vect{u}_j}{\vect{\lambda}^\ast}=\inner{2\vect{u}_j}{\vect{\lambda}^\ast}\in\Integers$, and the error probability analysis holds. 
    
    For $2$-level Construction C lattice-like packings of the form $\Gamma_{\textnormal{e}}=\phi(\code{C}_1)+2\phi(\code{C}_2)+4\Integers^n$~\cite{Forney88_1, BollaufZamirCosta19_1}, where $\code{C}_1,\code{C}_2\subseteq\Field_2^n$ are linear, the assumption required by Theorem~\ref{thm:2nd-bound_Eve-success-probability} does not hold in general. Indeed, we have $\Lambda_{\textnormal{e}}=4\Integers^n$, $\Lambda_{\textnormal{e}}^\star = \frac{1}{4} \Integers^n$, and for $\vect{u}_j = \phi(\vect{c}_1) + 2\phi(\vect{c}_2)$ with $\vect{c}_1\in\code{C}_1$, $\vect{c}_2\in\code{C}_2$, one can observe that it does not necessarily hold that $\inner{\vect{u}_j}{\vect{\lambda}^\ast}\in\frac{1}{2}\Integers$. This is the case of Example~\ref{ex:ex_ConstrC-LatticeLike_n2}.
\end{remark}

 For general lattice-like packings, it is unknown whether the result of Theorem~\ref{thm:2nd-bound_Eve-success-probability} is still valid. For the sake of brevity, we will always refer to a lattice-like packing under the premise of Theorem~\ref{thm:2nd-bound_Eve-success-probability} unless otherwise specified. % The challenge lies in the derivation of a Poisson formula for periodic packings, as well as the characterization of the dual of a lattice-like packing. For the time being, we can just state our beliefs as a conjecture.

% \begin{conjecture}
%   \label{conj:LatticeLike-packings_2nd-bound_Eve-Pc}
%   {\r Let $\Gamma_{\textnormal{e}}=\bigcup_{j=1}^K(\vect{u}_j+\Lambda_\textnormal{e})$ be a lattice-like packing. Then,
%   \begin{IEEEeqnarray*}{c}
%   P_{\textnormal{c,e}}\leq\frac{\vol{\Lambda_\textnormal{b}}}{(2\pi\sigma^2_\textnormal{e})^{\nicefrac{n}{2}}}\sum_{\vect{r}\in\Gamma_{\textnormal{e}}}e^{-\frac{\enorm{\vect{r}}^2}{2\sigma^2_\textnormal{e}}}.
% \end{IEEEeqnarray*}}
% \end{conjecture}

\begin{definition}[Secrecy function and secrecy gain~{\cite[Defs.~1 and~2]{OggierSoleBelfiore16_1}}]
  \label{def:secrecy_function}
  Let $\Gamma$ be a lattice-like packing with volume $\vol{\Gamma}=\nu^n$. The secrecy function of $\Gamma$ is defined by
  \begin{IEEEeqnarray*}{c}
    \Xi_{\Gamma}(\tau)\eqdef\frac{\Theta_{\nu\Integers^n}(i\tau)}{\Theta_{\Gamma}(i\tau)},
    \label{eq:def_secrecy-function}
  \end{IEEEeqnarray*} 
  for $\tau\eqdef -i z>0$. % As maximizing $\Xi_{\Lambda}(\tau)$ is equivalent to minimizing $\Theta_{\Gamma}(z)$, t
  The (strong) secrecy gain of a lattice is given by
  \begin{IEEEeqnarray*}{c}
    \xi_{\Gamma}\eqdef\sup_{\tau>0}\Xi_{\Gamma}(\tau).
    \label{eq:def_secrecy-gain}
  \end{IEEEeqnarray*}
\end{definition}

Ideally, the goal is to determine $\xi_{\Gamma}$. However, since the global maximum of a secrecy function of an arbitrary $\Gamma$ is in general not always easy to calculate, a weaker definition is introduced. We start by defining the \emph{symmetry point}.
\begin{definition}[Symmetry point]
  A point $\tau_0\in \Reals$ is said to be a symmetry point if for all $\tau >0$,
  \begin{IEEEeqnarray}{c}
    \Xi(\tau_0 \cdot \tau) = \Xi\Bigl(\frac{\tau_0}{\tau}\Bigr).
    \label{eq:symmetry-point_tau0}
  \end{IEEEeqnarray}
\end{definition}

\begin{definition}[Weak secrecy gain~{\cite[Def.~3]{OggierSoleBelfiore16_1}}]
  If the secrecy function of a lattice-like packing $\Gamma$ has a symmetry point $\tau_0$, then the weak secrecy gain $\chi_\Gamma$ is defined as $\chi_\Gamma=\Xi_\Gamma(\tau_0)$.
  % \begin{IEEEeqnarray}{c}
  %   = \frac{\Theta_{\nu\Integers^n}(\tau_0)}{\Theta_{\Lambda}(\tau_0)},
  %   \label{eq:weak-secrecy-gain}
  % \end{IEEEeqnarray}
  % where $\nu = \vol{\Lambda}^{\nicefrac{1}{n}}$.
\end{definition}

\section{Weak Secrecy Gain of Formally Unimodular Packings}
\label{sec:weak-secrecy-gain_FU-packings}

This section shows that formally unimodular packings also hold the same secrecy function property as unimodular, isodual lattices, and formally unimodular lattices~\cite{OggierSoleBelfiore16_1,BollaufLinYtrehus22_1}. Theorem~\ref{thm:weak_secrecy} gives a necessary and sufficient condition for a packing $\Gamma$ to achieve its weak secrecy gain at $\tau=1$.
\begin{theorem}
  \label{thm:weak_secrecy}
  Consider a lattice-like packing $\Gamma$ with $\vol{\Gamma}=1$. Then, $\Gamma$ achieves its weak secrecy gain at $\tau=1$, if and only if $\Gamma$ is formally unimodular.
\end{theorem}
\begin{IEEEproof}
  By the definition of the secrecy function and weak secrecy gain, we have
  \begin{IEEEeqnarray}{c}
    \frac{\Theta_{\Integers^n}(i\tau)}{\Theta_{\Gamma}(i\tau)}=\Xi_\Gamma(\tau) = \Xi_\Gamma\Bigl(\frac{1}{\tau}\Bigr)=\frac{\Theta_{\Integers^n}(\nicefrac{i}{\tau})}{\Theta_{\Gamma}(\nicefrac{i}{\tau})}.
    \label{eq:symmetry-point_at-tau1}
  \end{IEEEeqnarray}
  Consider that $z=i\tau$. Since the cubic lattice $\Integers^n$ is formally unimodular, it follows from~\eqref{eq:Jacobi-formula-FU-packings} that
  \begin{IEEEeqnarray}{c} 
    \Theta_{\Integers^n}(i\tau)=\Theta_{\Integers^n}(z) = \Bigl(\frac{i}{z}\Bigr)^{\frac{n}{2}}\Theta_{\Integers^n}\Bigl(-\frac{1}{z}\Big)=\Bigl(\frac{1}{\tau}\Bigr)^{\frac{n}{2}}\Theta_{\Integers^n}\Bigl(\frac{i}{\tau}\Big).
    \label{eq:Jacobi-formula-cubic-lattice}
  \end{IEEEeqnarray}
  
  Therefore, given that $\vol{\Gamma}=1$, we can obtain from~\eqref{eq:symmetry-point_at-tau1} and~\eqref{eq:Jacobi-formula-cubic-lattice} that
  \begin{IEEEeqnarray*}{rCl}
    \Theta_{\Gamma}(i\tau)& = &\frac{\Theta_{\Integers^n}(i\tau)}{\Xi_\Gamma(\tau)}=\frac{\Theta_{\Integers^n}(i\tau)}{\Xi_\Gamma\bigl(\frac{1}{\tau}\bigr)}=\frac{\Theta_{\Integers^n}(i\tau)\Theta_\Gamma(\nicefrac{i}{\tau})}{\Theta_{\Integers^n}(\nicefrac{i}{\tau})}
    \\
    & \stackrel{\eqref{eq:Jacobi-formula-cubic-lattice}}{=} &\Bigl(\frac{1}{\tau}\Bigr)^{\frac{n}{2}}\Theta_{\Gamma}\Bigl(\frac{i}{\tau}\Big),
  \end{IEEEeqnarray*}
  which also gives~\eqref{eq:Jacobi-formula-FU-packings} for $z=i\tau$. Hence, $\Gamma$ is formally unimodular.  
  
  Conversely, if 
  \begin{IEEEeqnarray*}{c}
    {\Theta_{\Gamma}(z)}  = \Bigl(\frac{i}{z}\Bigr)^{\frac{n}{2}}\Theta_{\Gamma}\Bigl(-\frac{1}{z}\Big),
  \end{IEEEeqnarray*}
  then for $z=i\tau$, we have
  \begin{IEEEeqnarray*}{c}
    \Xi_\Gamma(\tau) = \frac{\Theta_{\Integers^n}(i\tau)}{\Theta_{\Gamma}(i\tau)} = \frac{(\nicefrac{1}{\tau})^{\frac{n}{2}}\Theta_{\Integers^n}(\nicefrac{i}{\tau})}{(\nicefrac{1}{\tau})^{\frac{n}{2}}\Theta_{\Gamma}(\nicefrac{i}{\tau})} = \Xi_\Gamma\Bigl(\frac{1}{\tau}\Bigr).
  \end{IEEEeqnarray*}
  Thus, $\Gamma$ achieves its weak secrecy gain at $\tau=1$.
\end{IEEEproof}

Note that Theorem~\ref{thm:weak_secrecy} holds for isodual lattices as well, which yields to \cite[Prop.~1]{OggierSoleBelfiore16_1}.

% \begin{corollary}{\cite[Prop.~1]{OggierSoleBelfiore16_1}}
%   The secrecy function of an isodual lattice has a multiplicative symmetry point at $\tau_0 = 1$.
% \end{corollary}
\begin{corollary}
  \label{cor:weak_secrecy_not-at-tau1}
  Consider a lattice-like packing $\Gamma$ with $\vol{\Gamma}=\nu^n$. Then, $\Gamma$ achieves its weak secrecy gain at $\tau=\nu^{-2}$, % i.e.,
  % \begin{IEEEeqnarray}{c}
  %   \Xi_\Lambda(\nu^{-2} \cdot \tau) = \Xi_\Lambda\Bigl(\frac{\nu^{-2}}{\tau}\Bigr),
  %   \label{eq:symmetry-point_not-at-tau1}
  % \end{IEEEeqnarray}
  if and only if $\inv{\nu}\Gamma$ is formally unimodular.
\end{corollary}
\begin{IEEEproof}
  Consider a scaled packing $\widetilde{\Gamma}= \inv{\nu}\Gamma$. Then, we have $\evol{\widetilde{\Gamma}}=1$. Now, observe that
  \begin{IEEEeqnarray*}{rCl}
    \Xi_{\widetilde{\Gamma}}(\tau)& = &\frac{\Theta_{\Integers^n}(i\tau)}{\Theta_{\inv{\nu}\Gamma}(i\tau)}
    =\frac{\Theta_{\nu\Integers^n}(\nu^{-2}\cdot i\tau)}{\Theta_{\Gamma}(\nu^{-2}\cdot i\tau)}
    =\Xi_{\Gamma}(\nu^{-2}\cdot\tau), \textnormal{ and}
    \\
    \Xi_{\widetilde{\Gamma}}\Bigl(\frac{1}{\tau}\Bigr)& = &\frac{\Theta_{\Integers^n}(\nicefrac{i}{\tau})}{\Theta_{\inv{\nu}\Gamma}(\nicefrac{i}{\tau})}=\frac{\Theta_{\nu\Integers^n}(\nu^{-2}\cdot\nicefrac{i}{\tau})}{\Theta_{\Gamma}(\nu^{-2}\cdot\nicefrac{i}{\tau})}=\Xi_\Gamma\Bigl(\frac{\nu^{-2}}{\tau}\Bigr).
  \end{IEEEeqnarray*}
  A direct application of Theorem~\ref{thm:weak_secrecy} completes the proof.
\end{IEEEproof}

Equation~\eqref{eq:symmetry-point_tau0} with $\tau_0=\nu^{-2}$ holds for a lattice equivalent to its dual. See~\cite[Prop.~2]{OggierSoleBelfiore16_1}.

\section{Secrecy Gain of Formally Unimodular Packings}
\label{sec:secrecy-gain_FU-lattices}

Our goal in this section is to investigate the following conjecture.
\begin{conjecture}[{The {B}elfiore-{S}ol{\'e} Secrecy Function Conjecture~\cite{BelfioreSole10_1,BollaufLinYtrehus22_1}}]
  \label{conj:secrecy-gain_FU-lattices}
  The secrecy function of a formally unimodular lattice $\Lambda$ achieves its maximum at $\tau=1$, i.e., $\xi_{\Lambda}=\Xi_{\Lambda}(1)$.
\end{conjecture}

By Theorem~\ref{thm:weak_secrecy} a formally unimodular packing $\Gamma$ also achieves its weak secrecy gain at $\tau=1$. Hence, we focus on a generalized version of this conjecture, as follows.
\begin{conjecture}
  \label{conj:secrecy-gain_FU-packings}
  The secrecy function of a formally unimodular packing $\Gamma$ achieves its maximum at $\tau=1$, i.e., $\xi_{\Gamma}=\Xi_{\Gamma}(1)$.
\end{conjecture}

Although we cannot completely prove Conjectures~\ref{conj:secrecy-gain_FU-lattices} and  \ref{conj:secrecy-gain_FU-packings}, we proceed to study the secrecy gain for formally unimodular packings obtained from formally self-dual codes via Construction A (see Remark~\ref{rmk:FSD-codes_ConstructionA}). Note that for linear codes, it is known that formally self-dual codes that are not self-dual can outperform self-dual codes in some cases, as they comprise a wider class and hence may allow a better minimum Hamming distance or an overall more favorable weight enumerator. This leads us to look for improved results on the secrecy gain compared to unimodular lattices~\cite{Ernvall-Hytonen12_1,LinOggier13_1,Pinchak13_1}.
\begin{lemma}
  \label{lem:ratio_theta4-theta3}
  Let $s(\tau)\eqdef\nicefrac{\vartheta_4(i\tau)}{\vartheta_3(i\tau)}$. Then,
  % \begin{IEEEeqnarray*}{c}
  %   s(\tau)\eqdef\frac{\vartheta_4(i\tau)}{\vartheta_3(i\tau)}=\frac{1-2e^{-\pi\tau}+2e^{-4\pi\tau}-2e^{-9\pi\tau}+\cdots}{1+2e^{-\pi\tau}+2e^{-4\pi\tau}+2e^{-9\pi\tau}+\cdots}
  % \end{IEEEeqnarray*}
  $0< s(\tau) < 1$, and $s(\tau)$ is strictly increasing and a bijection for $\tau>0$.
\end{lemma}
\begin{IEEEproof}
  By definition, the fact that $0<s(\tau)< 1$ is trivial.  Next, consider the following product representation of Jacobi theta functions~\cite[p.~105]{ConwaySloane99_1}.
  \begin{IEEEeqnarray*}{c}
    \vartheta_3(\tau)=\prod_{m=1}^\infty(1-q^{2m})(1+q^{2m-1})^2,\quad
    \vartheta_4(\tau)=\prod_{m=1}^\infty(1-q^{2m})(1-q^{2m-1})^2,
  \end{IEEEeqnarray*}
  where $q=\ope^{-\pi\tau}$ for $\tau>0$. Hence,
  \begin{IEEEeqnarray*}{rCl}
    s(\tau)& = &\frac{\vartheta_4(\tau)}{\vartheta_3(\tau)}=\frac{\prod_{m=1}^\infty(1-\ope^{-(2m-1)\pi\tau})^2}{\prod_{m=1}^\infty(1+\ope^{-(2m-1)\pi\tau})^2}=\prod_{m=1}^\infty\left(\frac{\ope^{(m-\frac{1}{2})\pi\tau}-\ope^{-(m-\frac{1}{2})\pi\tau}}{\ope^{(m-\frac{1}{2})\pi\tau}+\ope^{-(m-\frac{1}{2})\pi\tau}}\right)^2
    \\
    & = &\prod_{m=1}^\infty\tanh^2\biggl(\Bigl(m-\frac{1}{2}\Bigr)\pi\tau\biggr).
  \end{IEEEeqnarray*}
  Observe that for any $m\in\Naturals$, $\tanh^{2}\bigl((m-\frac{1}{2})\pi\tau\bigr)$ is continuous and strictly increasing from $0$ to $1$ on $\tau>0$. Then, $s(\tau)$ is also continuous and ranges from $0$ to $1$, which indicates that $s(\tau)$ is an injection on $\tau>0$. Therefore, we have for any $s\in(0,1)$, $s(\tau)$ must pass through the value $s$ exactly once while $\tau$ increases, which shows that $s(\tau)$ is surjective.
\end{IEEEproof}
\begin{remark}
  \label{rem:bijective_t-tau}
  Let $t(\tau)\eqdef s(\tau)^2$. Then, $0<t(\tau) < 1$ and $t(\tau)$ is also an increasing function for $\tau>0$. Hence, according to Lemma~\ref{lem:ratio_theta4-theta3}, given any $t\in(0,1)$, there always exists a unique $\tau>0$ such that $t(\tau)=\nicefrac{\vartheta^2_4(i\tau)}{\vartheta^2_3(i\tau)}$. Moreover, we have $t(1)=\nicefrac{1}{\sqrt{2}}$ by using the identity of $\vartheta_3(i)=2^{\nicefrac{1}{4}}\vartheta_4(i)$ from~\cite{Weisstein_3}. 
\end{remark}
% {\r
%   \begin{remark}
%   \label{rem:range_t-tau}
%   Define $s(\tau)\eqdef\nicefrac{\vartheta_4(i\tau)}{\vartheta_3(i\tau)}$. Then, $0< s(\tau) < 1$. Let $t(\tau)\eqdef s(\tau)^2$. Then, $0<t(\tau) < 1$. Moreover, we have $t(1)=\nicefrac{1}{\sqrt{2}}$ by using the identity of $\vartheta_3(i)=2^{\nicefrac{1}{4}}\vartheta_4(i)$ from~\cite{Weisstein_3}. 
% \end{remark}}
Now, we are able to give a new universal approach to derive the strong secrecy gain of a Construction A packing obtained from a formally self-dual code. We define $f_{\code{C}}(t)\eqdef W_\code{C}(\sqrt{1+t},\sqrt{1-t})$ for $0< t < 1$.
\begin{theorem}
  \label{thm:inv_secrecy-function_WeightEnumerator}
  Consider a geometrically uniform $(n,M)$ code $\code{C}$ where $\vect{0}\in\code{C}$, and its Construction A packing has $\vol{\GammaA{\code{C}}}=1$. Then
  \begin{IEEEeqnarray*}{c}
    \inv{\left[\Xi_{\GammaA{\code{C}}}(\tau)\right]}=\frac{W_{\code{C}}\bigl(\sqrt{1+t(\tau)},\sqrt{1-t(\tau)}\bigr)}{2^\frac{n}{2}}=\frac{f_{\code{C}}(t(\tau))}{2^\frac{n}{2}},\label{eq:Xi-ft_ConstructionA_FSDcodes}\IEEEeqnarraynumspace
  \end{IEEEeqnarray*}
  where $0<t(\tau)=\nicefrac{\vartheta_4^2(i\tau)}{\vartheta^2_3(i\tau)} < 1$. % {\r Moreover, if there exists a $\tau^\ast>0$ such that $t^\ast\eqdef\argmin\limits_{t\in(0,1)}f_{\code{C}}(t)=t(\tau^\ast)$, then $\tau^\ast=\argmax\limits_{\tau>0}\bigl[\Xi_{\GammaA{\code{C}}}(\tau)\bigr]$.}
  Moreover, maximizing the secrecy function $\Xi_{\GammaA{\code{C}}}(\tau)$ is equivalent to determining the minimum of $f_{\code{C}}(t)$ on $t\in(0,1)$.
\end{theorem}
\begin{IEEEproof}
  From Lemma~\ref{lem:theta-series_non-lattice} and~\eqref{eq:identity_squares}, the theta series $\Theta_{\Gamma_{\textnormal{A}}(\code{C})}$ becomes
  \begin{IEEEeqnarray}{rCl}
    \Theta_{\Gamma_{\textnormal{A}}(\code{C})}(z)
    & = & \we{\code{C}}\Biggl(\sqrt{\frac{\vartheta^2_3(z)+\vartheta^2_4(z)}{2}}, \sqrt{\frac{\vartheta^2_3(z)-\vartheta^2_4(z)}{2}}\Biggr)
    \nonumber\\
    & \stackrel{(a)}{=} &\frac{1}{2^{\frac{n}{2}}}\we{\code{C}}\left(\sqrt{\vartheta^2_3(z)+\vartheta^2_4(z)},\sqrt{\vartheta^2_3(z)-\vartheta^2_4(z)}\right).\label{eq:theta-series_GammaA_code}
  \end{IEEEeqnarray}
  where $(a)$ follows by the definition of weight enumerator~\eqref{eq:weight-enumerator}.
  
  From Definition~\ref{def:secrecy_function}, the secrecy function of $\GammaA{\code{C}}$ with volume $1$ becomes
  \begin{IEEEeqnarray*}{rCl}
    \inv{\Bigl[\Xi_{\GammaA{\code{C}}}(\tau)\Bigr]}& = &\frac{\Theta_{\GammaA{\code{C}}}(z)}{\Theta_{\Integers^n}(z)}
    \stackrel{(b)}{=}\frac{1}{2^{\frac{n}{2}}}\frac{\we{\code{C}}\left(\sqrt{\vartheta^2_3(z)+\vartheta^2_4(z)},\sqrt{\vartheta^2_3(z)-\vartheta^2_4(z)}\right)}{\vartheta^n_3(z)}
    \\[1mm]
    & = &\frac{1}{2^{\frac{n}{2}}}\frac{\sum_{w=0}^n A_w(\code{C})\bigl(\sqrt{\vartheta^2_3(z)+\vartheta^2_4(z)}\bigr)^{n-w}\bigl(\sqrt{\vartheta^2_3(z)-\vartheta^2_4(z)}\bigr)^w}{\vartheta^{n-w}_3(z)\vartheta^{w}_3(z)},
    \\[1mm]
    & = &\frac{1}{2^{\frac{n}{2}}}\we{\code{C}}\left(\sqrt{1+\frac{\vartheta^2_4(z)}{\vartheta^2_3(z)}},\sqrt{1-\frac{\vartheta^2_4(z)}{\vartheta^2_3(z)}}\right)=\frac{W_{\code{C}}\bigl(\sqrt{1+t(\tau)},\sqrt{1-t(\tau)}\bigr)}{2^\frac{n}{2}},
  \end{IEEEeqnarray*}
  where $(b)$ holds because of $\Theta_{\Integers^n}(z)=\vartheta^n_3(z)$ and~\eqref{eq:theta-series_GammaA_code}. Lastly, the second part of the theorem follows directly from Remark~\ref{rem:bijective_t-tau}. % {\r the following observation.
  % \begin{IEEEeqnarray*}{c}
  %   \min_{\tau>0}\inv{\Bigl[\Xi_{\GammaA{\code{C}}}(\tau)\Bigr]}=\min_{\tau>0}\left[\frac{f_{\code{C}}(t(\tau))}{2^{\frac{n}{2}}}\right]\geq\min_{t\in(0,1)}\left[\frac{f_{\code{C}}(t)}{2^{\frac{n}{2}}}\right].
  % \end{IEEEeqnarray*}}
\end{IEEEproof}

We remark that it turns out that our universal approach is similar to the technique proposed by~\cite{Ernvall-Hytonen12_1} and adapted for~\cite{LinOggier13_1,Pinchak13_1}, but it is fundamentally different. The two fundamental differences are: i) We work on formally unimodular packings, which are not necessarily unimodular and isodual lattices. ii) The technique that the previous works rely on is to express the theta series of a unimodular lattice in terms of $\nicefrac{\vartheta^4_2(i\tau)\vartheta^4_4(i\tau)}{\vartheta_3^8(i\tau)}$. However, we derive the theta series of formally unimodular packings depending on $t(\tau)=\nicefrac{\vartheta^2_4(i\tau)}{\vartheta^2_3(i\tau)}$.

We can apply Theorem~\ref{thm:inv_secrecy-function_WeightEnumerator} for any formally unimodular packing as its volume is always equal to $1$.
\begin{example}
  \label{ex:ex_n6k3d3}
  Consider a $[6,3,3]$ odd formally self-dual code $\code{C}$ with $W_{\code{C}}(x,y) = x^6+4x^3y^3+3x^2y^4$~\cite{BetsumiyaHarada01_1}. Thus, we can get $f_{\code{C}}(t) = W_{\code{C}}(\sqrt{1+t},\sqrt{1-t})=4[1+t^3+(1-t^2)^{\nicefrac{3}{2}}]$ and $f'_{\code{C}}(t)=12t(t-\sqrt{1-t^2})$ by performing some simple calculations. Observe that for $0<t<\nicefrac{1}{\sqrt{2}}$, we have $\sqrt{1-t^2}>\nicefrac{1}{\sqrt{2}}$. Then, $t-\sqrt{1-t^2}<\nicefrac{1}{\sqrt{2}}-\nicefrac{1}{\sqrt{2}}=0$. This indicates that the derivative $f'_{\code{C}}(t)<0$ on $t\in(0,\nicefrac{1}{\sqrt{2}})$.
  % $f'_{\code{C}}(t)$ satisfies that
  % \begin{IEEEeqnarray*}{c}
  %   \begin{cases}
  %     <0 & \textnormal{as } t-\sqrt{1-t^2}< \frac{1}{\sqrt{2}} - \frac{1}{\sqrt{2}} =0,~t \in (0, \nicefrac{1}{\sqrt{2}}),
  %     \\
  %     =0 & \textnormal{if },~t=\frac{1}{\sqrt{2}},
  %     \\
  %     >0 & \textnormal{as } t-\sqrt{1-t^2}> \frac{1}{\sqrt{2}} + \frac{1}{\sqrt{2}} =0,~t \in (\nicefrac{1}{\sqrt{2}}, 1).
  %   \end{cases}
  % \end{IEEEeqnarray*}
  Similarly, one can also show that $f_{\code{C}}'(t)>0$ on $t\in(\nicefrac{1}{\sqrt{2}},1)$, and $t=\nicefrac{1}{\sqrt{2}}$ is the minimizer of $f_{\code{C}}(t)$. Hence, Remark~\ref{rem:bijective_t-tau} and Theorem~\ref{thm:inv_secrecy-function_WeightEnumerator} indicate that the maximum of $\Xi_{\ConstrA{\code{C}}}(\tau)$ is achieved at $\tau=1$. Also, one can get $\xi_{\ConstrA{\code{C}}}\approx 1.172$.\hfill\exampleend
  % \begin{figure}[h!]
  %   \centering
  %   \includegraphics[scale=0.4]{./Secrecy_Gain_Length6.png}
  %   \caption{Secrecy function of the Construction A lattice obtained from the $[6,3,3]-$code.} \label{fig:secrecy_gain_6}
  % \end{figure}
\end{example}   

\begin{example}
  \label{ex:NR16}
  Applying the result of Theorem~\ref{thm:inv_secrecy-function_WeightEnumerator} to $\code{N}_{16}$ (see Example~\ref{ex:nr_properties}), we have that $f_{\code{N}_{16}}(t) = -64 (-4 + 8 t^2 - 5 t^4 - 6 t^6 + 3 t^8),$ $f_{\code{N}_{16}}'(t)= -256 t (4 - 5 t^2 - 9 t^4 + 6 t^6)$. The value of $t$ that minimizes $f_{\code{N}_{16}}(t)$ and consequently maximizes the secrecy function is $t=\nicefrac{1}{\sqrt{2}}$, yielding  a strong secrecy gain of $2.207$.
\end{example}

\begin{example}
  \label{ex:Nonlattice_12_20}
  A similar approach can be carried out for the nonlinear but formally self-dual codes obtained from~\cite[Tab.~VII]{YooLeeKim17_1} in dimensions $12$ and $20$ (these codes are constructed under the Gray map $\psi$, thus are geometrically uniform and contain $\vect{0}$). For dimension $12$, among the four weight enumerators presented,
  \begin{IEEEeqnarray*}{c}
    W_{\code{C}_{12}}(x,y) = x^{12} + 6 x^8 y^4 +24 x^7 y^5+ 16 x^6 y^6+ 9 x^4 y^8+ 8 x^3 y^9
  \end{IEEEeqnarray*}
  gives the highest secrecy gain. Here, we have 
  \begin{IEEEeqnarray*}{rCl}
    f_{\code{C}}(t) & = & W_{\code{C}_{12}}(\sqrt{1+t},\sqrt{1-t})
    \\
    & = & 8\bigl[4-6t^2+4t^3+6t^4+(1-t)^{\nicefrac{9}{2}}(1+t)^{\nicefrac{3}{2}} + 3(1-t)^{\nicefrac{5}{2}}(1+t)^{\nicefrac{7}{2}}\bigr]
  \end{IEEEeqnarray*}  
  % $f_{\code{C}}(t)=W_{\code{C}_{12}}(\sqrt{1+t},\sqrt{1-t})=8\left(4-6t^2+4t^3+6t^4+(1-t)^{\nicefrac{9}{2}}(1+t)^{\nicefrac{3}{2}} + 3(1-t)^{\nicefrac{5}{2}}(1+t)^{\nicefrac{7}{2}}\right)$ 
  and $f'_{\code{C}}(t)=96t\cdot h_1(t)$, where $h_1(t)=-1+t+2t^2+\sqrt{1-t^2}~(2t^3-t-1)$. Observe that $h_1(t)$ has only one root in $[0,1]$, since $h_1(0)=-2$, $h_1(1)=2$, and $h_1'(t)>0$ on $t\in (0,1)$. This unique root is $t=\nicefrac{1}{\sqrt{2}}$, which gives $\xi_{\eGammaA{\code{C}_{12}}}\approx 1.657$ and coincides with the best in Table~\ref{tab:table_secrecy-gains_FU-lattices} below. 
  
  For dimension $20$, we consider the weight enumerator
  \begin{IEEEeqnarray*}{c}
    W_{\code{C}_{20}}(x,y) = x^{20}+90 x^{14} y^6+255 x^{12} y^8+332 x^{10} y^{10}+255 x^8 y^{12}+90 x^6 y^{14}+y^{20}.
  \end{IEEEeqnarray*}
  We have $f_{\code{C}}(t)=W_{\code{C}_{20}}(\sqrt{1+t},\sqrt{1-t})=-64(5t^8-10t^6-35t^4+40t^2-16)$ and $f'_\code{C}(t)=-1280t\cdot h_2(t)$, where $h_2(t)= 2t^6 -3t^4-7t^2+4$. Note that $h_2(t)$ has only one root in $[0,1]$, as $h_2(0)=-4$, $h_2(1)=4$, and $h'(t)<0$ on $t\in (0,1)$. This unique root is $t=\nicefrac{1}{\sqrt{2}}$, which yields $\xi_{\eGammaA{\code{C}_{20}}} \approx 2.813$, and coincides with the second best value tabulated in Table~\ref{tab:table_secrecy-gains_FU-lattices} below. Unlike dimension $16$, the weight enumerator coincides with the linear case, therefore there is the no improvement.
\end{example}

In summary, interesting results (and improvements) concerning the secrecy gain could also be accomplished by considering nonlinear codes.
    
The following lemma shows a general expression of $f_{\code{C}}(t)$ if $\code{C}$ is an even formally self-dual code.
\begin{lemma}
  \label{lem:f_C_even-FSDcodes}
  If $\code{C}$ is an $(n,2^{\nicefrac{n}{2}})$ even formally self-dual code, then we have
  \begin{IEEEeqnarray}{c}
    % W_{\code{C}}(\sqrt{1+t},\sqrt{1-t})=
    f_{\code{C}}(t)=2^{\frac{n}{2}}\sum_{r=0}^{\lfloor\frac{n}{8}\rfloor}a_r (t^4-t^2+1)^r,
    \label{eq:ThetaSeries_evenFSD-codes}
  \end{IEEEeqnarray}
  where $a_r\in\Rationals$ and $\sum_{r=0}^{\lfloor\frac{n}{8}\rfloor}a_r=1$.
\end{lemma}
\begin{IEEEproof}
  Using the \emph{invariant theory} (cf.~\cite[Ch.~19]{MacWilliamsSloane77_1}), one can show that the weight enumerator $W_{\code{C}}(x,y)$ of any even code $\code{C}$ (linear or nonlinear) that satisfies \eqref{eq:FSD_MacWilliams-identity} can be expressed as
  \begin{IEEEeqnarray}{c}
    W_{\code{C}}(x,y)=\sum_{r=0}^{\lfloor\nicefrac{n}{8}\rfloor} a_r g_1(x,y)^{\frac{n}{2}-4r}g_2(x,y)^{r}
    \label{eq:we_g1-g2}
  \end{IEEEeqnarray}
  where $g_1(x,y)\eqdef x^2+y^2$, $g_2(x,y)\eqdef x^8+14x^4y^4+y^8$, $a_r\in\Rationals$, and $\sum_{r=0}^{\lfloor\frac{n}{8}\rfloor}a_r=1$.\footnote{The result follows directly from the statement of~\cite[Ch.~19, Problem.~(3)]{MacWilliamsSloane77_1}.} Then, by performing some simple calculations, we obtain
  \begin{IEEEeqnarray*}{rCl}
    g_1(\sqrt{1+t},\sqrt{1-t})& = &2,
    \\
    g_2(\sqrt{1+t},\sqrt{1-t})& = &16(t^4-t^2+1).
  \end{IEEEeqnarray*}
  Therefore,~\eqref{eq:ThetaSeries_evenFSD-codes} follows from~\eqref{eq:we_g1-g2}.
\end{IEEEproof}
%%%%%%%%%%%%% Gleason coefficients section %%%%%%%%%%%%%

%\section{Determining Gleason's Coefficients from the Weight Enumerator}

%The 
% Let $\code{C}$ be an $(n,2^{\nicefrac{n}{2}})$ even formally self-dual code. Gleason's Theorem~\cite[Th.~9.2.1]{HuffmanPless03_1} states that 
% \begin{IEEEeqnarray}{c}
%   \label{eq:gleason}
%   W_{\code{C}}(x,y)=\sum_{r=0}^{\lfloor\nicefrac{n}{8}\rfloor} a_r g_1(x,y)^{\tfrac{n}{2}-4r}g_2(x,y)^{r},
% \end{IEEEeqnarray}
% where $g_1(x,y)=x^2+y^2$,  $g_2(x,y)=x^8+14x^4y^4+y^8$, $a_r \in \mathbb{Q}$, and $\sum_{r=0}^{\lfloor \tfrac{n}{8} \rfloor} a_r=1.$

Consider the weight enumerator $W_\code{C}(x,y)$ as in~\eqref{eq:weight-enumerator}. We need to determine the coefficients $a_r$ in~\eqref{eq:we_g1-g2} in terms of $A_w(\code{C})$, $w\in[0:n]$, if the coefficients $A_w(\code{C})$ are known, and will provide numerical details in Example~\ref{ex:n18k9d6}.  %, $w$ even.~\eqref{eq:weight-enumerator}.
%Let's first start to 
First, we expand $g_1(x,y)^{\frac{n}{2}-4r}$ and $g_2(x,y)^r$. Observe that
% \begin{IEEEeqnarray*}{rCl}
% g_1(x,y)^{\frac{n}{2}-4r} & = & (x^2+y^2)^{\frac{n}{2}-4r} \nonumber \\
% & = & \sum_{j=0}^{\frac{n}{2}-4r} \binom{\nicefrac{n}{2}-4r}{j} (x^2)^{(\frac{n}{2}-4r-j)} (y^2)^{j},
% \IEEEeqnarraynumspace
% \end{IEEEeqnarray*}
\begin{IEEEeqnarray*}{c}
g_1(x,y)^{\frac{n}{2}-4r}  =  (x^2+y^2)^{\frac{n}{2}-4r} = \sum_{j=0}^{\frac{n}{2}-4r} \binom{\nicefrac{n}{2}-4r}{j} (x^2)^{(\frac{n}{2}-4r-j)} (y^2)^{j}, ~ \text{and} ~
\IEEEeqnarraynumspace
\end{IEEEeqnarray*}
%and
% \begin{IEEEeqnarray*}{rCl}
%   \IEEEeqnarraymulticol{3}{l}{%
%     g_2(x,y)^{r}}\nonumber\\*%
%   & = &(x^8+14x^4y^4+y^8)^r=[(x^4+7y^4)^2-48y^8]^r
%   \nonumber \\
%   & = & \sum_{h=0}^{r}\binom{r}{h}\biggl[\sum_{\ell=0}^{2r-2h}\binom{2r-2h}{\ell}(x^4)^{2r-2h-\ell}(7y^4)^{\ell}\biggr](-48y^8)^h.
%   \nonumber\IEEEeqnarraynumspace
% \end{IEEEeqnarray*}
\begin{IEEEeqnarray*}{rCl}
    g_2(x,y)^{r} & = & (x^8+14x^4y^4+y^8)^r=[(x^4+7y^4)^2-48y^8]^r
  \nonumber \\
  & = & \sum_{h=0}^{r}\binom{r}{h}\biggl[\sum_{\ell=0}^{2r-2h}\binom{2r-2h}{\ell}(x^4)^{2r-2h-\ell}(7y^4)^{\ell}\biggr](-48y^8)^h.
  \nonumber\IEEEeqnarraynumspace
\end{IEEEeqnarray*}

Given $w\in[0:n]$, by collecting the terms of $y^{2j+8h+4\ell}$ for $2j+8h+4\ell=w$, we get
% \begin{IEEEeqnarray}{rCl}
%   \IEEEeqnarraymulticol{3}{l}{%
%     g_1(x,y)^{\frac{n}{2}-4r} g_2(x,y)^r
%   }\nonumber\\*\quad%
%   & = &\sum_{\substack{2j+8h+4\ell =w\\ j,h,\ell \in\Integers_{\geq 0}}} 7^\ell(-48)^h\binom{\nicefrac{n}{2}-4r}{j}\nonumber \\
%   \nonumber\\
%   && \>\times\binom{r}{h}\binom{2r-2h}{\ell}x^{n-2j-8h-4\ell} y^{2j+8h+4\ell},
%   \label{eq:expand_multiplication}
% \end{IEEEeqnarray}
% \begin{IEEEeqnarray}{rCl}
%   \IEEEeqnarraymulticol{3}{l}{%
%     g_1(x,y)^{\frac{n}{2}-4r} g_2(x,y)^r = \sum_{\substack{2j+8h+4\ell =w\\ j,h,\ell \in\Integers_{\geq 0}}} 7^\ell(-48)^h\binom{\nicefrac{n}{2}-4r}{j} \>\times\binom{r}{h}\binom{2r-2h}{\ell}x^{n-2j-8h-4\ell} y^{2j+8h+4\ell},
%   \label{eq:expand_multiplication}
% \end{IEEEeqnarray}
\begin{IEEEeqnarray*}{rCl}
  \IEEEeqnarraymulticol{3}{l}{%
    g_1(x,y)^{\frac{n}{2}-4r} g_2(x,y)^r}\nonumber\\*\quad%
  & = &\sum_{\substack{2j+8h+4\ell =w\\ j,h,\ell \in\Integers_{\geq 0}}} 7^\ell(-48)^h\binom{\nicefrac{n}{2}-4r}{j} \times\binom{r}{h}\binom{2r-2h}{\ell}x^{n-2j-8h-4\ell} y^{2j+8h+4\ell},\IEEEeqnarraynumspace\label{eq:expand_multiplication}
\end{IEEEeqnarray*}
where % $w\in[0:n]$, refers to the weight in the weight enumerator and 
we define $\binom{p}{q}=0,$ if $p<q$.

By comparing the coefficients of~\eqref{eq:weight-enumerator} and~\eqref{eq:we_g1-g2}, we get % a linear equation of such system is given by
\begin{IEEEeqnarray}{rCl}
  A_w(\code{C}) & = & \sum_{r=0}^{\lfloor\nicefrac{n}{8} \rfloor} a_r  \sum_{\substack{2j+8h+4\ell =w \\ j,k,\ell \in \Integers_{\geq 0}}} 7^\ell(-48)^h \binom{\nicefrac{n}{2}-4r}{j}\binom{r}{h} \times\binom{2r-2h}{\ell} %x^{n-2j-8h-4\ell} y^{2j+8h+4\ell}.
  \label{eq:Aw_general-ar}\IEEEeqnarraynumspace
\end{IEEEeqnarray}

For an even formally self-dual code, according to~\cite[p.~378]{HuffmanPless03_1}, we know that $A_w(\code{C})=A_{n-w}(\code{C})$ for $w$ even and $A_w(\code{C})=0$ for $w$ odd, in~\eqref{eq:weight-enumerator}. Thus, there are at most $\bigl\lfloor\frac{n}{4}\bigr\rfloor+1$ nonzero coefficients $A_w(\code{C})$. For instance, if we want to determine the coefficients of the term corresponding to $A_4$, this would only be possible if we set $j=2$, $h=\ell=0$ or $j=h=0$, $\ell=1$ in~\eqref{eq:Aw_general-ar}, which yields
\begin{IEEEeqnarray*}{rCl}
  A_4 & = &\Biggl(\sum_{r=0}^{\lfloor \nicefrac{n}{8} \rfloor} a_r\biggl({\binom{\nicefrac{n}{2} - 4r}{2}+ \underbrace{7 \binom{2r}{1}}_{14r}}\biggr) \Biggr) \\ %x^{n-4}y^4 \\
  & = & a_0 \binom{\nicefrac{n}{2}}{2} +  a_1 \left(  \binom{\nicefrac{n}{2}-4}{2} +14 \right)
  +a_2\biggl(\binom{\nicefrac{n}{2}-8}{2} + 28 \biggr) + a_3 \biggl(\binom{\nicefrac{n}{2}-12}{2} +42 \biggr)+\cdots.
  \IEEEeqnarraynumspace
\end{IEEEeqnarray*}

For ease of illustration, we compute more terms of~\eqref{eq:Aw_general-ar}: % The expression of the weight enumerator $W_{\code{C}}$ is then
\begin{IEEEeqnarray*}{rCl}
  % \IEEEeqnarraymulticol{3}{l}{%
  % W_{\code{C}}(x,y)}\nonumber\\*\!%
  A_0 & = &\sum_{r=0}^{\lfloor\nicefrac{n}{8}\rfloor} a_r,\quad A_2 =\sum_{r=0}^{\lfloor \nicefrac{n}{8} \rfloor} a_r \left(\tfrac{n}{2}-4r\right),
  % & = &\underbrace{\sum_{r=0}^{\lfloor \nicefrac{n}{8} \rfloor} a_r}_{A_0}x^n + \underbrace{\sum_{r=0}^{\lfloor \nicefrac{n}{8} \rfloor} a_r \left(\tfrac{n}{2}-4r\right)}_{A_2}x^{n-2}y^{2}
  \nonumber\\[1mm]
  A_6 & = &\sum_{r=0}^{\lfloor \nicefrac{n}{8} \rfloor} a_r \left({\binom{ \nicefrac{n}{2} - 4r}{3}+14r\left(\tfrac{n}{2} - 4r\right)} \right),
  \nonumber\\[1mm]
  A_8 & = & \sum_{r=0}^{\lfloor \nicefrac{n}{8} \rfloor} a_r \Biggl(\binom{\nicefrac{n}{2}-4r}{4}+14r\binom{\nicefrac{n}{2}-4r}{2}+\>49\binom{2r}{2}-48 r)\Biggr).\IEEEeqnarraynumspace
  % &&\!+\>\underbrace{  \sum_{r=0}^{\lfloor \nicefrac{n}{8} \rfloor} a_r \left({\binom{\nicefrac{n}{2} - 4r}{2}+14r}\right) }_{A_4}x^{n-4}y^{4}
  % \nonumber\\
  % &&\!+\>\underbrace{ \sum_{r=0}^{\lfloor \nicefrac{n}{8} \rfloor} a_r \left({\binom{ \nicefrac{n}{2} - 4r}{3}+14r\left(\tfrac{n}{2} - 4r\right)} \right)}_{A_6}x^{n-6}y^{6}\nonumber\\
  % &&\!\!\!+\>\underbrace{  \sum_{r=0}^{\lfloor \nicefrac{n}{8} \rfloor} a_r \Biggl(\binom{\nicefrac{n}{2}-4r}{4}+14r\binom{\nicefrac{n}{2}-4r}{2}+49\binom{2r}{2}-48 r)\Biggr)}_{A_8}
  % \nonumber\\
  % && \hspace{6.0cm}\>\times x^{n-8}y^{8}+\cdots.
%~ ~ \underbrace{ \left( \sum_{r=0}^{\lfloor \nicefrac{n}{8} \rfloor} a_r \left({\binom{ \nicefrac{n}{2} - 4r} {5}+14r\binom{ \nicefrac{n}{2} - 4r}{3}} + \left(49 \binom{2 r}{2}-48 r\right)\left(\tfrac{n}{2}-4r\right)  \right)\right)}_{A_{10}}x^{n-10}y^{10}+ \\
%~ ~ {\tiny \underbrace{ \left( \sum_{r=0}^{\lfloor \nicefrac{n}{8} \rfloor} a_r \left({\binom{ \nicefrac{n}{2} - 4r} {6}+14r\binom{ \nicefrac{n}{2} - 4r}{4}} + \left(49 \binom{2 r}{2}-48 r\right)\binom{ \nicefrac{n}{2} - 4r}{2}  -336r(2r-2)+343\binom{2r}{3} \right)\right)}_{A_{12}}x^{n-12}y^{12}}+ \dots \\
\end{IEEEeqnarray*}
	
% \begin{IEEEeqnarray*}{lCl}
% W_{\code{C}}(x,y) & = & \underbrace{(a_0+a_1+a_2+a_3+\dots)}_{A_0}x^n + \underbrace{\sum_{r=0}^{\lfloor \nicefrac{n}{8} \rfloor} a_r \left(\tfrac{n}{2}-4r\right)}_{A_2}x^{n-2}y^{2} + \\
% & & \underbrace{ \left( \sum_{r=0}^{\lfloor \nicefrac{n}{8} \rfloor} a_r \left({\binom{\nicefrac{n}{2} - 4r}{2}+14r}\right) \right)}_{A_4}x^{n-4}y^{4} + \underbrace{ \left( \sum_{r=0}^{\lfloor \nicefrac{n}{8} \rfloor} a_r \left({\binom{ \nicefrac{n}{2} - 4r}{3}+14r\left(\tfrac{n}{2} - 4r\right)} \right)\right)}_{A_6}x^{n-6}y^{6}+ \\
% & & \underbrace{ \left( \sum_{r=0}^{\lfloor \nicefrac{n}{8} \rfloor} a_r \left({\binom{ \nicefrac{n}{2} - 4r} {4}+14r\binom{ \nicefrac{n}{2} - 4r}{2}} + 49 \binom{2 r}{2}-48 r)  \right)\right)}_{A_8}x^{n-8}y^{8}+ \\
% & & \underbrace{ \left( \sum_{r=0}^{\lfloor \nicefrac{n}{8} \rfloor} a_r \left({\binom{ \nicefrac{n}{2} - 4r} {5}+14r\binom{ \nicefrac{n}{2} - 4r}{3}} + \left(49 \binom{2 r}{2}-48 r\right)\left(\tfrac{n}{2}-4r\right)  \right)\right)}_{A_{10}}x^{n-10}y^{10}+ \\
% & & {\tiny \underbrace{ \left( \sum_{r=0}^{\lfloor \nicefrac{n}{8} \rfloor} a_r \left({\binom{ \nicefrac{n}{2} - 4r} {6}+14r\binom{ \nicefrac{n}{2} - 4r}{4}} + \left(49 \binom{2 r}{2}-48 r\right)\binom{ \nicefrac{n}{2} - 4r}{2}  -336r(2r-2)+343\binom{2r}{3} \right)\right)}_{A_{12}}x^{n-12}y^{12}}+ \dots \\
% \end{IEEEeqnarray*}

As a result, we can obtain % Note that there are    
the $\bigl\lfloor\frac{n}{8}\bigr\rfloor+1$ unknown coefficients $a_r$, $r\in [0:\bigl\lfloor\frac{n}{8}\bigr\rfloor]$ by solving the system of $\bigl\lfloor\frac{n}{4}\bigr\rfloor+1$ linear equations in~\eqref{eq:Aw_general-ar}. %Since $\bigl\lfloor\frac{n}{4}\bigr\rfloor+1 > \bigl\lfloor\frac{n}{8}\bigr\rfloor+1$, 
% We obtain the solution of $a_r$ by numerically solving the $\bigl\lfloor\frac{n}{4}\bigr\rfloor+1$ linear equations in~\eqref{eq:Aw_general-ar}. % always have more equations than the unknown variables, which indicates that $a_r$ can be determined by considering the  
The uniqueness of the set of coefficients $a_r$ follows from Gleason's Theorem~\cite[Th.~9.2.1]{HuffmanPless03_1}.

%%%%%%%%%%%%%%%%%%%%%%%%%%%%%%%%%%%%%%%%%%%%%%%%%%%%%%%%
Next, we provide a sufficient condition for a Construction A packing obtained from an even formally self-dual code to achieve its strong secrecy gain at $\tau=1$, or, equivalently, $t=\nicefrac{1}{\sqrt{2}}$. Note that unless otherwise specified, for the rest of the paper, the formally self-dual codes we consider are geometrically uniform and contain $\vect{0}$.

\begin{theorem}% xxx
  \label{thm:strong-secrecy-gain_unimodular-lattices}
  Consider $n \geq 8$ and an $(n,2^{\nicefrac{n}{2}})$ even formally self-dual code $\code{C}$. Let $h(t)\eqdef t^4-t^2+1$. If the coefficients $a_r$ of $f_{\code{C}}(t)$ expressed in terms of \eqref{eq:ThetaSeries_evenFSD-codes} satisfy
  \begin{IEEEeqnarray}{c}
    \sum_{r=1}^{\lfloor\frac{n}{8}\rfloor} r a_r h(t)^{r-1} > 0 
    \label{eq:condition_ai-ht_evenFSD}
  \end{IEEEeqnarray}
  on $t\in(0,1)$, then the secrecy gain of $\GammaA{\code{C}}$ is achieved at $\tau=1$.
\end{theorem}
\begin{IEEEproof}
  It is enough to show that the function $f_{\code{C}}(t)$ as in \eqref{eq:ThetaSeries_evenFSD-codes} defined for $0 < t < 1$ achieves its minimum at $t=\nicefrac{1}{\sqrt{2}}$.
  
  The derivative of $f_{\code{C}}(t)$ satisfies
  \begin{IEEEeqnarray*}{c}
    \frac{\dd f_{\code{C}}(t)}{\dd t} = 2^{\frac{n}{2}} h'(t) \sum_{r=1}^{\lfloor\frac{n}{8}\rfloor} r a_r h(t)^{r-1},
  %\label{eq:derivative_fcode}
  \end{IEEEeqnarray*}  
  and $h'(t)=4t^3-2t=2t(2t^2-1)$. As the hypothesis holds, the behavior of the derivative is dominated by $h'(t)$. Since
  \begin{IEEEeqnarray*}{c}
    h'(t)
    \begin{cases}
      <0 & \textnormal{if }0< t<\frac{1}{\sqrt{2}},
      \\
      =0 & \textnormal{if }t=\frac{1}{\sqrt{2}},
      \\
      >0 & \textnormal{if }\frac{1}{\sqrt{2}}< t < 1,
    \end{cases}
  \end{IEEEeqnarray*}
  which implies that $f_{\code{C}}(t)$ is decreasing in $t\in(0,\nicefrac{1}{\sqrt{2}})$ and increasing in $t\in(\nicefrac{1}{\sqrt{2}},1)$. This completes the proof.
\end{IEEEproof}

\begin{example}
  \label{ex:n18k9d6}
  Consider an $[18,9,6]$ even formally self-dual code $\code{C}$ with weight enumerator
%   \begin{IEEEeqnarray*}{rCl}
%     W_{\code{C}}(x,y) & = &x^{18}+102 x^{12} y^6+153 x^{10} y^8
%     \nonumber\\
%     && \>+153 x^8 y^{10}+102 x^6 y^{12}+y^{18}. 
%   \end{IEEEeqnarray*}
  \begin{IEEEeqnarray*}{c}
    W_{\code{C}}(x,y)  = x^{18}+102 x^{12} y^6+153 x^{10} y^8+153 x^8 y^{10}+102 x^6 y^{12}+y^{18}. 
  \end{IEEEeqnarray*}
  By solving $f_{\code{C}}(t)=W_{\code{C}}(\sqrt{1+t},\sqrt{1-t})$ with~\eqref{eq:ThetaSeries_evenFSD-codes}, %(see the details of derivation provided in~Appendix~\ref{sec:coefficients_gleason})
  we find that $a_0=-\nicefrac{29}{16}, a_1=\nicefrac{27}{8}$ and $a_2=-\nicefrac{9}{16}$. Observe that the left-hand side of \eqref{eq:condition_ai-ht_evenFSD} gives $\sum_{r=1}^2 r a_r h(t)^{r-1}=-\frac{9}{8}(t^4-t^2-2)=-\frac{9}{8}(t^2-\frac{1}{2})^2+\frac{81}{32}$. One can easily verify that $0<\frac{9}{4}\leq-\frac{9}{8}(t^2-\frac{1}{2})^2+\frac{81}{32}\leq\frac{81}{32}$ on $t\in(0,1)$, which indicates that the condition~{\m\eqref{eq:condition_ai-ht_evenFSD}} in Theorem~\ref{thm:strong-secrecy-gain_unimodular-lattices} for those coefficients is satisfied.
  % Observe that $p'(t) = -\frac{9}{8} (4 t^3-2 t)>0$, for $t \in (0, \nicefrac{1}{\sqrt{2}})$ and $p'(t)<0$, for $t \in (\nicefrac{1}{\sqrt{2}},1)$, meaning that $p(t)$ is increasing for $t \in (0, \nicefrac{1}{\sqrt{2}})$ and decreasing for $t \in (\nicefrac{1}{\sqrt{2}},1)$. We can then conclude that the image of $p(t)$ lies on the interval $[\nicefrac{9}{4},\nicefrac{81}{32}]$ and $p(t)>0$, for $t \in (0,1)$. 
  Thus, the secrecy gain conjecture is true for the formally unimodular lattice $\ConstrA{\code{C}}$.\hfill\exampleend
\end{example}

\subsection{A Necessary Condition for Best Formally Unimodular Construction~A Packings}
\label{sec:necessary-condition_best-ConstrA-FU-packings}

We have studied how to determine the (strong) secrecy gain of Construction~A formally unimodular packings obtained from formally self-dual codes. In this subsection, we take a further step to investigate the structure of the corresponding formally self-dual codes that maximize the secrecy gain.

\begin{definition}
  \label{def:optimal-FSD-code}
  A formally self-dual code $\code{C}^\ast$ of length $n$ is called strongly secrecy-optimal if
  \begin{IEEEeqnarray*}{c}
    \code{C}^\ast=\argmax_{\code{C}\colon\textnormal{formally self-dual}}\xi_{\GammaA{\code{C}}}.
    % \xi_{\GammaA{\code{C}^\ast}}\geq\xi_{\GammaA{\code{C}}}
  \end{IEEEeqnarray*}  
\end{definition}

Based on Conjecture~\ref{conj:secrecy-gain_FU-packings}, Remark~\ref{rem:bijective_t-tau}, and Theorem~\ref{thm:inv_secrecy-function_WeightEnumerator}, we further conjecture the following condition for a formally self-dual code to be secrecy-optimal.
\begin{conjecture}
  \label{conj:conjecture_optimal-fsd-codes}
  For a given formally self-dual code $\code{C}^\ast$ of length $n$, if
  \begin{IEEEeqnarray*}{c}
    \code{C}^\ast=\argmin_{\code{C}\colon\textnormal{formally self-dual}}f_{\code{C}}\biggl(\frac{1}{\sqrt{2}}\biggr),
    % f_{\code{C}^\ast}\biggl(\frac{1}{\sqrt{2}}\biggr)\leq f_{\code{C}}\biggl(\frac{1}{\sqrt{2}}\biggr),
    \label{eq:conjecture_optimal-fsd-codes}
  \end{IEEEeqnarray*}  
  then the code $\code{C}^\ast$ is strongly secrecy-optimal.
\end{conjecture}

As in general, it is hard to confirm whether a Construction A packing obtained from a formally self-dual code $\code{C}$ achieves its strong secrecy gain at $t=\nicefrac{1}{\sqrt{2}}$, we then use the secrecy function $\Xi_{\GammaA{\code{C}}}(\tau)$ to give a slightly weaker definition of a secrecy-optimal formally self-dual code.
\begin{definition}
  \label{def:good-FSD-code}
  A formally self-dual code $\code{C}^\diamond$ of length $n$ is said to be weakly secrecy-optimal if for all $\tau>0$,
  \begin{IEEEeqnarray*}{c}
    \Xi_{\GammaA{\code{C}^\diamond}}(\tau)\geq\Xi_{\GammaA{\code{C}}}(\tau)
  \end{IEEEeqnarray*}
  for any $n$-dimensional formally self-dual code $\code{C}$.
\end{definition}

We seek to establish a necessary condition for a formally self-dual code $\code{C}$ to be weakly secrecy-optimal by considering its weight distribution $\{A_w(\code{C})\}_{w=0}^n$.
\begin{theorem}
  \label{thm:necessary-condition_optimal-FSDcode}
  Given a length $n\geq 2$, if $\code{C}^\diamond$ is weakly secrecy-optimal, then
  \begin{IEEEeqnarray*}{c}
    \code{C}^\diamond=\argmin_{\code{C}\colon\textnormal{formally self-dual}}\left\{\sum_{w=0}^n\frac{A_w(\code{C})}{w+1}\right\}.
    %\sum_{w=0}^n\biggl(\frac{A_w(\code{C})-A_w(\code{C}^\diamond)}{w+1}\biggr)\geq 0,
    \label{eq:necessary-condtion_good-fsd-codes}
  \end{IEEEeqnarray*}
\end{theorem}
\begin{IEEEproof}
  By Definition~\ref{def:good-FSD-code} and applying Theorem~\ref{thm:inv_secrecy-function_WeightEnumerator}, one can see that if a formally self-dual code $\code{C}^\diamond$ is weakly secrecy-optimal, then for all $t\in(0,1)$, $f_{\code{C}^\diamond}(t)-f_{\code{C}}(t)\leq 0$ for any $n$-dimensional formally self-dual code $\code{C}$. Expressed in terms of the weight enumerators, we can obtain
  \begin{IEEEeqnarray}{rCl}
    f_{\code{C}}(t)-f_{\code{C}^\diamond}(t)& = &
    \sum_{w=0}^n A_w(\code{C})\bigl(\sqrt{1+t}\bigr)^{n-w}\bigl(\sqrt{1-t}\bigr)^w-\sum_{w=0}^n A_w(\code{C}^\diamond)\bigl(\sqrt{1+t}\bigr)^{n-w}\bigl(\sqrt{1-t}\bigr)^w
    \nonumber\\
    & = &(\sqrt{1+t})^{n}\left[\sum_{w=0}^n A_w(\code{C})\biggl(\sqrt{\frac{1-t}{1+t}}\biggr)^w-\sum_{w=0}^n A_w(\code{C}^\diamond)\biggl(\sqrt{\frac{1-t}{1+t}}\biggr)^w\right]\geq 0\label{eq:difference_f_codes}.
  \end{IEEEeqnarray}
  Now, we define $u(t)\eqdef\sqrt{\frac{1-t}{1+t}}$ over $t\in(0,1)$. It can be shown that $u(t)$ is a decreasing function for $0<t<1$, and we have $0<u(t)<1$. Hence, \eqref{eq:difference_f_codes} implies that for any formally self-dual code $\code{C}$,
  \begin{IEEEeqnarray*}{c}
    \Delta(u)\eqdef\sum_{w=0}^n A_w(\code{C})u^w-\sum_{w=0}^n A_w(\code{C}^\diamond)u^w\geq 0.
  \end{IEEEeqnarray*}
  Integrating $\Delta(u)$ over $u\in(0,1)$ results in 
  \begin{IEEEeqnarray*}{c}
    \sum_{w=0}^n\biggl(\frac{A_w(\code{C})-A_w(\code{C}^\diamond)}{w+1}\biggr)\geq 0
  \end{IEEEeqnarray*}
  for any formally self-dual code $\code{C}$. This completes the proof.
\end{IEEEproof}

\begin{figure}[t!]
  \centering
  % \subfloat[$n=18$.]{
  \input{\Figs/secrecy_gains_ConstrA_unimodular_n18.tex}
  % \includegraphics[scale=0.4]{n18k9.png}
  % }
  %   \hfill
  %   \subfloat[$n=12$.]{
  %   \input{\Figs/secrecy_gains_ConstrA_unimodular_n12.tex}
  %   %   \includegraphics[scale=0.4]{n12k6d4.png}
  % }
  \caption{$n=18$. Since $\varphi_{\code{C}_\textnormal{efsd}^{(6)}}(u)>\varphi_{\code{C}_\textnormal{ofsd}^{(5)}}(u)>\varphi_{\code{C}_\textnormal{sd}^{(4)}}(u)$ for all $u\in(0,1)$, it implies that $\Xi_{\eConstrA{\code{C}_\textnormal{efsd}^{(6)}}}(\tau)>\Xi_{\eConstrA{\code{C}_\textnormal{ofsd}^{(5)}}}(\tau)>\Xi_{\eConstrA{\code{C}_\textnormal{sd}^{(4)}}}(\tau)$ for all $\tau>0$.}
  \label{fig:weakly-secrecy-optimal-fsd-codes}
\end{figure}

\begin{example}
  \label{ex:FSD-codes_n18}
  For $n=18$, we consider the Construction A lattices obtained from the codes listed in Table~\ref{tab:table_secrecy-gains_FU-lattices} below. We use the function $\varphi_{\code{C}}(u)\eqdef \sum_{w=0}^n A_w(\code{C})u^w$ for a formally self-dual code $\code{C}$, to indicate the secrecy performance. Note that $u(t)=\sqrt{\frac{1-t}{1+t}}$, $f_{\code{C}}(t)=(\sqrt{1+t})^{n}\sum_{w=0}^n A_w(\code{C})u^w(t)$, and $u(\nicefrac{1}{\sqrt{2}})=\sqrt{2}-1\approx 0.4142$. Figure~\ref{fig:weakly-secrecy-optimal-fsd-codes} indicates that $\Xi_{\eConstrA{\code{C}_\textnormal{efsd}^{(6)}}}(\tau)>\Xi_{\eConstrA{\code{C}_\textnormal{ofsd}^{(5)}}}(\tau)>\Xi_{\eConstrA{\code{C}_\textnormal{sd}^{(4)}}}(\tau)$ for all $\tau>0$, and one can verify that $\code{C}_\textnormal{efsd}^{(6)}=\argmin_{\code{C}\in\{\code{C}_\textnormal{sd}^{(4)},\code{C}_\textnormal{ofsd}^{(5)},\code{C}_\textnormal{efsd}^{(6)}\}}\left\{\sum_{w=0}^n\frac{A_w(\code{C})}{w+1}\right\}$ (see Appendix~\ref{sec:all-weight-enumerators-codes} for the corresponding weight enumerators).
\end{example}
\begin{remark}\leavevmode  
  \label{rem:weakly-secrecy-optimal-fsd-codes}
 \begin{itemize}
  \item In Example~\ref{ex:FSD-codes_n18}, only the three best-known formally self-dual codes are verified for Theorem~\ref{thm:necessary-condition_optimal-FSDcode}. However, to exactly determine the weakly secrecy-optimal code $\code{C}^\diamond$, we need to check all the possible weight enumerators of formally self-dual codes for a given dimension $n$.
 \item It is intuitive to believe that the secrecy gain can be improved with higher minimum Hamming distance and lower kissing number (i.e., the number of codewords with minimum weight $d$) of a given dimension $n$. However, Theorem~\ref{thm:necessary-condition_optimal-FSDcode} indicates that one needs to rely on the entire weight enumerators to determine the weakly secrecy-optimal code. In the following, we present two counterexamples. 
 \end{itemize}  
\end{remark}

To generate the two counterexamples, we use the \emph{direct sum construction}~\cite[p.~76]{MacWilliamsSloane77_1} of codes. Given two binary codes $\code{C}_1$, $\code{C}_2$, with parameters $(n_1,M_1, d_1)$ and $(n_2, M_2,d_2)$ respectively, the direct sum $\code{C}\subseteq\Field_2^{n_1+n_2}$ consists of the concatenation of vectors $\vect{c} = (\vect{c}_1\mid\vect{c}_2)\in\code{C}$, where $\vect{c}_1\in\code{C}_1$ and $\vect{c}_2\in\code{C}_2$. This idea can be generalized for more codes, and we apply next this construction for two and three codes. Since the underlying codes to be considered are formally self-dual, geometrically uniform, and contain $\vect{0}$, it can be shown that their direct sum also preserves such properties.

\begin{example}
  \label{ex:counterexample_dim18}
  Consider two formally self-dual codes, the $[2,1,2]$ repetition code and the $(16,256,8)$ Nordstrom-Robinson code $\code{N}_{16}$. Their direct sum, denoted by $\code{C}_\textnormal{efsd}^{(2)}$, is an $(18, 2^9, 2)$ code, which is also formally self-dual. This code has weight enumerator
  \begin{IEEEeqnarray*}{c}
    W_{\code{C}_\textnormal{efsd}^{(2)}}(x,y)  = x^{18}+x^{16} y^2+112 x^{12} y^6+142 x^{10} y^8+142 x^8 y^{10}+112 x^6 y^{12}+x^2 y^{16}+y^{18}.
  \end{IEEEeqnarray*}
  The secrecy gain of its respective Construction A packing is $\xi_{\eGammaA{\code{C}_\textnormal{efsd}^{(2)}}}=2.207$. 
  
  Now, we consider three $[6,3,3]$ formally self-dual codes, their direct sum gives an $[18, 9, 3]$ odd formally self-dual code $\code{C}_\textnormal{efsd}^{(3)}$ with weight enumerator
  \begin{IEEEeqnarray*}{rCl}
    W_{\code{C}_\textnormal{ofsd}^{(3)}}(x,y) & = & x^{18}+12 x^{15} y^3+9 x^{14} y^4+48 x^{12} y^6+72 x^{11} y^7\nonumber\\
    && +\>27 x^{10} y^8+ 64 x^9 y^9 +144 x^8 y^{10}+108 x^7 y^{11}+27 x^6 y^{12},
  \end{IEEEeqnarray*}
  which induces a Construction A lattice such that $\xi_{\eGammaA{\code{C}_\textnormal{ofsd}^{(3)}}} = 1.608$. Hence, comparing the two direct sum formally self-dual codes $\code{C}_\textnormal{efsd}^{(2)}$ and $\code{C}_\textnormal{efsd}^{(3)}$, we obtain a better secrecy gain from a formally self-dual code with a smaller minimum Hamming distance. However, by verifying the condition of Theorem~\ref{thm:necessary-condition_optimal-FSDcode}, it can be seen that  $\code{C}_\textnormal{efsd}^{(2)}$ outperforms $\code{C}_\textnormal{efsd}^{(3)}$ in terms of secrecy gain.
\end{example}

\section{Construction of Isodual Codes from Rate \texorpdfstring{$\nicefrac{1}{2}$}{1/2} Binary Convolutional codes}
\label{sec:isodual_coco}

Before presenting the numerical results, we introduce in this session the construction of some even and odd formally self-dual codes obtained via convolutional codes. The construction allows us to consider codes of parameters larger than
those found in the current literature on formally self dual codes. The performance of the corresponding Construction A lattices is presented in Table~\ref{tab:table_secrecy-gains_FU-lattices}. % whose secrecy function of the respective Construction A lattice is of interest in this paper.

An $(n,k,m)$ binary convolutional code $\code{C}$ is a $k$-dimensional subspace of $\Field_2(D)^{n}$, where $D$ is an indeterminate variable, $\Field_2(D)$ consists of all rational functions in $D$, and $m$ is the \emph{memory}, \emph{i.e.,}  the maximum degree of the generator polynomials for $\code{C}$. For a background on convolutional codes, please see, \emph{e.g.},~\cite{LinCostello04_1}. It is well known~\cite{BocharovaJohannessonKudryashovStahl02_1} that \emph{tail-biting convolutional codes} often produce very competitive linear codes. We point out the following property of the linear block codes obtained by tail-biting technique applied to convolutional codes of rate $\nicefrac{1}{2}$.

\begin{proposition}
  Let $\code{C}$ be a $(2,1,m)$ binary convolutional code. Then, any $[2k,k ]$ linear code $\code{C}_{\textnormal{tb}}$ obtained from $\code{C}$ by tail-biting is isodual, where $k\geq (m+1)$.
   %(2) the minimum distance $d_{\textnormal{tb}}$ is bounded as $d_{\textnormal{tb}} \leq d_{\textnormal{free}}$, where $d_{\textnormal{free}}$ is the free distance of $\cal C$
\end{proposition}
\begin{IEEEproof}
  For brevity, we prove this by an example of the convolutional code generated by the \emph{minimal} generator matrix
  \begin{IEEEeqnarray*}{c}
    \mat{G}(D)=
    \begin{pmatrix}
      g_1(D) & g_2(D)
    \end{pmatrix}
    =
    \begin{pmatrix}
      a + c D + e D^2 & b + d D + f D^2
    \end{pmatrix}\IEEEeqnarraynumspace
  \end{IEEEeqnarray*}
and its associated $[2\times 5,1\times 5]=[10,5]$ linear code $\code{C}_{\textnormal{tb}}$  by tail-biting for $k=5$. The proof is easily adapted to other tail-biting codes for different code dimensions $k$ and to other convolutional codes with different  memory length, but the matrices involved tend to not fit nicely in a page.

It is well known \cite{SolomonVanTilborg79_1},~\cite[p.~107]{MaWolf86_1} that the matrix $\mat{G}_{\textnormal{tb}}$ is a generator matrix of the linear code $\code{C}_{\textnormal{tb}}$, where 
\begin{IEEEeqnarray}{c}
  \mat{G}_{\textnormal{tb}} =
  \begin{pmatrix}
    a & b & c & d & e & f & & & &
    \\
    & & a & b & c & d & e & f & &
    \\
    & & & & a & b & c & d & e & f
    \\
    e & f & & & & & a & b & c & d
    \\
    c & d & e & f & & & & & a & b     
  \end{pmatrix},\IEEEeqnarraynumspace\label{eq:genmat_tb}
\end{IEEEeqnarray}
and that the matrix $\mat{H}_{\textnormal{tb}}$ is a parity check matrix for $\code{C}_{\textnormal{tb}}$, where
% \[
\begin{IEEEeqnarray*}{c}
  \mat{H}_{\textnormal{tb}} =
  \begin{pmatrix}
    b & a &   &   &   &   & f & e & d & c
    \\
    d & c & b & a &   &   &  &  & f & e
    \\
    f & e & d & c & b & a &   &   &  &
    \\
    &   & f & e & d & c & b & a &  &
    \\
    &   &   &   & f & e & d & c & b & a     
  \end{pmatrix}.\IEEEeqnarraynumspace
\end{IEEEeqnarray*}
% \]

Clearly, for binary codes, $\mat{G}_{\textnormal{tb}}\trans{\mat{H}_{\textnormal{tb}}}= \mat{O}$, where $\mat{O}$ represents an all-zero matrix. $\mat{G}_{\textnormal{tb}}$ and $ \mat{H}_{\textnormal{tb}}$ generate $[10,5]$ linear codes that are mutually reversed with respect to order of coordinates, and hence they are isodual (thus, they share the same weight enumerator as well).
\end{IEEEproof}
% Please note that the proposition also holds for convolutional codes over a field of characteristic 2, although we will apply it only to binary codes in this paper.

\begin{remark}\leavevmode
  \begin{itemize}
  \item A $[2k,k]$ tail-biting code for any integer  $k \geq (m+1)$, % where $m = \max \{\deg g_1(D), \deg  g_2(D) \}$,
    is generated by a matrix constructed like the one in (\ref{eq:genmat_tb}), with the first $k-m$ rows containing successive  two-coordinate shifts of the generator polynomial's coefficients and the last $m$ rows wrapping around like in (\ref{eq:genmat_tb}).
  \item Consider a convolution code $\code{C}$ with %its 
  free distance $d_\textnormal{free}$. It is well known that the minimum distance $d_{\textnormal{tb}}$ of the tail-biting code $\code{C}_\textnormal{tb}$ is bounded as $d_{\textnormal{tb}}\leq d_{\textnormal{free}}$, % where $d_{\textnormal{free}}$ is the free distance of the convolution code $\code{C}$,
    and that $d_\textnormal{tb} = d_{\textnormal{free}}$ for any dimension $k \geq k_{\code{C}}$, where $k_{\code{C}}$ is a modest lower threshold that depends only on $\code{C}$. 
  \item The exact weight enumerators, as presented in~\ifthenelse{\boolean{short_version}}{~\cite[App.~D]{BollaufLinYtrehus21_2sub}}{Appendix~\ref{sec:all-weight-enumerators-codes}} of this paper, of isodual tail-biting codes, indicated by ``tb'', are conveniently computed by a modified Viterbi algorithm. A straightforward application of this algorithm has a complexity of $O(k \cdot 2^{2m})$.
  \end{itemize}
\end{remark}

\section{Numerical Results and Comparisons}
\label{sec:numerical-results}

Even though the result of Theorem~\ref{thm:strong-secrecy-gain_unimodular-lattices} is restricted to formally unimodular packings obtained from even formally self-dual codes, we have numerical evidence showing that Conjectures~\ref{conj:secrecy-gain_FU-lattices} and~\ref{conj:secrecy-gain_FU-packings} also hold for formally unimodular lattices (and lattice-like packings in general) obtained from odd formally self-dual codes. The secrecy gains of some formally unimodular Construction A lattices obtained from (even and odd) formally self-dual codes are summarized in Table~\ref{tab:table_secrecy-gains_FU-lattices}.

\begin{table*}[t]
  \centering
  % added the optional argument to \caption with ifthenelse:
  \caption[]{Comparison of (strong) secrecy gains of formally unimodular Construction A lattice-like packings for several values of even dimensions $n$. Codes without references are obtained by tail-biting the rate $\nicefrac{1}{2}$ convolutional codes (Section~\ref{sec:isodual_coco}).
    % Starred values correspond to the best ones in the respective dimension, while the entries of the second column are the best results from~\cite[Tables~I and II]{LinOggier13_1}.
  }
  \label{tab:table_secrecy-gains_FU-lattices}
  \vskip -2.0ex
  \Scale[0.98]{\begin{IEEEeqnarraybox}[
    \IEEEeqnarraystrutmode
    \IEEEeqnarraystrutsizeadd{3.5pt}{3.0pt}]{V/c/V/c/V/c/V/c/V/c/V/c/V/c/V/c/V/c/V/c/V}
    \IEEEeqnarrayrulerow\\
    & n
    && \textnormal{Upper bound}~\textnormal{\cite{LinOggier12_1}}
    && \code{C}_\textnormal{sd}^{(d)}
    && \xi_{\eConstrA{\code{C}_\textnormal{sd}}}
    && \code{C}_\textnormal{efsd}^{(d)}
    && \xi_{\eConstrA{\code{C}_\textnormal{efsd}}}
    && \code{C}_\textnormal{ofsd}^{(d)}%~\textnormal{\cite{BetsumiyaHarada01_1}, Appendix~\ref{sec:isodual_coco}}
    && \xi_{\eConstrA{\code{C}_\textnormal{ofsd}}}
    && \code{C}_\textnormal{nfsd}^{(d)}
    && \xi_{\eGammaA{\code{C}_\textnormal{nfsd}}}
    &\\    
    \hline\hline
    &6 && 1 && -  && - && \code{C}_\textnormal{efsd}^{(2)}~ \textnormal{\cite{HuffmanPless03_1}} && 1 && \code{C}_\textnormal{ofsd}^{(3)}~\textnormal{\cite{BetsumiyaHarada01_1}} && ~\mathbf{1.172} && - && - &\\
    \IEEEeqnarrayrulerow\\
    &8 && 1.33 && \code{C}_\textnormal{sd}^{(4)}~\textnormal{\cite{HuffmanPless03_1}}  && \mathbf{1.333} && - && - && \code{C}_\textnormal{ofsd}^{(3)}~\textnormal{\cite{BetsumiyaHarada01_1}} && ~1.282  && - && - &\\
    \IEEEeqnarrayrulerow\\
    &10 && 1.45 && - && - && \code{C}_\textnormal{efsd}^{(4)}~\textnormal{\cite{KennedyPless94_1}}  && 1.455 && \code{C}_\textnormal{ofsd}^4~\textnormal{\cite{BetsumiyaHarada01_1}} && \mathbf{1.478} && - && - &\\  
    \IEEEeqnarrayrulerow\\
    &12 && 1.6 && \code{C}_\textnormal{sd}^{(4)}~\textnormal{\cite{LinOggier13_1}} && 1.6 && \code{C}_\textnormal{efsd}^{(4)}~\textnormal{\cite{BetsumiyaGulliverHarada99_1}}  && 1.6 && \code{C}_\textnormal{ofsd}^{(4)}~\textnormal{\cite{BetsumiyaHarada01_1}}  && \mathbf{1.657} && \code{C}_\textnormal{nfsd}^{(4)}~\textnormal{\cite{YooLeeKim17_1}} && \mathbf{1.657} & \\
    \IEEEeqnarrayrulerow\\
    &14 && 1.78 && \code{C}_\textnormal{sd}^{(4)}~\textnormal{\cite{LinOggier13_1}} && 1.778 && \code{C}_\textnormal{efsd}^{(4)}~\textnormal{\cite{BetsumiyaGulliverHarada99_1}} && 1.825 && \code{C}_\textnormal{ofsd}^{(4)}~\textnormal{\cite{BetsumiyaHarada01_1}}  && \mathbf{1.875} && - && - & \\
    \IEEEeqnarrayrulerow\\
    &16 && 2.21 && \code{C}_\textnormal{sd}^{(4)}~\textnormal{\cite{LinOggier13_1}} && 2 && \code{C}_\textnormal{efsd}^{(4)}~\textnormal{\cite{BetsumiyaHarada01_2}}  && 2.133 && \code{C}_\textnormal{ofsd}^{(5)}~\textnormal{\cite{BetsumiyaHarada01_1}}  && 2.141 && \code{C}_\textnormal{nfsd}^{(6)}~\textnormal{\cite{MacWilliamsSloane77_1}} && \mathbf{2.207} & \\
    \IEEEeqnarrayrulerow\\
    &18 && 2.49 &&\code{C}_\textnormal{sd}^{(4)}~\textnormal{\cite{LinOggier13_1}} && 2.286 && \code{C}_\textnormal{efsd}^{(6)}~\textnormal{\cite{OEIS-we-list}}   && \mathbf{2.485} && \code{C}_\textnormal{ofsd}^{(5)}  && 2.424 && - && - & \\
    \IEEEeqnarrayrulerow\\
    &20 && 2.81 && \code{C}_\textnormal{sd}^{(4)}~\textnormal{\cite{LinOggier13_1}} && 2.667 && \code{C}_\textnormal{efsd}^{(6)}~ \textnormal{\cite{FieldsGaboritHuffmanPless01_1}}  && 2.813 && \code{C}_\textnormal{ofsd}^{(6)}~\textnormal{\cite{BetsumiyaGulliverHarada99_1}} && \mathbf{2.868} && \code{C}_\textnormal{nfsd}^{(6)}~\textnormal{\cite{YooLeeKim17_1}} && 2.813 & \\
    \IEEEeqnarrayrulerow\\
    &22 && 3.2 && \code{C}_\textnormal{sd}^{(6)}~\textnormal{\cite{LinOggier13_1}} && 3.2 &&  \code{C}_\textnormal{efsd}^{(6)}  && 3.2  && \code{C}_\textnormal{ofsd}^{(7)}~\textnormal{\cite{BetsumiyaHarada01_1}}  && \mathbf{3.335} && - && - & \\
    \IEEEeqnarrayrulerow\\
     &30 && 5.84 && \code{C}_\textnormal{sd}^{(6)}~\textnormal{\cite{ConwaySloane90_1}} && 5.689 && \code{C}_\textnormal{efsd}^{(8)}~\textnormal{\cite{BouyuklievaBouyukliev10_1}}   && \mathbf{5.843} && \code{C}_\textnormal{ofsd}^{(7)}  && 5.785 && - && - & \\
    \IEEEeqnarrayrulerow\\
    &32 && 7.00 && \code{C}_\textnormal{sd}^{(8)}~\textnormal{\cite{ConwaySloane90_1}} && 6.737 && %\prescript{\textnormal{tb}}{}{\code{C}}_\textnormal{efsd}^{(8)}%\textnormal{,~Appendix~\ref{sec:isodual_coco}}
    {\code{C}}_\textnormal{efsd}^{(8)}
    && 6.748 && \code{C}_\textnormal{ofsd}^{(7)}  && 6.628 && - && - & \\
    \IEEEeqnarrayrulerow\\
    &40 && 12.81 && \code{C}_\textnormal{sd}^{(8)}~\textnormal{\cite{ConwaySloane90_1}} && 12.191 && %\prescript{\textnormal{tb}}{}{\code{C}}_\textnormal{efsd}^{(8)}% \textnormal{,~Appendix~\ref{sec:isodual_coco}}
    {\code{C}}_\textnormal{efsd}^{(8)}
    && 12.134 && \code{C}_\textnormal{ofsd}^{(9)}  && \mathbf{12.364} && - && - & \\
    \IEEEeqnarrayrulerow \\
    &70 && 130.15 && \code{C}_\textnormal{sd}^{(12)}~\textnormal{\cite{Harada97_1}} && 127.712 && %\prescript{\textnormal{tb}}{}{\code{C}}_\textnormal{efsd}^{(12)}% \textnormal{,~Appendix~\ref{sec:isodual_coco}}
    {\code{C}}_\textnormal{efsd}^{(12)}
    && 128.073 && \code{C}_\textnormal{ofsd}^{(13)}  && \mathbf{128.368} && - && - & \\
    \IEEEeqnarrayrulerow
  \end{IEEEeqnarraybox}}
\end{table*}

Note that all codes have parameters $[n, \nicefrac{n}{2}]$ or $(n,2^{\nicefrac{n}{2}})$. The superscript ``$(d)$'' refers to the minimum Hamming distance $d$ of the code. Their exact weight enumerators can be found in Appendix~\ref{sec:all-weight-enumerators-codes}. The highlighted values represent the best values found in the respective dimensions, when comparing self-dual (sd), linear even and odd formally self-dual (efsd and ofsd) codes, and nonlinear formally self-dual (nfsd) codes. In some cases (\emph{e.g.} $[12,6]$, $[22,11]$) we were unable to find efsd codes different from sd codes. ``[$\cdot$]'' indicates the reference number.

\begin{figure}[t!]
  \centering
  % This file was created by tikzplotlib v0.9.1.
\begin{tikzpicture}

\begin{axis}[
width=12.25cm,
height=6.75cm,
legend cell align={left},
legend style={legend style={draw=none,fill=none}, font=\scriptsize,
minimum height=0.15in, at={(axis cs: 1.50,2.25)}, anchor=north west},
xmin=-6,
xmax=6,
%xtick distance=0.1,
xlabel={$\tau$ (dB)},
grid style={gray,opacity=0.5,dotted},
xmajorgrids,
ymajorgrids,
ymin=1,
max space between ticks=20pt,
ymax=2.3,
ylabel={$\Xi_{\Gamma}$},
ylabel style={rotate=-90},
]

\addplot [color=black!50!green, solid, line width=1pt, mark=diamond, mark options={solid, mark size=1.25pt}]
table {%
6.0 1.00012
5.9 1.00016
5.8 1.00021
5.7 1.00027
5.6 1.00036
5.5 1.00046
5.4 1.00059
5.3 1.00076
5.2 1.00097
5.1 1.00123
5.0 1.00155
4.9 1.00195
4.8 1.00243
4.7 1.00301
4.6 1.00372
4.5 1.00458
4.4 1.0056
4.3 1.00683
4.2 1.00828
4.1 1.01
4.0 1.01203
3.9 1.01441
3.8 1.0172
3.7 1.02044
3.6 1.02421
3.5 1.02857
3.4 1.03361
3.3 1.03939
3.2 1.04602
3.1 1.0536
3.0 1.06224
2.9 1.07205
2.8 1.08319
2.7 1.09578
2.6 1.10998
2.5 1.12597
2.4 1.14394
2.3 1.16407
2.2 1.18658
2.1 1.21168
2.0 1.23962
1.9 1.27063
1.8 1.30494
1.7 1.34278
1.6 1.38435
1.5 1.42983
1.4 1.47932
1.3 1.53283
1.2 1.59024
1.1 1.65127
1.0 1.7154
0.9 1.78185
0.8 1.8495
0.7 1.9169
0.6 1.98223
0.5 2.04336
0.4 2.09794
0.3 2.1436
0.2 2.17809
0.1 2.19959
0.0 2.2069
-0.1 2.19959
-0.2 2.17809
-0.3 2.1436
-0.4 2.09794
-0.5 2.04336
-0.6 1.98223
-0.7 1.9169
-0.8 1.8495
-0.9 1.78185
-1.0 1.7154
-1.1 1.65127
-1.2 1.59024
-1.3 1.53283
-1.4 1.47932
-1.5 1.42983
-1.6 1.38435
-1.7 1.34278
-1.8 1.30494
-1.9 1.27063
-2.0 1.23962
-2.1 1.21168
-2.2 1.18658
-2.3 1.16407
-2.4 1.14394
-2.5 1.12597
-2.6 1.10998
-2.7 1.09578
-2.8 1.08319
-2.9 1.07205
-3.0 1.06224
-3.1 1.0536
-3.2 1.04602
-3.3 1.03939
-3.4 1.03361
-3.5 1.02857
-3.6 1.02421
-3.7 1.02044
-3.8 1.0172
-3.9 1.01441
-4.0 1.01203
-4.1 1.01
-4.2 1.00828
-4.3 1.00683
-4.4 1.0056
-4.5 1.00458
-4.6 1.00372
-4.7 1.00301
-4.8 1.00243
-4.9 1.00195
-5.0 1.00155
-5.1 1.00123
-5.2 1.00097
-5.3 1.00076
-5.4 1.00059
-5.5 1.00046
-5.6 1.00036
-5.7 1.00027
-5.8 1.00021
-5.9 1.00016
-6.0 1.00012
};
\addlegendentry{$\GammaA{\code{N}_{16}}$}

\addplot [color=red,solid,line width=1pt, mark=-*, mark options={solid, line width = 0.5pt, fill=white}]
table {%
-6.0 1.00463
-5.9 1.00566
-5.8 1.0069
-5.7 1.00836
-5.6 1.0101
-5.5 1.01215
-5.4 1.01455
-5.3 1.01736
-5.2 1.02064
-5.1 1.02444
-5.0 1.02884
-4.9 1.03391
-4.8 1.03975
-4.7 1.04644
-4.6 1.05408
-4.5 1.06279
-4.4 1.0727
-4.3 1.08393
-4.2 1.09664
-4.1 1.11097
-4.0 1.12711
-3.9 1.14524
-3.8 1.16556
-3.7 1.18827
-3.6 1.21361
-3.5 1.24179
-3.4 1.27307
-3.3 1.30766
-3.2 1.3458
-3.1 1.38768
-3.0 1.43347
-2.9 1.48324
-2.8 1.53702
-2.7 1.59465
-2.6 1.65583
-2.5 1.72002
-2.4 1.78642
-2.3 1.85389
-2.2 1.92096
-2.1 1.9858
-2.0 2.04629
-1.9 2.10011
-1.8 2.14491
-1.7 2.1785
-1.6 2.19911
-1.5 2.20561
-1.4 2.19763
-1.3 2.17561
-1.2 2.14078
-1.1 2.09495
-1.0 2.04033
-0.9 1.97929
-0.8 1.91412
-0.7 1.84693
-0.6 1.77951
-0.5 1.71329
-0.4 1.64937
-0.3 1.58854
-0.2 1.5313
-0.1 1.47793
0.0 1.42857
0.1 1.38319
0.2 1.3417
0.3 1.30394
0.4 1.2697
0.5 1.23875
0.6 1.21087
0.7 1.18582
0.8 1.16336
0.9 1.14328
1.0 1.12536
1.1 1.10942
1.2 1.09526
1.3 1.08271
1.4 1.07162
1.5 1.06184
1.6 1.05325
1.7 1.0457
1.8 1.03911
1.9 1.03336
2.0 1.02835
2.1 1.02402
2.2 1.02027
2.3 1.01705
2.4 1.01428
2.5 1.01192
2.6 1.00991
2.7 1.0082
2.8 1.00676
2.9 1.00555
3.0 1.00453
3.1 1.00368
3.2 1.00298
3.3 1.0024
3.4 1.00192
3.5 1.00153
3.6 1.00122
3.7 1.00096
3.8 1.00075
3.9 1.00059
4.0 1.00046
4.1 1.00035
4.2 1.00027
4.3 1.00021
4.4 1.00016
4.5 1.00012
4.6 1.00009
4.7 1.00006
4.8 1.00005
4.9 1.00003
5.0 1.00003
5.1 1.00002
5.2 1.00001
5.3 1.00001
5.4 1.00001
5.5 1.0
5.6 1.0
5.7 1.0
5.8 1.0
5.9 1.0
6.0 1.0
};
\addlegendentry{$\lattice{BW}_{16}$}

\addplot [color=blue, densely dashed, line width=1pt, mark=o, mark options={solid, mark size=0.5pt}]
table {%
-6.0 1.00012
-5.9 1.00016
-5.8 1.00021
-5.7 1.00027
-5.6 1.00036
-5.5 1.00046
-5.4 1.00059
-5.3 1.00076
-5.2 1.00097
-5.1 1.00123
-5.0 1.00155
-4.9 1.00195
-4.8 1.00243
-4.7 1.00301
-4.6 1.00372
-4.5 1.00458
-4.4 1.0056
-4.3 1.00683
-4.2 1.00828
-4.1 1.01
-4.0 1.01203
-3.9 1.01441
-3.8 1.01719
-3.7 1.02044
-3.6 1.0242
-3.5 1.02856
-3.4 1.03358
-3.3 1.03936
-3.2 1.04597
-3.1 1.05353
-3.0 1.06214
-2.9 1.07192
-2.8 1.08299
-2.7 1.09551
-2.6 1.10961
-2.5 1.12547
-2.4 1.14325
-2.3 1.16315
-2.2 1.18534
-2.1 1.21005
-2.0 1.23746
-1.9 1.2678
-1.8 1.30127
-1.7 1.33804
-1.6 1.37828
-1.5 1.42211
-1.4 1.46958
-1.3 1.52063
-1.2 1.5751
-1.1 1.63266
-1.0 1.69277
-0.9 1.75465
-0.8 1.81723
-0.7 1.87917
-0.6 1.93881
-0.5 1.99427
-0.4 2.04351
-0.3 2.0845
-0.2 2.11535
-0.1 2.13452
0.0 2.14103
0.1 2.13452
0.2 2.11535
0.3 2.0845
0.4 2.04351
0.5 1.99427
0.6 1.93881
0.7 1.87917
0.8 1.81723
0.9 1.75465
1.0 1.69277
1.1 1.63266
1.2 1.5751
1.3 1.52063
1.4 1.46958
1.5 1.42211
1.6 1.37828
1.7 1.33804
1.8 1.30127
1.9 1.2678
2.0 1.23746
2.1 1.21005
2.2 1.18534
2.3 1.16315
2.4 1.14325
2.5 1.12547
2.6 1.10961
2.7 1.09551
2.8 1.08299
2.9 1.07192
3.0 1.06214
3.1 1.05353
3.2 1.04597
3.3 1.03936
3.4 1.03358
3.5 1.02856
3.6 1.0242
3.7 1.02044
3.8 1.01719
3.9 1.01441
4.0 1.01203
4.1 1.01
4.2 1.00828
4.3 1.00683
4.4 1.0056
4.5 1.00458
4.6 1.00372
4.7 1.00301
4.8 1.00243
4.9 1.00195
5.0 1.00155
5.1 1.00123
5.2 1.00097
5.3 1.00076
5.4 1.00059
5.5 1.00046
5.6 1.00036
5.7 1.00027
5.8 1.00021
5.9 1.00016
6.0 1.00012
};
\addlegendentry{$\bigConstrA{\code{C}^{(5)}_{\textnormal{ofsd}}}$,~\textnormal{see Table}~\ref{tab:table_secrecy-gains_FU-lattices}}

% \addplot[mark=none, black] coordinates {(2,2.2069) (-6,2.2069)};
% \addlegendentry{}

\end{axis}

\end{tikzpicture}
  \vspace{-1ex}
  \caption{The secrecy functions of $\GammaA{\code{N}_{16}}$, $\lattice{BW}_{16}$, and $\bigConstrA{\code{C}^{(5)}_{\textnormal{ofsd}}}$, as a function of $\tau$ in dB.}
  \label{fig:secrecy-function_16}
  \vspace{-2ex}
\end{figure}

\begin{remark} We remark the following about Table~\ref{tab:table_secrecy-gains_FU-lattices}:
  \begin{description}
    \label{rem:remark_TableII}
  \item[Regarding Theorem~\ref{thm:strong-secrecy-gain_unimodular-lattices}:] We use the sufficient condition~\eqref{eq:condition_ai-ht_evenFSD} in Theorem~\ref{thm:strong-secrecy-gain_unimodular-lattices} for the even codes and the numerical derivative analysis with Wolfram Mathematica~\cite{Mathematica} for the odd codes to confirm the strong secrecy gain in Table~\ref{tab:table_secrecy-gains_FU-lattices}.
  \item[Regarding Theorem~\ref{thm:necessary-condition_optimal-FSDcode}:] For a given dimension $n$, the condition of Theorem~\ref{thm:necessary-condition_optimal-FSDcode} was checked for all codes in Table~\ref{tab:table_secrecy-gains_FU-lattices}, and the best one among them satisfies the condition. 
  \item[Improvements:] For most dimensions $n>8$, the secrecy gain of formally unimodular lattices that are not unimodular outperform the unimodular lattices (obtained from self-dual codes), presented in~\cite[Tables~I and II]{LinOggier13_1}. %In some cases (\emph{e.g.} [12,6], [22,11]) we were unable to find good efsd codes with secrecy gains different from the sd codes. 
    Also, to highlight the comparison with unimodular lattices, the second column refers to the upper bound on the secrecy gain of unimodular lattices obtained from Construction~A in~\cite[Tab.~III]{LinOggier12_1} and not all of the values are known to be achieved. Improvements can be observed in dimensions $10$, $12$, $14$, $20$, and $22$.
  \item[Dimension $32$:] It is known that the Barnes-Wall lattice of dimension $32$, denoted by $\lattice{BW}_{32}$, achieves its secrecy gain of $\nicefrac{64}{9}\approx 7.11$~\cite[Sec.~IV-C]{OggierSoleBelfiore16_1}, which is better than all the tabulated values in dimension $32$. However, because $\lattice{BW}_{32}$ is not obtained via Construction~A, we did not include its secrecy gain here.
    % \item[Type I/II:] Comparing the secrecy gain of lattices obtained from even self-dual codes (namely, the type I codes) and doubly even self-dual codes (Type II codes) in dimensions $32$ and $40$ (see Section~\ref{sec:all-weight-enumerators-codes}), we get $\xi_{\Lambda_\textnormal{A}(\code{C}_\textnormal{sd})} \approx 6.564 < 6.748 = \xi_{\Lambda_\textnormal{A}(\code{C}_\textnormal{efsd})}$ in dimension $32$ and $\xi_{\Lambda_\textnormal{A}(\code{C}_\textnormal{sd})} \approx 11.977 < 12.364 = \xi_{\Lambda_\textnormal{A}(\code{C}_\textnormal{ofsd})}$ in dimension $40$. Note that with respect to the approach in~\cite{Pinchak13_1}, where the secrecy gain conjecture for Construction A lattices obtained from doubly even self-dual codes was demonstrated, our Theorem~\ref{thm:inv_secrecy-function_WeightEnumerator} also applies for type II codes.
    %   xxx: not clear if the type II codes are among the best?
  \item[Nonlattice packings:] We have presented nonlattice packings, generated via Construction~A from nonlinear formally self-dual codes, in dimensions $12, 16$, and $20$ (see Examples~\ref{ex:NR16} and~\ref{ex:Nonlattice_12_20}). Among these, a gain is observed only in dimension $16$, illustrated in Figure~\ref{fig:secrecy-function_16}. The functions being represented are respectively, the best formally unimodular lattice from Table~\ref{tab:table_secrecy-gains_FU-lattices}, the Construction~A packing generated from the Nordstrom-Robinson code $\code{N}_{16}$, and the Barnes-Wall lattice $\lattice{BW}_{16}$, whose theta series and construction can be found at \cite[pp.~129--131]{ConwaySloane99_1}. In terms of comparison, we have $\xi_{\GammaA{\code{N}_{16}}} = 2.2069 > \xi_{\lattice{BW}_{16}} = 2.2056 > \xi_{\eConstrA{\code{C}^{(5)}_{\textnormal{ofsd}}}} = 2.141$.
  \end{description}
\end{remark}

\section{Conclusion and Future work}
\label{sec:conclusion-future-work}

A new class consisting of nonlattice packings that are periodic and geometrically uniform was introduced, namely the \emph{lattice-like packings}. Its subclass, called \emph{formally unimodular (lattice-like) packings}, which is analogous to isodual and unimodular lattices, was further studied. We showed several fundamental properties of formally unimodular packings and built a coset coding scheme for a lattice containing a lattice-like packing as a subset. Their secrecy function behavior over the Gaussian WTC was shown to be the same as that of unimodular and isodual lattices.

In particular, we investigated the Construction~A formally unimodular packings obtained from formally self-dual codes and gave a universal approach to determine their secrecy gain. Furthermore, we provided a necessary condition for Construction~A formally unimodular packings to be weakly secrecy-optimal of a given dimension. The necessary condition is based on the weight distribution of the underlying formally self-dual code. We found formally unimodular packings/lattices of better secrecy gain than the best-known unimodular lattices from the literature.

Note that we mainly focus on the secrecy performance comparison between formally unimodular packings of the \emph{same} dimension. The desired properties of the weight enumerator of Theorem~\ref{thm:necessary-condition_optimal-FSDcode} showed that the rough rule-of-thumb suggesting that codes with larger minimum Hamming distances yield a better secrecy gain is generally not true. However, even comparing the secrecy gains of formally unimodular packings for \emph{different} dimensions, this simple rule can fail. %such indication could also be failed. 
To illustrate this, consider two pairs of code examples in Appendix~\ref{sec:all-weight-enumerators-codes}: the $[72,36,16]$ code $\code{C}_\textnormal{sd}^{(16)}$ versus the $[78,39,14]$ code $\code{C}_\textnormal{efsd}^{(14)}$ and the $[104,52,20]$ code versus the $[108,54,14]$ code. It can be seen that the $[72,36,16]$ self-dual code with a larger minimum Hamming distance is worse than the $[78,39,14]$ even formally self-dual code in terms of secrecy gain ($\xi_{\eGammaA{\code{C}_{\textnormal{sd}}^{(16)}}}=146.844<\xi_{\eGammaA{\code{C}_{\textnormal{efsd}}^{(14)}}}=241.042$). The same observation can be made for the $[104,52,20]$ self-dual code and the $[108,54,14]$ even formally self-dual code. These two examples illustrate that providing sufficient or necessary conditions to verify the optimality of secrecy gain for different dimensions remains an interesting avenue for future work. It would also be interesting to investigate whether the result of Theorem~\ref{thm:2nd-bound_Eve-success-probability} applies to more general lattice-like packings.

Finally, in order to limit the scope of the paper, we have chosen not to address several interesting issues. We save these for future research. First of all, we plan to address the asymptotic behavior of the secrecy gain of lattice-like packings, as well as bounds on the mutual information leakage to the eavesdropper. Likewise, the practical application of the kind of schemes addressed in this work necessitates a discussion of the complexity of  encoding and decoding. Schemes based on codes with manageable trellis complexity, like the Nordstrom-Robinson and tail-biting convolutional codes, may offer decoding benefits.

% The technique we used to construct the theta series of a formally unimodular lattice is based on Construction A from a formally self-dual code. Hence, only results of formally unimodular packings with even dimensions are discussed. However, obtaining the closed-form expression of the theta series of a formally unimodular lattice with odd dimension is possible, e.g., generalizing Hecke's theorem~\cite[Th.~7, Ch.~7]{ConwaySloane99_1}. This direction of study is of great interest for future research. We also observe that the secrecy gain is generally improved with a higher minimum Hamming distance and lower kissing number, and it appears to increase exponentially with the dimension. The precise relation with these parameters will be investigated in future work.

% add the caption={} option for itemize or minipage environment
% \todo[author=Lin,inline,caption={}]{Future work on odd-dimension packings, generalize Hecke's theorem?}

\section*{Acknowledgment}

The authors would like to thank the two anonymous reviewers and the Associate Editor Prof.~Moshe Schwartz for their thoughtful and valuable comments.

% \balance
%\ifthenelse{\boolean{short_version}}{\IEEEtriggeratref{10}}{\IEEEtriggeratref{20}}
% \IEEEtriggercmd{\enlargethispage{-5cm}}

\bibliographystyle{IEEEtran}
\bibliography{defshort1,biblioHY}

\clearpage
% change between short and long versions.
\appendices
%%%%%%%%%%%%%%%%%%%%%%%%%%%%%%%%%%%%%%%%%%%%%%%%%%%%%%%%%%%%%%%%%%%%%
\ifthenelse{\boolean{short_version}}{
  \makeatletter\afterpage{\if@firstcolumn \else\afterpage{
      \onecolumn
      \section{Weight Enumerators of Codes for Table~\ref{tab:table_secrecy-gains_FU-lattices}}
\label{sec:all-weight-enumerators-codes}

\begin{longtable}{|c|c|c|p{9.5cm}|c|c|}
  \caption{Codes and Their Weight Enumerators}
  \\*\hline
  $\code{C}$
  & Type
  & Reference
  & $W_\code{C}(x,y)$
  & $\xi_{\eGammaA{\code{C}}}$
  \\*\hline\hline
  $[6,3,2]$  & efsd & \cite{HuffmanPless03_1} &  $x^6+3x^4y^2+3x^2y^4+y^6$  & $1$
  \\*\hline
  $[6,3,3]$  & ofsd & \cite{BetsumiyaHarada01_1}%\textnormal{,~App.~\ref{sec:isodual_coco}}
  & $x^6+4x^3y^3+3x^2y^4$ & $\mathbf{1. 172}$
  \\*\hline
  $[8,4,4]$  & sd   & \cite{HuffmanPless03_1} & $x^8+14x^4y^4+y^8$ & $\mathbf{1.333}$ 
  \\*\hline
  $[8,4,3]$  & ofsd & \cite{BetsumiyaHarada01_1} & $x^8+3x^5y^3+7x^4y^4+4x^3y^5+xy^7$ & 1.282  
  \\*\hline
  $[8,4,3]$  & ofsd & \cite{BetsumiyaHarada01_1} & $x^8+4x^5y^3+5x^4y^4+4x^3y^5+2x^2y^6$ & $1.264$
  \\*\hline
  $[10,5,4]$ & efsd & \cite{HuffmanPless03_1} & $x^{10}+15x^6y^4+15x^4y^6+y^{10}$ & $1.455$
  \\*\hline
  $[10,5,4]$ & ofsd & \cite{BetsumiyaHarada01_1} & $x^{10}+10x^6y^4+16x^5y^5+5x^2y^8$ & $\mathbf{1.478}$
  \\*\hline  
  $[12,6,4]$ & sd & \cite{LinOggier13_1} & $x^{12}+15x^8y^4+32x^6y^6+15x^4y^8+y^{12}$ & $1.6$
  \\*\hline  
  $[12,6,4]$ & efsd & \cite{BetsumiyaGulliverHarada99_1} & $x^{12}+15x^8y^4+32x^6y^6+15x^4y^8+y^{12}$ & $1.6$
  \\*\hline
  $[12,6,4]$ & ofsd & \cite{BetsumiyaHarada01_1} & $x^{12}+6x^8y^4+24x^7y^5+16x^6y^6+9x^4y^8+8x^3y^9$ & $\mathbf{1.657}$
  \\*\hline
  $(12,64,4)$ & ofsd & \cite{YooLeeKim17_1} & $x^{12}+6x^8y^4+24x^7y^5+16x^6y^6+9x^4y^8+8x^3y^9$ & $\mathbf{1.657}$
  \\*\hline
  $[14,7,4]$  & sd & \cite{LinOggier13_1} & $x^{14}+14x^{10}y^4+49x^8y^6+49x^6y^8+14x^4y^{10}+y^{14}$ & $1.778$ 
  \\*\hline
  $[14,7,2]$  & efsd & \cite{BetsumiyaGulliverHarada99_1} & $x^{14}+x^{12} y^2+15 x^{10} y^4+47 x^8 y^6+47 x^6 y^8+15 x^4 y^{10}+x^2 y^{12}+y^{14}$ & $1.6$ 
  \\*\hline
  $[14,7,4]$  & ofsd & \cite{BetsumiyaHarada01_1} & $x^{14}+3x^{10}y^4+24x^9y^5+36x^8y^{6}+16x^7y^7 + 11x^6y^8 + 24x^5y^9+12x^4y^{10}+x^2y^{12}$ & $\mathbf{1.875}$ 
  \\*\hline
  $[16,8,4]$  & sd & \cite{LinOggier13_1} & $x^{16}+12x^{12}y^4+64x^{10}y^6+102x^8y^8+64x^6y^{10}+12x^4y^{12}+y^{16}$ & $2$ 
  \\*\hline
  $[16,8,4]$  & efsd & \cite{BetsumiyaHarada01_2} & $x^{16}+4 x^{12} y^4+96 x^{10} y^6+54 x^8 y^8+96 x^6 y^{10}+4 x^4 y^{12}+y^{16}$ & $2.133$ 
  \\*\hline
  $[16,8,5]$  & ofsd & \cite{BetsumiyaHarada01_1} & $x^{16}+24x^{11}y^5+44x^{10}y^6+40x^9y^7+45x^8y^{8}+40x^7y^9+28x^6y^{10}+24x^5y^{11}+10x^4y^{12}$ & $2.141$ 
  \\*\hline
  $(16,256,6)$  & efsd & \cite[p.~74]{MacWilliamsMallowsSloane72_1} & $x^{16}+112x^{10}y^6+30x^8y^8+112x^6y^{10}+y^{16}$ & $\mathbf{2.207}$  
  \\*\hline 
  \multirow{2}{*}{$(18, 512, 2)$}  & \multirow{2}{*}{efsd} & \multirow{2}{*}{Ex.~\ref{ex:counterexample_dim18}} & $x^{18}+x^{16} y^2+112 x^{12} y^6+142 x^{10} y^8+142 x^8 y^{10}+112 x^6 y^{12}+x^2 y^{16}+y^{18}$ & \multirow{2}{*}{$2.207$}
  \\*\hline
  \multirow{2}{*}{$[18, 9, 3]$}  & \multirow{2}{*}{ofsd} & \multirow{2}{*}{Ex.~\ref{ex:counterexample_dim18}} & $x^{18}+12 x^{15} y^3+9 x^{14} y^4+48 x^{12} y^6+72 x^{11} y^7+ 27 x^{10} y^8+ 64 x^9 y^9 +144 x^8 y^{10}+108 x^7 y^{11}+27 x^6 y^{12}$ & \multirow{2}{*}{$1.608$} 
  \\*\hline
  \multirow{2}{*}{$[18,9, 4]$}  & \multirow{2}{*}{sd} & \multirow{2}{*}{\cite{LinOggier13_1}} & $x^{18}+9x^{14}y^4+75x^{12}y^6+171x^{10}y^8+171x^{8}y^{10}+75x^{6}y^{12}+9x^4y^{14}+y^{18}$ & \multirow{2}{*}{$2.286$} 
  \\*\hline
  $[18,9, 6]$  & efsd & \cite{OEIS-we-list} &  $x^{18}+102x^{12}y^6+153x^{10}y^8+153x^{8}y^{10}+102x^{6}y^{12}+y^{18}$ & $\mathbf{2.485}$  
  \\*\hline
  \multirow{2}{*}{$[18,9, 5]$}  & \multirow{2}{*}{ofsd} & \multirow{2}{*}{tb} % \textnormal{,~App.~\ref{sec:isodual_coco}}
  & $x^{18}+18x^{13}y^5+48x^{12}y^6+63x^{11}y^7+81x^{10}y^{8}+100x^{9}y^{9}+72x^8y^{10}+54x^7y^{11}+54x^6y^{12}+18x^5y^{13}+3x^{3}y^{15}$ & \multirow{2}{*}{$2.424$} 
  \\*\hline
  \multirow{2}{*}{$[20, 10, 4]$}  & \multirow{2}{*}{sd} & \multirow{2}{*}{\cite{LinOggier13_1}} % \textnormal{,~App.~\ref{sec:isodual_coco}}
  & $x^{20}+5 x^{16} y^4+80 x^{14} y^6+250 x^{12} y^8+352 x^{10} y^{10}+250 x^8 y^{12}+80 x^6 y^{14}+5 x^4 y^{16}+y^{20}$ & \multirow{2}{*}{$2.667$} 
  \\*\hline
   \multirow{2}{*}{$[20, 10, 6]$}  & \multirow{2}{*}{efsd} & \multirow{2}{*}{\cite{FieldsGaboritHuffmanPless01_1}} % \textnormal{,~App.~\ref{sec:isodual_coco}}
  & $x^{20}+90x^{14}y^6+255x^{12}y^8+332x^{10}y^{10}+255x^{8}y^{12}+90x^{6}y^{14}+y^{20}$ & \multirow{2}{*}{$2.813$} 
  \\*\hline
  \multirow{2}{*}{$(20, 1024, 6)$}  & \multirow{2}{*}{efsd} & \multirow{2}{*}{\cite{YooLeeKim17_1}} % \textnormal{,~App.~\ref{sec:isodual_coco}}
  & $x^{20}+90x^{14}y^6+255x^{12}y^8+332x^{10}y^{10}+255x^{8}y^{12}+90x^{6}y^{14}+y^{20}$ & \multirow{2}{*}{$2.813$} 
  \\*\hline
  \multirow{2}{*}{$[20, 10, 6]$}  & \multirow{2}{*}{ofsd} & \multirow{2}{*}{\cite{BetsumiyaGulliverHarada99_1}} % \textnormal{,~App.~\ref{sec:isodual_coco}}
  & $x^{20}+40 x^{14} y^6+160 x^{13} y^7+130 x^{12} y^8+176 x^{10} y^{10}+320 x^9 y^{11}+120 x^8 y^{12}+40 x^6 y^{14}+32 x^5 y^{15}+5 x^4 y^{16}$ & \multirow{2}{*}{$\mathbf{2.868}$} 
  \\*\hline
  \multirow{2}{*}{$[22, 11, 6]$}  & \multirow{2}{*}{sd} & \multirow{2}{*}{\cite{LinOggier13_1}} % \textnormal{,~App.~\ref{sec:isodual_coco}}
  & $x^{22}+77x^{16}y^6+330x^{14}y^8+616x^{12}y^{10}+616x^{10}y^{12}+330x^{8}y^{14}+77x^{6}y^{16}+y^{22}$ & \multirow{2}{*}{$3.2$}
  \\*\hline
   \multirow{3}{*}{$[22, 11, 6]$}  & \multirow{3}{*}{ofsd} & \multirow{3}{*}{tb} % \textnormal{,~App.~\ref{sec:isodual_coco}}
  & $x^{22}+44x^{16}y^6+121x^{15}y^7+143x^{14}y^{8}+231x^{13}y^{9}+319x^{12}y^{10}+298x^{11}y^{11}+330x^{10}y^{12}+286x^{9}y^{13}+154x^{8}y^{14}+77x^{7}y^{15}+22x^{6}y^{16}+11x^{5}y^{17}+11x^{4}y^{18}$ & \multirow{3}{*}{$3.243$} 
  \\*\hline
   \multirow{2}{*}{$[22, 11, 7]$}  & \multirow{2}{*}{ofsd} & \multirow{2}{*}{\cite{BetsumiyaHarada01_1}} % \textnormal{,~App.~\ref{sec:isodual_coco}}
  & $x^{22}+176x^{15}y^7+330x^{14}y^8+672x^{11}y^{11}+616x^{10}y^{12}+176x^{7}y^{15}+77x^{6}y^{16}$ & \multirow{2}{*}{$\mathbf{3.335}$} 
  \\*\hline
  $[24, 12, 8]$  & sd & \cite{MacWilliamsSloane77_1} % \textnormal{,~App.~\ref{sec:isodual_coco}}
  & $x^{24} + 759 x^{16} y^{8} + 2576 x^{12} y^{12} + 759 x^{8} y^{16} + y^{24}$ & $\mathbf{3.879}$ 
  \\*\hline
  \multirow{2}{*}{$[24, 12, 6]$}  & \multirow{2}{*}{efsd} & \multirow{2}{*}{tb} % \textnormal{,~App.~\ref{sec:isodual_coco}}
  & $x^{24}+ 64 x^{18} y^6+375 x^{16} y^8+960 x^{14} y^{10}+1296 x^{12} y^{12}+960 x^{10} y^{14}+375 x^8 y^{16}+64 x^6 y^{18}+y^{24}$ & \multirow{2}{*}{$3.657$} 
  \\*\hline
  \multirow{3}{*}{$[30, 15, 6]$}  & \multirow{3}{*}{sd} & \multirow{3}{*}{\cite{ConwaySloane90_1}} % \textnormal{,~App.~\ref{sec:isodual_coco}}
  & $x^{30}+19 x^{24} y^6+393 x^{22} y^8+1848 x^{20} y^{10}+5192 x^{18} y^{12}+8931 x^{16} y^{14}+8931 x^{14} y^{16}+5192 x^{12} y^{18}+1848 x^{10} y^{20}+393 x^8 y^{22}+19 x^6 y^{24}+y^{30}$ & \multirow{3}{*}{$5.689$} 
  \\*\hline
  \multirow{2}{*}{$[30, 15, 8]$}  & \multirow{2}{*}{efsd} & \multirow{2}{*}{\cite{BouyuklievaBouyukliev10_1}} % \textnormal{,~App.~\ref{sec:isodual_coco}}
  & $x^{30}+ 450x^{22}y^8 + 1848x^{20}y^{10}+5040x^{18}y^{12}+9045x^{16}y^{14}+9045x^{14}y^{16}+5040x^{12}y^{18}+1848x^{10}y^{20}+450x^{8}y^{22}+y^{30}$ & \multirow{2}{*}{$\mathbf{5.843}$} 
  \\*\hline
  \multirow{5}{*}{$[30, 15, 7]$}  & \multirow{5}{*}{ofsd} & \multirow{5}{*}{tb} % \textnormal{,~App.~\ref{sec:isodual_coco}}
  & $x^{30} + 60 x^{23} y^7+210 x^{22} y^8+500 x^{21} y^9+930 x^{20} y^{10}+1560 x^{19} y^{11}+2570 x^{18} y^{12}+3660 x^{17} y^{13}+4530 x^{16} y^{14}+4824 x^{15} y^{15}+4335 x^{14} y^{16}+3660 x^{13} y^{17}+2710 x^{12} y^{18}+1560 x^{11} y^{19}+918 x^{10} y^{20}+500 x^9 y^{21}+150 x^8 y^{22}+60 x^7 y^{23}+30 x^6 y^{24}$ & \multirow{5}{*}{$5.785$} 
  \\*\hline
   \multirow{2}{*}{$[32, 16, 8]$}  & \multirow{2}{*}{sd} & \multirow{2}{*}{\cite{OEIS-we-list}} % \textnormal{,~App.~\ref{sec:isodual_coco}}
  & $x^{32}+ 620 x^{24} y^8+13888 x^{20} y^{12}+36518 x^{16} y^{16}+13888 x^{12} y^{20}+620 x^8 y^{24}+y^{32}$ & \multirow{2}{*}{$6.564$} 
  \\*\hline
  \multirow{3}{*}{$[32, 16, 8]$}  & \multirow{3}{*}{sd} & \multirow{3}{*}{\cite{OEIS-we-list}} % \textnormal{,~App.~\ref{sec:isodual_coco}}
  & $x^{32}+364 x^{24} y^8+2048 x^{22} y^{10}+6720 x^{20} y^{12}+14336 x^{18} y^{14}+18598 x^{16} y^{16}+14336 x^{14} y^{18}+6720 x^{12} y^{20}+2048 x^{10} y^{22}+364 x^8 y^{24}+y^{32}$ & \multirow{3}{*}{$6.737$} 
  \\*\hline
  \multirow{3}{*}{$[32, 16, 8]$}  & \multirow{3}{*}{efsd} & \multirow{3}{*}{tb} % \textnormal{,~App.~\ref{sec:isodual_coco}}
  & $x^{32}+348 x^{24} y^8+2176 x^{22} y^{10}+6272 x^{20} y^{12}+15232 x^{18} y^{14}+17478 x^{16} y^{16}+15232 x^{14} y^{18}+6272 x^{12} y^{20}+2176 x^{10} y^{22}+348 x^8 y^{24}+y^{32}$ & \multirow{3}{*}{$\mathbf{6.748}$}  
  \\*\hline
   \multirow{5}{*}{$[32, 16, 7]$}  & \multirow{5}{*}{ofsd} & \multirow{5}{*}{tb} % \textnormal{,~App.~\ref{sec:isodual_coco}}
  & $x^{32}+64 x^{25} y^7+176 x^{24} y^8+384 x^{23} y^9+984 x^{22} y^{10}+2096 x^{21} y^{11}+3500 x^{20} y^{12}+5136 x^{19} y^{13}+7096 x^{18} y^{14}+8624 x^{17} y^{15}+9133 x^{16} y^{16}+8848 x^{15} y^{17}+7384 x^{14} y^{18}+5136 x^{13} y^{19}+3292 x^{12} y^{20}+1968 x^{11} y^{21}+1032 x^{10} y^{22}+464 x^9 y^{23}+154 x^8 y^{24}+48 x^7 y^{25}+16 x^6 y^{26}$ & \multirow{5}{*}{$6.628$} 
  \\*\hline
  \multirow{3}{*}{$[40, 20, 8]$}  & \multirow{3}{*}{sd} & \multirow{3}{*}{\cite{ConwaySloane90_1}} % \textnormal{,~App.~\ref{sec:isodual_coco}}
  & $x^{40}+285 x^{32} y^8+21280 x^{28} y^{12}+239970 x^{24} y^{16}+525504 x^{20} y^{20}+239970 x^{16} y^{24}+21280 x^{12} y^{28}+285 x^8 y^{32}+y^{40}$ & \multirow{3}{*}{$11.977$} 
  \\*\hline
  \multirow{4}{*}{$[40, 20, 8]$}  & \multirow{4}{*}{sd} & \multirow{4}{*}{\cite{ConwaySloane90_1}} % \textnormal{,~App.~\ref{sec:isodual_coco}}
  & $x^{40}+125 x^{32} y^8+1664 x^{30} y^{10}+10720 x^{28} y^{12}+44160 x^{26} y^{14}+119810 x^{24} y^{16}+216320 x^{22} y^{18}+262976 x^{20} y^{20}+216320 x^{18} y^{22}+119810 x^{16} y^{24}+44160 x^{14} y^{26}+10720 x^{12} y^{28}+1664 x^{10} y^{30}+125 x^8 y^{32}+y^{40}$ & \multirow{4}{*}{$12.191$} 
  \\*\hline
  \multirow{4}{*}{$[40, 20, 8]$}  & \multirow{4}{*}{efsd} & \multirow{4}{*}{tb} % \textnormal{,~App.~\ref{sec:isodual_coco}}
  & $x^{40}+150 x^{32} y^8+1564 x^{30} y^{10}+10770 x^{28} y^{12}+44460 x^{26} y^{14}+119385 x^{24} y^{16}+216120 x^{22} y^{18}+263676 x^{20} y^{20}+216120 x^{18} y^{22}+119385 x^{16} y^{24}+44460 x^{14} y^{26}+10770 x^{12} y^{28}+1564 x^{10} y^{30}+150 x^8 y^{32}+y^{40}$ & \multirow{4}{*}{$12.134$} 
  \\*\hline
  \multirow{7}{*}{$[40, 20, 9]$}  & \multirow{7}{*}{ofsd} & \multirow{7}{*}{tb} % \textnormal{,~App.~\ref{sec:isodual_coco}}
  & $x^{40}+360 x^{31} y^9+922 x^{30} y^{10}+2060 x^{29} y^{11}+5775 x^{28} y^{12}+11340 x^{27} y^{13}+20980 x^{26} y^{14}+39064 x^{25} y^{15}+60185 x^{24} y^{16}+83680 x^{23} y^{17}+109740 x^{22} y^{18}+125640 x^{21} y^{19}+130046 x^{20} y^{20}+125640 x^{19} y^{21}+107680 x^{18} y^{22}+83680 x^{17} y^{23}+60830 x^{16} y^{24}+39064 x^{15} y^{25}+22250 x^{14} y^{26}+11340 x^{13} y^{27}+4755 x^{12} y^{28}+2060 x^{11} y^{29}+1084 x^{10} y^{30}+360 x^9 y^{31}+40 x^8 y^{32}$ & \multirow{7}{*}{$\mathbf{12.364}$} 
  \\*\hline
  \multirow{4}{*}{$[42, 21, 10]$}  & \multirow{4}{*}{efsd} & \multirow{4}{*}{tb} % \textnormal{,~App.~\ref{sec:isodual_coco}}
  & $x^{42}+1722 x^{32} y^{10}+10619 x^{30} y^{12}+49815 x^{28} y^{14}+157563 x^{26} y^{16}+341530 x^{24} y^{18}+487326 x^{22} y^{20}+487326 x^{20} y^{22}+341530 x^{18} y^{24}+157563 x^{16} y^{26}+49815 x^{14} y^{28}+10619 x^{12} y^{30}+1722 x^{10} y^{32}+y^{42}$ & \multirow{4}{*}{$\mathbf{14.482}$} 
  \\*\hline
  \multirow{6}{*}{$[56, 28, 12]$}  & \multirow{6}{*}{efsd} & \multirow{6}{*}{tb} % \textnormal{,~App.~\ref{sec:isodual_coco}}
  & $x^{56}+4634 x^{44} y^{12}+44828 x^{42} y^{14}+307650 x^{40} y^{16}+1575924 x^{38} y^{18}+5865384 x^{36} y^{20}+15969660 x^{34} y^{22}+32430013 x^{32} y^{24}+ 49502068 x^{30} y^{26}+57035132 x^{28} y^{28}+ 49502068 x^{26} y^{30}+32430013 x^{24} y^{32}+15969660 x^{22} y^{34}+5865384 x^{20} y^{36}+1575924 x^{18} y^{38}+307650 x^{16} y^{40}+44828 x^{14} y^{42}+ 4634 x^{12} y^{44}+y^{56}$ & \multirow{6}{*}{$\mathbf{42.838}$} 
  \\*\hline
  \multirow{10}{*}{$[70, 35, 12]$}  & \multirow{10}{*}{sd} & \multirow{10}{*}{\cite{Harada97_1}} % \textnormal{,~App.~\ref{sec:isodual_coco}}
  & $x^{70}+ 832 x^{58} y^{12}+10770 x^{56} y^{14}+142279 x^{54} y^{16}+1353320 x^{52} y^{18}+9437352 x^{50} y^{20}+49957193 x^{48} y^{22}+204165154 x^{46} y^{24}+650426976 x^{44} y^{26}+1627816992 x^{42} y^{28}+3221537516 x^{40} y^{30}+5066102223 x^{38} y^{32}+6348918576 x^{36} y^{34}+6348918576 x^{34} y^{36}+5066102223 x^{32} y^{38}+3221537516 x^{30} y^{40}+1627816992 x^{28} y^{42}+650426976 x^{26} y^{44}+204165154 x^{24} y^{46}+49957193 x^{22} y^{48}+9437352 x^{20} y^{50}+1353320 x^{18} y^{52}+142279 x^{16} y^{54}+10770 x^{14} y^{56}+832 x^{12} y^{58}+y^{70}$ & \multirow{10}{*}{$127.712$}  
  \\*\hline
  \multirow{10}{*}{$[70, 35, 12]$}  & \multirow{10}{*}{efsd} & \multirow{10}{*}{tb} % \textnormal{,~App.~\ref{sec:isodual_coco}}
  & $x^{70}  + 455  x^{58}  y^{12} + 11235  x^{56}  y^{14} + 145985  x^{54}  y^{16} + 1348130  x^{52}  y^{18} + 9430974  x^{50}  y^{20} + 49926695  x^{48}  y^{22} + 204318835  x^{46}  y^{24} + 650297655  x^{44}  y^{26} + 1627628010  x^{42}  y^{28} + 3221888194  x^{40}  y^{30} + 5066010495  x^{38}  y^{32} + 6348862520  x^{36}  y^{34} + 6348862520  x^{34}  y^{36} + 5066010495  x^{32}  y^{38} + 3221888194  x^{30}  y^{40} + 1627628010  x^{28}  y^{42} + 650297655  x^{26}  y^{44} + 204318835  x^{24}  y^{46} + 49926695  x^{22}  y^{48} + 9430974  x^{20}  y^{50} + 1348130  x^{18}  y^{52} + 145985  x^{16}  y^{54} + 11235  x^{14}  y^{56} + 455  x^{12}  y^{58} +  y^{70}$ & \multirow{10}{*}{$128.073$} 
  \\*\hline
   \multirow{18}{*}{$[70, 35, 13]$}  & \multirow{18}{*}{ofsd} & \multirow{18}{*}{tb} % \textnormal{,~App.~\ref{sec:isodual_coco}}
  & $x^{70}+1225 x^{57} y^{13}+6125 x^{56} y^{14}+21700 x^{55} y^{15}+72590 x^{54} y^{16}+232680 x^{53} y^{17}+676410 x^{52} y^{18}+1838375 x^{51} y^{19}+4711427 x^{50} y^{20}+11204975 x^{49} y^{21}+24964310 x^{48} y^{22}+52191335 x^{47} y^{23}+102128145 x^{46} y^{24}+187879531 x^{45} y^{25}+325261230 x^{44} y^{26}+529884495 x^{43} y^{27}+813742900 x^{42} y^{28}+1178595250 x^{41} y^{29}+1610725606 x^{40} y^{30}+2078727420 x^{39} y^{31}+2533396005 x^{38} y^{32}+2916830420 x^{37} y^{33}+3174375820 x^{36} y^{34}+3264970134 x^{35} y^{35}+3174028690 x^{34} y^{36}+2917093830 x^{33} y^{37}+2533383720 x^{32} y^{38}+2078410810 x^{31} y^{39}+1610915418 x^{30} y^{40}+1178784530 x^{29} y^{41}+813674900 x^{28} y^{42}+529809070 x^{27} y^{43}+325223220 x^{26} y^{44}+187929077 x^{25} y^{45}+102154885 x^{24} y^{46}+52153640 x^{23} y^{47}+24962700 x^{22} y^{48}+11215020 x^{21} y^{49}+4706842 x^{20} y^{50}+1841315 x^{19} y^{51}+682115 x^{18} y^{52}+232155 x^{17} y^{53}+69930 x^{16} y^{54}+20727 x^{15} y^{55}+5845 x^{14} y^{56}+1435 x^{13} y^{57}+350 x^{12} y^{58}+35 x^{11} y^{59}$ & \multirow{18}{*}{$\mathbf{128.368}$} 
  \\*\hline
  \multirow{6}{*}{$[72, 36, 16]$}  & \multirow{6}{*}{sd} & \multirow{6}{*}{\cite{OEIS-we-list}} % \textnormal{,~App.~\ref{sec:isodual_coco}}
  & $x^{72}+2982 x^{60} y^{12}+214065 x^{56} y^{16}+18303516 x^{52} y^{20}+462306915 x^{48} y^{24}+4398818490 x^{44} y^{28}+16600354155 x^{40} y^{32}+25759476488 x^{36} y^{36}+16600354155 x^{32} y^{40}+4398818490 x^{28} y^{44}+462306915 x^{24} y^{48}+18303516 x^{20} y^{52}+214065 x^{16} y^{56}+2982 x^{12} y^{60}+y^{72}$ & \multirow{6}{*}{$146.844$} 
  \\*\hline
  \multirow{11}{*}{$[78, 39, 14]$}  & \multirow{11}{*}{efsd} & \multirow{11}{*}{tb} % \textnormal{,~App.~\ref{sec:isodual_coco}}
  & $x^{78}+3471 x^{64} y^{14}+63336 x^{62} y^{16}+772980 x^{60} y^{18}+7219368 x^{58} y^{20}+51527346 x^{56} y^{22}+287551706 x^{54} y^{24}+1266693912 x^{52} y^{26}+4442835540 x^{50} y^{28}+12510913844 x^{48} y^{30}+28453167444 x^{46} y^{32}+52493946648 x^{44} y^{34}+78823802720 x^{42} y^{36}+96539408628 x^{40} y^{38}+96539408628 x^{38} y^{40}+78823802720 x^{36} y^{42}+52493946648 x^{34} y^{44}+28453167444 x^{32} y^{46}+12510913844 x^{30} y^{48}+4442835540 x^{28} y^{50}+1266693912 x^{26} y^{52}+287551706 x^{24} y^{54}+51527346 x^{22} y^{56}+7219368 x^{20} y^{58}+772980 x^{18} y^{60}+63336 x^{16} y^{62}+3471 x^{14} y^{64}+y^{78}$ & \multirow{11}{*}{$241.042$} 
  \\*\hline
  \multirow{9}{*}{$[104, 52, 20]$}  & \multirow{9}{*}{sd} & \multirow{9}{*}{\cite{OEIS-we-list}} % \textnormal{,~App.~\ref{sec:isodual_coco}}
  & $x^{104}+1138150 x^{84} y^{20}+206232780 x^{80} y^{24}+15909698064 x^{76} y^{28}+567725836990 x^{72} y^{32}+9915185041320 x^{68} y^{36}+88355709788905 x^{64} y^{40}+413543821457520 x^{60} y^{44}+1036378989344140 x^{56} y^{48}+1406044530294756 x^{52} y^{52}+1036378989344140 x^{48} y^{56}+413543821457520 x^{44} y^{60}+88355709788905 x^{40} y^{64}+9915185041320 x^{36} y^{68}+567725836990 x^{32} y^{72}+15909698064 x^{28} y^{76}+206232780 x^{24} y^{80}+1138150 x^{20} y^{84}+y^{104}$ & \multirow{9}{*}{$1885.06$} 
  \\*\hline
  \multirow{19}{*}{$[108, 54, 14]$}  & \multirow{19}{*}{efsd} & \multirow{19}{*}{tb} % \textnormal{,~App.~\ref{sec:isodual_coco}}
  & $x^{108}+756 x^{94} y^{14}+5022 x^{92} y^{16}+30354 x^{90} y^{18}+371223 x^{88} y^{20}+5418846 x^{86} y^{22}+71085987 x^{84} y^{24}+765738684 x^{82} y^{26}+6738702390 x^{80} y^{28}+48969093384 x^{78} y^{30}+296438923962 x^{76} y^{32}+1505875815558 x^{74} y^{34}+6456109668648 x^{72} y^{36}+23473804361040 x^{70} y^{38}+72678688668432 x^{68} y^{40}+192289983824466 x^{66} y^{42}+436005471914253 x^{64} y^{44}+849263560631748 x^{62} y^{46}+1423721807648100 x^{60} y^{48}+2057133110131674 x^{58} y^{50}+2564434300382478 x^{56} y^{52}+2759767104647972 x^{54} y^{54}+2564434300382478 x^{52} y^{56}+2057133110131674 x^{50} y^{58}+1423721807648100 x^{48} y^{60}+849263560631748 x^{46} y^{62}+436005471914253 x^{44} y^{64}+192289983824466 x^{42} y^{66}+72678688668432 x^{40} y^{68}+23473804361040 x^{38} y^{70}+6456109668648 x^{36} y^{72}+1505875815558 x^{34} y^{74}+296438923962 x^{32} y^{76}+48969093384 x^{30} y^{78}+6738702390 x^{28} y^{80}+765738684 x^{26} y^{82}+71085987 x^{24} y^{84}+5418846 x^{22} y^{86}+371223 x^{20} y^{88}+30354 x^{18} y^{90}+5022 x^{16} y^{92}+756 x^{14} y^{94}+y^{108}$ & \multirow{19}{*}{$2573.53$} 
  \\*\hline
\end{longtable}

      % \twocolumn}
    }
    \fi}\makeatother
}{}

\end{document}